\documentclass[onecolumn, longauth]{aa}
\pdfoutput=1

\usepackage{euclid}
\usepackage[T1]{fontenc}
\usepackage[utf8]{inputenc}
\usepackage{geometry}
\usepackage{float}
\usepackage[section]{placeins}
\usepackage{amsmath}
\usepackage{mathtools}
\usepackage[switch, modulo]{lineno}
\usepackage[usenames,dvipsnames]{xcolor} 
\usepackage[normalem]{ulem}
\usepackage{enumitem}
\usepackage{csquotes}
\usepackage{xspace}

\usepackage{listings}
\usepackage{color}
\definecolor{dkgreen}{rgb}{0,0.6,0}
\definecolor{gray}{gray}{0.9}
\definecolor{mauve}{rgb}{0.58,0,0.82}
\newcommand{\cosmicfish}{\texttt{CosmicFish}\xspace}

\newcommand{\montepython}{\texttt{MontePython}\xspace}

\newcommand{\class}{\texttt{CLASS}\xspace}
\newcommand{\camb}{\texttt{CAMB}\xspace}
\newcommand{\AP}{Alcock--Paczy\'nski\xspace}

\newcommand{\bcemu}{\texttt{BCemu}}
\newcommand{\GCsp}{\text{GC}\ensuremath{_\mathrm{sp}}\xspace}
\newcommand{\GCph}{\text{GC}\ensuremath{_\mathrm{ph}}\xspace}
\newcommand{\XCph}{\text{XC}\ensuremath{_\mathrm{ph}}\xspace}
\newcommand{\smnu}{\ensuremath{\sum m_{\nu}}}
\newcommand{\neff}{\ensuremath{N_{\rm eff}}}
\newcommand{\dneff}{\ensuremath{\Delta N_{\rm eff}}}

\newcommand{\de}{\mathrm{d}}

\defcitealias{Euclid:2019clj}{EC20}
\defcitealias{EUCLID:2023uep}{Casas et al. 2023}

\usepackage[colorlinks=true, allcolors=blue,breaklinks]{hyperref}
\usepackage[capitalise,nameinlink]{cleveref}
\usepackage{orcidlink}

\Crefname{section}{Section}{Sections}
\crefname{section}{Sect.}{Sects.}
\Crefname{figure}{Figure}{Figures}
\crefname{figure}{Fig.}{Figs.}

\usepackage{multirow}
\usepackage{tabularx}
\usepackage{colortbl}
\usepackage{graphicx}
\usepackage{footmisc}
\usepackage{amssymb}
\usepackage{amsfonts}
\usepackage{comment}
\usepackage{siunitx}
\bibpunct{(}{)}{;}{a}{}{,}
\makeatletter
\renewcommand*\aa@pageof{, page \thepage{} of \pageref*{LastPage}}
\makeatother

\begin{document}
\title{\Euclid preparation}
\subtitle{Sensitivity to neutrino parameters}    

\newcommand{\orcid}[1]{\orcidlink{#1}} 
\author{Euclid Collaboration: M.~Archidiacono\orcid{0000-0003-4952-9012}\thanks{\email{maria.archidiacono@unimi.it}}$^{1,2}$
\and J.~Lesgourgues\orcid{0000-0001-7627-353X}\thanks{\email{Julien.Lesgourgues@physik.rwth-aachen.de}}$^{3}$
\and S.~Casas\orcid{0000-0002-4751-5138}$^{3}$
\and S.~Pamuk\orcid{0009-0004-0852-8624}$^{3}$
\and N.~Sch\"oneberg\orcid{0000-0002-7873-0404}$^{4}$
\and Z.~Sakr\orcid{0000-0002-4823-3757}$^{5,6,7}$
\and G.~Parimbelli\orcid{0000-0002-2539-2472}$^{8,9,10}$
\and A.~Schneider\orcid{0000-0001-7055-8104}$^{11}$
\and F.~Hervas~Peters$^{12,11}$
\and F.~Pace\orcid{0000-0001-8039-0480}$^{13,14,15}$
% ...
}
										   
\institute{$^{1}$ Dipartimento di Fisica "Aldo Pontremoli", Universit\`a degli Studi di Milano, Via Celoria 16, 20133 Milano, Italy\label{aff1}
\\
$^{2}$ INFN-Sezione di Milano, Via Celoria 16, 20133 Milano, Italy\label{aff2}
\\
$^{3}$ Institute for Theoretical Particle Physics and Cosmology (TTK), RWTH Aachen University, 52056 Aachen, Germany\label{aff3}
\\
$^{4}$ Institut de Ci\`{e}ncies del Cosmos (ICCUB), Universitat de Barcelona (IEEC-UB), Mart\'{i} i Franqu\`{e}s 1, 08028 Barcelona, Spain\label{aff4}
\\
$^{5}$ Institut f\"ur Theoretische Physik, University of Heidelberg, Philosophenweg 16, 69120 Heidelberg, Germany\label{aff5}
\\
$^{6}$ Institut de Recherche en Astrophysique et Plan\'etologie (IRAP), Universit\'e de Toulouse, CNRS, UPS, CNES, 14 Av. Edouard Belin, 31400 Toulouse, France\label{aff6}
\\
$^{7}$ Universit\'e St Joseph; Faculty of Sciences, Beirut, Lebanon\label{aff7}
\\
$^{8}$ Institute of Space Sciences (ICE, CSIC), Campus UAB, Carrer de Can Magrans, s/n, 08193 Barcelona, Spain\label{aff8}
\\
$^{9}$ Dipartimento di Fisica, Universit\`a degli studi di Genova, and INFN-Sezione di Genova, via Dodecaneso 33, 16146, Genova, Italy\label{aff9}
\\
$^{10}$ SISSA, International School for Advanced Studies, Via Bonomea 265, 34136 Trieste TS, Italy\label{aff10}
\\
$^{11}$ Department of Astrophysics, University of Zurich, Winterthurerstrasse 190, 8057 Zurich, Switzerland\label{aff11}
\\
$^{12}$ AIM, CEA, CNRS, Universit\'{e} Paris-Saclay, Universit\'{e} de Paris, 91191 Gif-sur-Yvette, France\label{aff12}
\\
$^{13}$ Dipartimento di Fisica, Universit\`a degli Studi di Torino, Via P. Giuria 1, 10125 Torino, Italy\label{aff13}
\\
$^{14}$ INFN-Sezione di Torino, Via P. Giuria 1, 10125 Torino, Italy\label{aff14}
\\
$^{15}$ INAF-Osservatorio Astrofisico di Torino, Via Osservatorio 20, 10025 Pino Torinese (TO), Italy\label{aff15}
\\
$^{16}$ Hamburger Sternwarte, University of Hamburg, Gojenbergsweg 112, 21029 Hamburg, Germany\label{aff16}
\\
$^{17}$ Dipartimento di Fisica - Sezione di Astronomia, Universit\`a di Trieste, Via Tiepolo 11, 34131 Trieste, Italy\label{aff17}
\\
$^{18}$ INAF-Osservatorio Astronomico di Trieste, Via G. B. Tiepolo 11, 34143 Trieste, Italy\label{aff18}
\\
$^{19}$ IFPU, Institute for Fundamental Physics of the Universe, via Beirut 2, 34151 Trieste, Italy\label{aff19}
\\
$^{20}$ INAF-IASF Milano, Via Alfonso Corti 12, 20133 Milano, Italy\label{aff20}
\\
$^{21}$ Universit\'e Libre de Bruxelles (ULB), Service de Physique Th\'eorique CP225, Boulevard du Triophe, 1050 Bruxelles, Belgium\label{aff21}
\\
$^{22}$ Department of Physics "E. Pancini", University Federico II, Via Cinthia 6, 80126, Napoli, Italy\label{aff22}
\\
$^{23}$ Ludwig-Maximilians-University, Schellingstrasse 4, 80799 Munich, Germany\label{aff23}
\\
$^{24}$ INFN, Sezione di Trieste, Via Valerio 2, 34127 Trieste TS, Italy\label{aff24}
\\
$^{25}$ Department of Physics and Astronomy, University of British Columbia, Vancouver, BC V6T 1Z1, Canada\label{aff25}
\\
$^{26}$ ICSC - Centro Nazionale di Ricerca in High Performance Computing, Big Data e Quantum Computing, Via Magnanelli 2, Bologna, Italy\label{aff26}
\\
$^{27}$ School of Mathematics and Physics, University of Surrey, Guildford, Surrey, GU2 7XH, UK\label{aff27}
\\
$^{28}$ INAF-Osservatorio Astronomico di Brera, Via Brera 28, 20122 Milano, Italy\label{aff28}
\\
$^{29}$ INAF-Osservatorio di Astrofisica e Scienza dello Spazio di Bologna, Via Piero Gobetti 93/3, 40129 Bologna, Italy\label{aff29}
\\
$^{30}$ Dipartimento di Fisica e Astronomia, Universit\`a di Bologna, Via Gobetti 93/2, 40129 Bologna, Italy\label{aff30}
\\
$^{31}$ INFN-Sezione di Bologna, Viale Berti Pichat 6/2, 40127 Bologna, Italy\label{aff31}
\\
$^{32}$ Max Planck Institute for Extraterrestrial Physics, Giessenbachstr. 1, 85748 Garching, Germany\label{aff32}
\\
$^{33}$ Dipartimento di Fisica, Universit\`a di Genova, Via Dodecaneso 33, 16146, Genova, Italy\label{aff33}
\\
$^{34}$ INFN-Sezione di Genova, Via Dodecaneso 33, 16146, Genova, Italy\label{aff34}
\\
$^{35}$ INAF-Osservatorio Astronomico di Capodimonte, Via Moiariello 16, 80131 Napoli, Italy\label{aff35}
\\
$^{36}$ INFN section of Naples, Via Cinthia 6, 80126, Napoli, Italy\label{aff36}
\\
$^{37}$ Instituto de Astrof\'isica e Ci\^encias do Espa\c{c}o, Universidade do Porto, CAUP, Rua das Estrelas, PT4150-762 Porto, Portugal\label{aff37}
\\
$^{38}$ INAF-Osservatorio Astronomico di Roma, Via Frascati 33, 00078 Monteporzio Catone, Italy\label{aff38}
\\
$^{39}$ INFN-Sezione di Roma, Piazzale Aldo Moro, 2 - c/o Dipartimento di Fisica, Edificio G. Marconi, 00185 Roma, Italy\label{aff39}
\\
$^{40}$ Institut de F\'{i}sica d'Altes Energies (IFAE), The Barcelona Institute of Science and Technology, Campus UAB, 08193 Bellaterra (Barcelona), Spain\label{aff40}
\\
$^{41}$ Port d'Informaci\'{o} Cient\'{i}fica, Campus UAB, C. Albareda s/n, 08193 Bellaterra (Barcelona), Spain\label{aff41}
\\
$^{42}$ Dipartimento di Fisica e Astronomia "Augusto Righi" - Alma Mater Studiorum Universit\`a di Bologna, Viale Berti Pichat 6/2, 40127 Bologna, Italy\label{aff42}
\\
$^{43}$ Institute for Astronomy, University of Edinburgh, Royal Observatory, Blackford Hill, Edinburgh EH9 3HJ, UK\label{aff43}
\\
$^{44}$ Jodrell Bank Centre for Astrophysics, Department of Physics and Astronomy, University of Manchester, Oxford Road, Manchester M13 9PL, UK\label{aff44}
\\
$^{45}$ European Space Agency/ESRIN, Largo Galileo Galilei 1, 00044 Frascati, Roma, Italy\label{aff45}
\\
$^{46}$ ESAC/ESA, Camino Bajo del Castillo, s/n., Urb. Villafranca del Castillo, 28692 Villanueva de la Ca\~nada, Madrid, Spain\label{aff46}
\\
$^{47}$ Universit\'e Claude Bernard Lyon 1, CNRS/IN2P3, IP2I Lyon, UMR 5822, Villeurbanne, F-69100, France\label{aff47}
\\
$^{48}$ Institute of Physics, Laboratory of Astrophysics, Ecole Polytechnique F\'ed\'erale de Lausanne (EPFL), Observatoire de Sauverny, 1290 Versoix, Switzerland\label{aff48}
\\
$^{49}$ UCB Lyon 1, CNRS/IN2P3, IUF, IP2I Lyon, 4 rue Enrico Fermi, 69622 Villeurbanne, France\label{aff49}
\\
$^{50}$ Departamento de F\'isica, Faculdade de Ci\^encias, Universidade de Lisboa, Edif\'icio C8, Campo Grande, PT1749-016 Lisboa, Portugal\label{aff50}
\\
$^{51}$ Instituto de Astrof\'isica e Ci\^encias do Espa\c{c}o, Faculdade de Ci\^encias, Universidade de Lisboa, Campo Grande, 1749-016 Lisboa, Portugal\label{aff51}
\\
$^{52}$ Department of Astronomy, University of Geneva, ch. d'Ecogia 16, 1290 Versoix, Switzerland\label{aff52}
\\
$^{53}$ Universit\'e Paris-Saclay, CNRS, Institut d'astrophysique spatiale, 91405, Orsay, France\label{aff53}
\\
$^{54}$ Department of Physics, Oxford University, Keble Road, Oxford OX1 3RH, UK\label{aff54}
\\
$^{55}$ INFN-Padova, Via Marzolo 8, 35131 Padova, Italy\label{aff55}
\\
$^{56}$ INAF-Istituto di Astrofisica e Planetologia Spaziali, via del Fosso del Cavaliere, 100, 00100 Roma, Italy\label{aff56}
\\
$^{57}$ Universit\'e Paris-Saclay, Universit\'e Paris Cit\'e, CEA, CNRS, AIM, 91191, Gif-sur-Yvette, France\label{aff57}
\\
$^{58}$ Istituto Nazionale di Fisica Nucleare, Sezione di Bologna, Via Irnerio 46, 40126 Bologna, Italy\label{aff58}
\\
$^{59}$ INAF-Osservatorio Astronomico di Padova, Via dell'Osservatorio 5, 35122 Padova, Italy\label{aff59}
\\
$^{60}$ Universit\"ats-Sternwarte M\"unchen, Fakult\"at f\"ur Physik, Ludwig-Maximilians-Universit\"at M\"unchen, Scheinerstrasse 1, 81679 M\"unchen, Germany\label{aff60}
\\
$^{61}$ Institute of Theoretical Astrophysics, University of Oslo, P.O. Box 1029 Blindern, 0315 Oslo, Norway\label{aff61}
\\
$^{62}$ Leiden Observatory, Leiden University, Einsteinweg 55, 2333 CC Leiden, The Netherlands\label{aff62}
\\
$^{63}$ von Hoerner \& Sulger GmbH, Schlossplatz 8, 68723 Schwetzingen, Germany\label{aff63}
\\
$^{64}$ Technical University of Denmark, Elektrovej 327, 2800 Kgs. Lyngby, Denmark\label{aff64}
\\
$^{65}$ Cosmic Dawn Center (DAWN), Denmark\label{aff65}
\\
$^{66}$ Max-Planck-Institut f\"ur Astronomie, K\"onigstuhl 17, 69117 Heidelberg, Germany\label{aff66}
\\
$^{67}$ Department of Physics and Astronomy, University College London, Gower Street, London WC1E 6BT, UK\label{aff67}
\\
$^{68}$ Department of Physics and Helsinki Institute of Physics, Gustaf H\"allstr\"omin katu 2, 00014 University of Helsinki, Finland\label{aff68}
\\
$^{69}$ Aix-Marseille Universit\'e, CNRS/IN2P3, CPPM, Marseille, France\label{aff69}
\\
$^{70}$ Jet Propulsion Laboratory, California Institute of Technology, 4800 Oak Grove Drive, Pasadena, CA, 91109, USA\label{aff70}
\\
$^{71}$ Mullard Space Science Laboratory, University College London, Holmbury St Mary, Dorking, Surrey RH5 6NT, UK\label{aff71}
\\
$^{72}$ Universit\'e de Gen\`eve, D\'epartement de Physique Th\'eorique and Centre for Astroparticle Physics, 24 quai Ernest-Ansermet, CH-1211 Gen\`eve 4, Switzerland\label{aff72}
\\
$^{73}$ Department of Physics, P.O. Box 64, 00014 University of Helsinki, Finland\label{aff73}
\\
$^{74}$ Helsinki Institute of Physics, Gustaf H{\"a}llstr{\"o}min katu 2, University of Helsinki, Helsinki, Finland\label{aff74}
\\
$^{75}$ NOVA optical infrared instrumentation group at ASTRON, Oude Hoogeveensedijk 4, 7991PD, Dwingeloo, The Netherlands\label{aff75}
\\
$^{76}$ Universit\"at Bonn, Argelander-Institut f\"ur Astronomie, Auf dem H\"ugel 71, 53121 Bonn, Germany\label{aff76}
\\
$^{77}$ Aix-Marseille Universit\'e, CNRS, CNES, LAM, Marseille, France\label{aff77}
\\
$^{78}$ Dipartimento di Fisica e Astronomia "Augusto Righi" - Alma Mater Studiorum Universit\`a di Bologna, via Piero Gobetti 93/2, 40129 Bologna, Italy\label{aff78}
\\
$^{79}$ Department of Physics, Institute for Computational Cosmology, Durham University, South Road, DH1 3LE, UK\label{aff79}
\\
$^{80}$ Universit\'e C\^{o}te d'Azur, Observatoire de la C\^{o}te d'Azur, CNRS, Laboratoire Lagrange, Bd de l'Observatoire, CS 34229, 06304 Nice cedex 4, France\label{aff80}
\\
$^{81}$ Institut d'Astrophysique de Paris, UMR 7095, CNRS, and Sorbonne Universit\'e, 98 bis boulevard Arago, 75014 Paris, France\label{aff81}
\\
$^{82}$ Universit\'e Paris Cit\'e, CNRS, Astroparticule et Cosmologie, 75013 Paris, France\label{aff82}
\\
$^{83}$ Institut d'Astrophysique de Paris, 98bis Boulevard Arago, 75014, Paris, France\label{aff83}
\\
$^{84}$ European Space Agency/ESTEC, Keplerlaan 1, 2201 AZ Noordwijk, The Netherlands\label{aff84}
\\
$^{85}$ School of Mathematics, Statistics and Physics, Newcastle University, Herschel Building, Newcastle-upon-Tyne, NE1 7RU, UK\label{aff85}
\\
$^{86}$ Department of Physics and Astronomy, University of Aarhus, Ny Munkegade 120, DK-8000 Aarhus C, Denmark\label{aff86}
\\
$^{87}$ Waterloo Centre for Astrophysics, University of Waterloo, Waterloo, Ontario N2L 3G1, Canada\label{aff87}
\\
$^{88}$ Department of Physics and Astronomy, University of Waterloo, Waterloo, Ontario N2L 3G1, Canada\label{aff88}
\\
$^{89}$ Perimeter Institute for Theoretical Physics, Waterloo, Ontario N2L 2Y5, Canada\label{aff89}
\\
$^{90}$ Universit\'e Paris-Saclay, Universit\'e Paris Cit\'e, CEA, CNRS, Astrophysique, Instrumentation et Mod\'elisation Paris-Saclay, 91191 Gif-sur-Yvette, France\label{aff90}
\\
$^{91}$ Space Science Data Center, Italian Space Agency, via del Politecnico snc, 00133 Roma, Italy\label{aff91}
\\
$^{92}$ Centre National d'Etudes Spatiales -- Centre spatial de Toulouse, 18 avenue Edouard Belin, 31401 Toulouse Cedex 9, France\label{aff92}
\\
$^{93}$ Institute of Space Science, Str. Atomistilor, nr. 409 M\u{a}gurele, Ilfov, 077125, Romania\label{aff93}
\\
$^{94}$ Instituto de Astrof\'isica de Canarias, Calle V\'ia L\'actea s/n, 38204, San Crist\'obal de La Laguna, Tenerife, Spain\label{aff94}
\\
$^{95}$ Departamento de Astrof\'isica, Universidad de La Laguna, 38206, La Laguna, Tenerife, Spain\label{aff95}
\\
$^{96}$ Dipartimento di Fisica e Astronomia "G. Galilei", Universit\`a di Padova, Via Marzolo 8, 35131 Padova, Italy\label{aff96}
\\
$^{97}$ Departamento de F\'isica, FCFM, Universidad de Chile, Blanco Encalada 2008, Santiago, Chile\label{aff97}
\\
$^{98}$ Universit\"at Innsbruck, Institut f\"ur Astro- und Teilchenphysik, Technikerstr. 25/8, 6020 Innsbruck, Austria\label{aff98}
\\
$^{99}$ Institut d'Estudis Espacials de Catalunya (IEEC),  Edifici RDIT, Campus UPC, 08860 Castelldefels, Barcelona, Spain\label{aff99}
\\
$^{100}$ Satlantis, University Science Park, Sede Bld 48940, Leioa-Bilbao, Spain\label{aff100}
\\
$^{101}$ Centro de Investigaciones Energ\'eticas, Medioambientales y Tecnol\'ogicas (CIEMAT), Avenida Complutense 40, 28040 Madrid, Spain\label{aff101}
\\
$^{102}$ Instituto de Astrof\'isica e Ci\^encias do Espa\c{c}o, Faculdade de Ci\^encias, Universidade de Lisboa, Tapada da Ajuda, 1349-018 Lisboa, Portugal\label{aff102}
\\
$^{103}$ Universidad Polit\'ecnica de Cartagena, Departamento de Electr\'onica y Tecnolog\'ia de Computadoras,  Plaza del Hospital 1, 30202 Cartagena, $^{}$ Spain\label{aff103}
\\
$^{104}$ INFN-Bologna, Via Irnerio 46, 40126 Bologna, Italy\label{aff104}
\\
$^{105}$ Infrared Processing and Analysis Center, California Institute of Technology, Pasadena, CA 91125, USA\label{aff105}
\\
$^{106}$ INAF, Istituto di Radioastronomia, Via Piero Gobetti 101, 40129 Bologna, Italy\label{aff106}
\\
$^{107}$ Astronomical Observatory of the Autonomous Region of the Aosta Valley (OAVdA), Loc. Lignan 39, I-11020, Nus (Aosta Valley), Italy\label{aff107}
\\
$^{108}$ Institut de Ciencies de l'Espai (IEEC-CSIC), Campus UAB, Carrer de Can Magrans, s/n Cerdanyola del Vall\'es, 08193 Barcelona, Spain\label{aff108}
\\
$^{109}$ Center for Computational Astrophysics, Flatiron Institute, 162 5th Avenue, 10010, New York, NY, USA\label{aff109}
\\
$^{110}$ School of Physics and Astronomy, Cardiff University, The Parade, Cardiff, CF24 3AA, UK\label{aff110}
\\
$^{111}$ Centre de Calcul de l'IN2P3/CNRS, 21 avenue Pierre de Coubertin 69627 Villeurbanne Cedex, France\label{aff111}
\\
$^{112}$ Junia, EPA department, 41 Bd Vauban, 59800 Lille, France\label{aff112}
\\
$^{113}$ Instituto de F\'isica Te\'orica UAM-CSIC, Campus de Cantoblanco, 28049 Madrid, Spain\label{aff113}
\\
$^{114}$ CERCA/ISO, Department of Physics, Case Western Reserve University, 10900 Euclid Avenue, Cleveland, OH 44106, USA\label{aff114}
\\
$^{115}$ Laboratoire Univers et Th\'eorie, Observatoire de Paris, Universit\'e PSL, Universit\'e Paris Cit\'e, CNRS, 92190 Meudon, France\label{aff115}
\\
$^{116}$ Dipartimento di Fisica e Scienze della Terra, Universit\`a degli Studi di Ferrara, Via Giuseppe Saragat 1, 44122 Ferrara, Italy\label{aff116}
\\
$^{117}$ Istituto Nazionale di Fisica Nucleare, Sezione di Ferrara, Via Giuseppe Saragat 1, 44122 Ferrara, Italy\label{aff117}
\\
$^{118}$ Institut de Physique Th\'eorique, CEA, CNRS, Universit\'e Paris-Saclay 91191 Gif-sur-Yvette Cedex, France\label{aff118}
\\
$^{119}$ Minnesota Institute for Astrophysics, University of Minnesota, 116 Church St SE, Minneapolis, MN 55455, USA\label{aff119}
\\
$^{120}$ Institute Lorentz, Leiden University, Niels Bohrweg 2, 2333 CA Leiden, The Netherlands\label{aff120}
\\
$^{121}$ Institute for Astronomy, University of Hawaii, 2680 Woodlawn Drive, Honolulu, HI 96822, USA\label{aff121}
\\
$^{122}$ Department of Physics \& Astronomy, University of California Irvine, Irvine CA 92697, USA\label{aff122}
\\
$^{123}$ Department of Astronomy \& Physics and Institute for Computational Astrophysics, Saint Mary's University, 923 Robie Street, Halifax, Nova Scotia, B3H 3C3, Canada\label{aff123}
\\
$^{124}$ Departamento F\'isica Aplicada, Universidad Polit\'ecnica de Cartagena, Campus Muralla del Mar, 30202 Cartagena, Murcia, Spain\label{aff124}
\\
$^{125}$ Institute of Cosmology and Gravitation, University of Portsmouth, Portsmouth PO1 3FX, UK\label{aff125}
\\
$^{126}$ Department of Computer Science, Aalto University, PO Box 15400, Espoo, FI-00 076, Finland\label{aff126}
\\
$^{127}$ Ruhr University Bochum, Faculty of Physics and Astronomy, Astronomical Institute (AIRUB), German Centre for Cosmological Lensing (GCCL), 44780 Bochum, Germany\label{aff127}
\\
$^{128}$ Universit\'e Paris-Saclay, CNRS/IN2P3, IJCLab, 91405 Orsay, France\label{aff128}
\\
$^{129}$ Department of Physics and Astronomy, Vesilinnantie 5, 20014 University of Turku, Finland\label{aff129}
\\
$^{130}$ Serco for European Space Agency (ESA), Camino bajo del Castillo, s/n, Urbanizacion Villafranca del Castillo, Villanueva de la Ca\~nada, 28692 Madrid, Spain\label{aff130}
\\
$^{131}$ ARC Centre of Excellence for Dark Matter Particle Physics, Melbourne, Australia\label{aff131}
\\
$^{132}$ Centre for Astrophysics \& Supercomputing, Swinburne University of Technology, Victoria 3122, Australia\label{aff132}
\\
$^{133}$ W.M. Keck Observatory, 65-1120 Mamalahoa Hwy, Kamuela, HI, USA\label{aff133}
\\
$^{134}$ School of Physics and Astronomy, Queen Mary University of London, Mile End Road, London E1 4NS, UK\label{aff134}
\\
$^{135}$ Department of Physics and Astronomy, University of the Western Cape, Bellville, Cape Town, 7535, South Africa\label{aff135}
\\
$^{136}$ ICTP South American Institute for Fundamental Research, Instituto de F\'{\i}sica Te\'orica, Universidade Estadual Paulista, S\~ao Paulo, Brazil\label{aff136}
\\
$^{137}$ Oskar Klein Centre for Cosmoparticle Physics, Department of Physics, Stockholm University, Stockholm, SE-106 91, Sweden\label{aff137}
\\
$^{138}$ Astrophysics Group, Blackett Laboratory, Imperial College London, London SW7 2AZ, UK\label{aff138}
\\
$^{139}$ Univ. Grenoble Alpes, CNRS, Grenoble INP, LPSC-IN2P3, 53, Avenue des Martyrs, 38000, Grenoble, France\label{aff139}
\\
$^{140}$ INAF-Osservatorio Astrofisico di Arcetri, Largo E. Fermi 5, 50125, Firenze, Italy\label{aff140}
\\
$^{141}$ Dipartimento di Fisica, Sapienza Universit\`a di Roma, Piazzale Aldo Moro 2, 00185 Roma, Italy\label{aff141}
\\
$^{142}$ Centro de Astrof\'{\i}sica da Universidade do Porto, Rua das Estrelas, 4150-762 Porto, Portugal\label{aff142}
\\
$^{143}$ Dipartimento di Fisica, Universit\`a di Roma Tor Vergata, Via della Ricerca Scientifica 1, Roma, Italy\label{aff143}
\\
$^{144}$ INFN, Sezione di Roma 2, Via della Ricerca Scientifica 1, Roma, Italy\label{aff144}
\\
$^{145}$ Institute of Astronomy, University of Cambridge, Madingley Road, Cambridge CB3 0HA, UK\label{aff145}
\\
$^{146}$ Theoretical astrophysics, Department of Physics and Astronomy, Uppsala University, Box 515, 751 20 Uppsala, Sweden\label{aff146}
\\
$^{147}$ Department of Astrophysical Sciences, Peyton Hall, Princeton University, Princeton, NJ 08544, USA\label{aff147}
\\
$^{148}$ Niels Bohr Institute, University of Copenhagen, Jagtvej 128, 2200 Copenhagen, Denmark\label{aff148}
\\
$^{149}$ Cosmic Dawn Center (DAWN)\label{aff149}
\\
$^{150}$ Center for Cosmology and Particle Physics, Department of Physics, New York University, New York, NY 10003, USA\label{aff150}}    

\date{}
\authorrunning{M.~Archidiacono et al.}
\abstract
{The \Euclid mission of the European Space Agency will deliver weak gravitational lensing and galaxy clustering surveys that can be used to constrain the standard cosmological model and extensions thereof.
}
{We present forecasts from the combination of the \Euclid photometric galaxy surveys (weak lensing, galaxy clustering and their cross-correlation) and its spectroscopic redshift survey on the sensitivity to cosmological parameters including the summed neutrino mass $\smnu$ and the effective number of relativistic species $\neff$ in the standard $\Lambda$CDM scenario and in a scenario with dynamical dark energy ($w_0 w_a$CDM).
}
{We compare the accuracy of different algorithms predicting the nonlinear matter power spectrum for such models. We then validate several pipelines for Fisher matrix and Markov Chain Monte Carlo (MCMC) forecasts, using different theory codes, algorithms for numerical derivatives, and assumptions concerning the non-linear cut-off scale.
}
{The \Euclid primary probes alone will reach a sensitivity of $\sigma(\smnu=60\,\mathrm{meV})=56\,\mathrm{meV}$ in the $\Lambda$CDM+$\smnu$ model, whereas the combination with cosmic microwave background (CMB) data from \Planck is expected to achieve $\sigma(\smnu)=23\,\mathrm{meV}$ and raise the evidence for a non-zero neutrino mass to at least the $2.6\,\sigma$ level. This can be pushed to a $4\,\sigma$ detection if future CMB data from LiteBIRD and CMB Stage-IV are included. In combination with \Planck, \Euclid will also deliver tight constraints on $\dneff< 0.144$ (95\%CL) in the $\Lambda$CDM+$\smnu$+$\neff$ model, or $\dneff< 0.063$ when future CMB data are included.
When floating the dark energy parameters, we find that the sensitivity to $\neff$ remains stable, while that to $\smnu$ degrades at most by a factor two.
}
{This work illustrates the complementarity between the \Euclid spectroscopic and imaging/photometric surveys and between \Euclid and CMB constraints. \Euclid will have a great potential for measuring the neutrino mass and excluding well-motivated scenarios with additional relativistic particles. 
}

\maketitle

\section{Introduction}
Over the last two decades, the continuous improvement of the precision and accuracy of cosmological observations has opened a window to constrain neutrino properties via cosmological data. In particular, the sum of the three neutrino masses can be well constrained due to its effect of suppressing the clustering of cold dark matter after the neutrino non-relativistic transition.

Neutrino oscillation experiments demonstrated that at least two neutrinos are massive by measuring two squared mass differences $\Delta m^2_{ij}=m_i^2-m_j^2$:
\begin{equation}
\begin{split}
\Delta m^2_{21} &=\left(7.42^{+0.21}_{-0.20}\right)\times 10^{-5} \,{\rm eV}^2 \,, \\
|\Delta m^2_{3i}|&= \left\{
\begin{array}{ll}
\left(2.517^{+0.026}_{-0.028}\right)\times 10^{-3}\, {\rm eV}^2 &\quad \rm (NO) \,, \\
\left(- 2.498^{+0.028}_{-0.028}\right)\times 10^{-3}\, {\rm eV}^2 &\quad \rm (IO) \,,
\end{array}
\right.
\end{split}
\label{eq:split}
\end{equation}
where the sign of $\Delta m_{3i}^2$ depends on the mass ordering, which can be either normal (NO: $m_1<m_2<m_3$, $\Delta m_{3i}^2=\Delta m_{31}^2>0$) or inverted (IO: $m_3<m_1<m_2$, $\Delta m_{3i}^2=\Delta m_{32}^2<0$), see \cite{Esteban:2020cvm, ParticleDataGroup:2022pth2}. According to these results, the minimum value of the neutrino mass sum is either $0.06\,{\rm eV}$\footnote{Hereafter for neutrino masses we assume natural units and report the values in ${\rm eV}$, rather than in ${\rm eV}/c^2$.} in NO or $0.10\,{\rm eV}$ in IO. However, neutrino oscillation experiments cannot constrain the absolute neutrino mass sum. On the other hand, $\beta$-decay experiments are sensitive to the effective electron anti-neutrino mass. Recently, the Karlsruhe Tritium Neutrino Experiment (KATRIN) constrained $m_{\nu} < 0.8\,{\rm eV}$ \citep[90\% CL, see][]{KATRIN:2021uub}. Future $\beta$-decay experiments, such as for example Project 8 \citep{Project8:2022wqh}, can potentially reach a sensitivity of $40\,{\rm meV}$.

Despite the great progress in the precision of $\beta$-decay experiments,\footnote{We note that, if neutrinos are Majorana particles, their mass can also be constrained by means of neutrinoless double-$\beta$-decay experiments \citep{Cirigliano:2022oqy}; however, there is no evidence for lepton number violation yet.} cosmology provides the most stringent -- though model-dependent -- constraints to date on the absolute neutrino mass scale. The \Planck Collaboration reported an upper bound of \mbox{$\smnu<0.24\,{\rm eV}$ (95\% CL)} by combining temperature, polarisation, and lensing of the cosmic microwave background (CMB); adding external baryon acoustic oscillation (BAO) data further improves the bound to $\smnu<0.12\,{\rm eV}$ \citep[95\% CL, see][]{Planck:2018vyg}. Including Pantheon type Ia supernovae, baryon acoustic oscillations and redshift-space distortions from the Sloan Digital Sky Survey, and weak lensing measurements from the Dark Energy Survey leads to an upper bound of $0.11\,{\rm eV}$ \citep{eBOSS:2020yzd}.
Recently, the DESI Collaboration reported an upper limit of $\smnu<0.072 \, {\rm eV}$ at the 95\% CL from the combination of DESI BAO and CMB data. This bound mildly favours the NO scenario and shows that there might be a possibility of indirectly constraining the neutrino mass hierarchy using cosmological data.
Current data on the full-shape galaxy power spectrum do not yet improve the constraint on $\smnu$ compared to the combination of CMB and the BAO geometric information \citep{Ivanov:2019hqk}, but future galaxy surveys are expected to improve it. Finally, it is important to reiterate that changing the underlying cosmological model can potentially relax the constraints obtained from many of the cosmological probes \citep{Lambiase:2018ows, diValentino:2022njd}.

The ability of cosmology to constrain the neutrino mass paved the way to investigate, via cosmological data, additional neutrino properties, foremost the number density of neutrinos. The standard model of particles physics predicts three active neutrino species, as confirmed by accelerator experiments \citep{Mele:2015etc}. In cosmology, this would correspond to an effective neutrino number $\neff=3$ in the instantaneous decoupling limit, or, sticking to the same definition of this effective number, to $\neff=3.044$ when neutrino decoupling is modelled accurately
\citep{Froustey:2020mcq,Akita:2020szl,Bennett:2020zkv}. Any deviation from this value is a hint at non-standard physics in the neutrino sector \citep[such as low-temperature reheating, non-thermal corrections to the neutrino distribution generated by new physics, non-standard neutrino interactions or sterile neutrinos,][]{Dvorkin:2022jyg, Archidiacono:2022ich, Dasgupta:2021ies} or exotic radiation components \citep[such as axions or other forms of dark radiation,][]{Marsh:2015xka, Kawasaki:2013ae, DiLuzio:2020wdo}.

The continuous improvement of cosmological constraints on neutrino properties opened the question of whether upcoming cosmological surveys will be able to deliver a detection of a non-zero neutrino mass, to distinguish between the IO and NO scenarios due to a tight upper limit, or to investigate additional neutrino properties. One of the primary science goals of \Euclid is to improve the cosmological constraints on the neutrino mass, possibly delivering evidence for a non-zero value. 
The aim of this work is to assess the sensitivity of \Euclid to the neutrino mass, as well as its robustness against variations of the number of neutrinos or the modelling of dark energy.

\Euclid \citep{EUCLID:2011zbd, Euclid:2021icp} is a medium-class mission of the European Space Agency, which will map the local Universe to improve our understanding of the expansion history and of the growth of structures. The satellite will observe roughly $15\,000\,{\rm deg^2}$ of the sky through two instruments, a visible imager \citep[VIS,][]{2016SPIE.9904E..0QC} and Near-Infrared Spectrometer and Photometer \citep[NISP,][]{2022SPIE12180E..1KM}, delivering the images of more than one billion galaxies and the spectra of between 20 and 30 million galaxies out to redshift of about $2$. The combination of spectroscopy and photometry will allow us to measure galaxy clustering and weak gravitational lensing, aiming at a $1\%$ accuracy in the corresponding power spectra.

The paper is organised as follows. In \cref{sec:nu-cosmo} we review the main effects of massive neutrinos in cosmology, and we discuss their impact on the theoretical predictions for \Euclid observables in \cref{sec:euclid-observables}. In \cref{sec:add_prob} we introduce additional probes like cluster number counts from \Euclid itself and external CMB data, which are potentially very synergistic with \Euclid primary (weak lensing and galaxy clustering) probes. The methodology used to perform the forecast is described in \cref{sec:method}, the code is validated in \cref{sec:validation}, and the results are presented in \cref{sec:results_euclid}. Finally, we draw our conclusions on the sensitivity of \Euclid to neutrino physics in \cref{sec:conclusions}.
Along the main text, we point the reader to several Appendices~\ref{app:der}, \ref{app:EBS}, \ref{app:step}, \ref{app:halofit}, \ref{app:cb}, \ref{app:res}, \ref{app:nuisance}, \ref{app:bf},
providing technical details on the analysis, as well as additional figures and consistency checks.

\section{Neutrino cosmology\label{sec:nu-cosmo}}

In the following, we briefly review the main effects of the neutrino masses and number density on structure formation. For a thorough description, we refer the readers to \cite{Bashinsky:2003tk}, \cite{Hannestad:2006zg}, \cite{Lesgourgues:2006nd}, \cite{Lesgourgues:2012uu}, \cite{Lesgourgues:2013sjj}, \cite{Lattanzi:2017ubx}, \cite{Archidiacono:2016lnv}, \cite{Archidiacono:2020dvx} and \cite{ParticleDataGroup:2022pth}.

\subsection{Massive neutrinos\label{sec:mnu}}
Massive neutrinos play an important role in the distribution of large-scale structures. They decouple from the primordial plasma around $T \sim 1\,{\rm MeV}$, while still being relativistic. The redshift $z_{\rm nr}$ at which they enter the non-relativistic regime depends on the individual mass $m_\nu$ of each neutrino as $(1+z_{\rm nr}) \sim 2 \times 10^3 (m_\nu / 1 \, {\rm eV})$.  After decoupling, neutrinos travel with a thermal velocity $v_{\rm th}$ that defines the neutrino free-streaming wavenumber \citep{Lesgourgues:2006nd}
\begin{equation}
 k_{\rm FS} = \left(\frac{4\pi G_{\sfont{N}}\,\bar{\rho}\,a^2}{v_{\rm th}^2}\right)^{1/2}\,,
\end{equation}
where $\bar{\rho}$ is the total background density, $a$ is the scale factor, and $G_{\sfont{N}}$ is the Newton constant.
After the non-relativistic transition the thermal velocity decays as $v_{\rm th} = \langle p \rangle \, m_\nu^{-1} \propto a^{-1}$, such that the free-streaming wavenumber goes through a minimum value $k_{\rm nr}=k_{\rm FS}(z=z_{\rm nr}) \sim {\cal O}(10^{-2})\, h\, {\rm Mpc}^{-1}$ at the time of the non-relativistic transition, where $h=H_0/(100\,\kmsMpc)$ is the reduced Hubble parameter. Neutrinos cannot cluster in regions smaller than their free-streaming length ($\lambda_{\rm FS} = 2\pi a/k_{\rm FS}$). On scales larger than the maximum free-streaming length, corresponding to $k < k_{\rm nr}$, massive neutrinos always behave as cold dark matter, and the power spectrum of matter density fluctuations $P_{\rm mm}(k)$ does not change with respect to the one of a cosmology in which the neutrinos are massless, but their non-relativistic density is added to that of cold dark matter. On the other hand, on scales smaller than the current value of the free-streaming length ($k > k_{\rm FS}^0$), massive neutrinos induce a suppression of the linear matter spectrum $P_{\rm mm}(k)$ at redshift $z=0$ by approximately $\Delta P_{\rm mm}(k)/P_{\rm mm}(k) \simeq -8f_{\nu}$, where $f_{\nu} = \Omega_{\nu,0}/\Omega_{{\rm m},0}$ is the fraction of neutrino with respect to matter density \citep{Hu:1997mj,Lesgourgues:2013sjj}. The reason is twofold: on scales where neutrinos are free-streaming, they do not contribute to gravitational clustering; and they slow down the growth of cold dark matter perturbations, which is given by $a^{1-3f_{\nu}/5}$ during matter domination \citep{Bond:1980ha}.
The second effect is responsible for the majority of the overall suppression, and the scale-dependence of the linear growth factor induced by massive neutrinos must be taken into account in the modelling of large-scale structure observables (see \cref{sec:euclid-observables}).

An important consideration is that the suppression of the power spectrum (as well as the effect of neutrino masses on the CMB spectrum) is almost the same independently of which neutrino ordering (IO or NO) is present in nature. This has been shown for example in \cite{Lesgourgues:2004ps}, \cite{Lesgourgues:2006nd} and \cite{Archidiacono:2020dvx}. Indeed, the cosmological observables are mainly sensitive to $f_\nu$, that is, to $\Omega_{\nu,0}\, h^2 \approx \smnu/(93.12\,\mathrm{eV})$, and thus, to the summed neutrino mass $\smnu$ \cite[see][]{Mangano:2005cc}.\footnote{However, the effect of the ordering is slightly enhanced by the nonlinear evolution \citep{Wagner:2012sw} and the detection might become possible in the future, especially with upcoming line-intensity mapping surveys, as suggested in \cite{Pritchard:2008wy} and \cite{Bernal:2021ylz}.}

\begin{figure}
    \centering
    \includegraphics[width=0.48\textwidth]{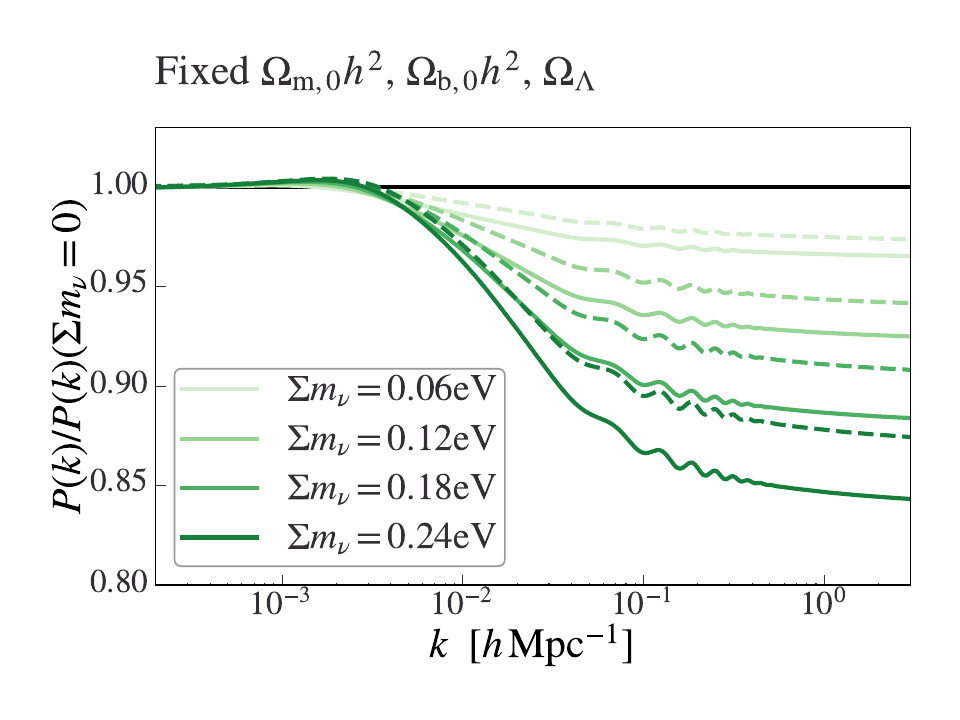}
    \includegraphics[width=0.48\textwidth]{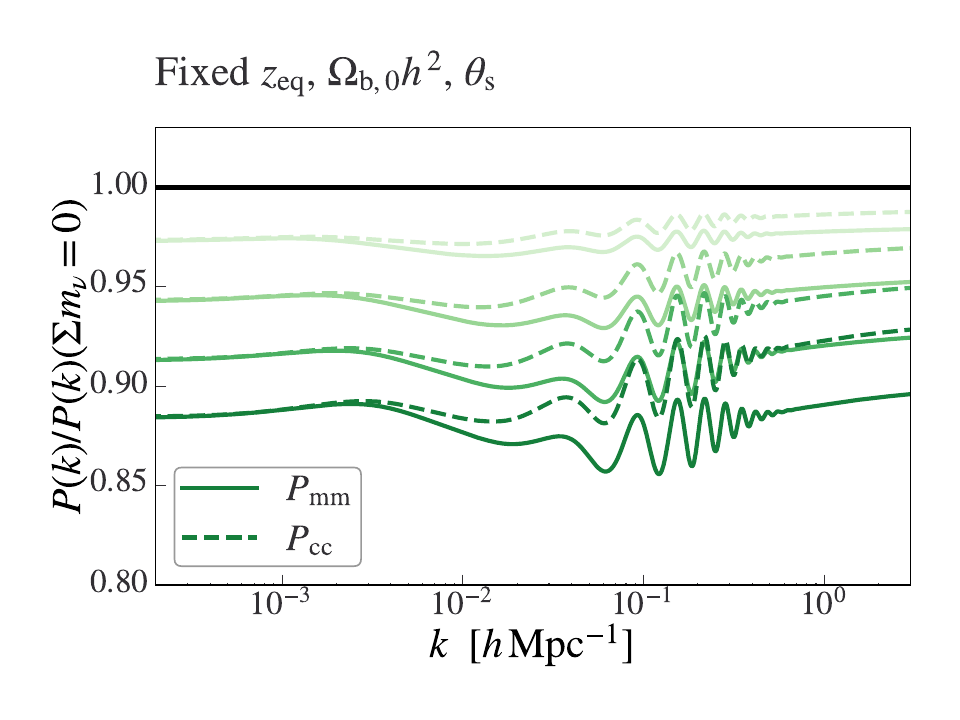}
    \caption{Effects of the neutrino mass on the linear total matter (solid) or cold-plus-baryonic matter (dashed) power spectrum, presented as ratios with respect to the $\Lambda$CDM spectrum with massless neutrinos for four different values of the summed neutrino mass. \emph{Left}: to single out the effect of neutrino free-streaming, we keep the parameters $\{ \Omega_{\mathrm{m},0} h^2, \Omega_{\mathrm{b},0}h^2, \Omega_\Lambda \}$ fixed. \emph{Right}: to show what is left for \Euclid to measure, we fix the quantities best constrained by CMB data, that is, $\{z_{\rm eq}, \Omega_{\mathrm{b},0} h^2, \theta_\mathrm{s}\}$. The mass is always equally split between the three neutrino species.}
    \label{fig:Pk_Mnu_linear_suppression}
\end{figure}

The step-like suppression induced by neutrino free streaming is best seen when varying \smnu{} while fixing the density of total matter, the density of baryonic matter, and the fractional density of the cosmological constant, that is, the parameters $\{\Omega_{{\rm m},0}h^2, \Omega_{{\rm b},0}h^2, \Omega_\Lambda\}$. This is illustrated in the left panel of \cref{fig:Pk_Mnu_linear_suppression}.
However, such a transformation also changes characteristic times and scales that are strongly constrained by CMB experiments, such as the redshift of equality between radiation and matter or the angular diameter distance to the last-scattering surface. In order to understand what is left for experiments like \Euclid to measure given our knowledge of the CMB spectrum, it makes more sense to look at the variation of the matter power spectrum when floating \smnu{} while fixing the redshift of equality, the baryonic matter density, and the angular scale of the sound horizon at decoupling \citep{Archidiacono:2016lnv,ParticleDataGroup:2022pth}, that is, fixing the parameters $\{\Omega_{{\rm c},0}h^2, \Omega_{{\rm b},0}h^2, \theta_{\rm s}\}$, where $\Omega_{{\rm c},0}$ accounts for the fractional density of cold dark matter only. Note that neutrinos with a realistic mass were still relativistic at the time of radiation-to-matter equality, such that for a fixed CMB temperature and effective neutrino number the redshift of equality is fixed by $(\Omega_{{\rm c},0} + \Omega_{{\rm b},0})h^2$ -- and not by $\Omega_{{\rm m},0} h^2=(\Omega_{{\rm b},0}+\Omega_{{\rm c},0}+\Omega_{\nu,0})h^2$.
In that case, the effect of the neutrino mass is displayed in the right panel of \cref{fig:Pk_Mnu_linear_suppression}. The increase of \smnu{} with a fixed $\theta_{\rm s}$ implies a decrease of $h$ that suppresses the large-scale power spectrum roughly by the same amount as neutrino free-streaming suppresses the small-scale power spectrum. As a result, in a combined fit to \Euclid data and CMB data from (for instance) \Planck, the neutrino mass would manifest itself mainly as an overall decrease of the amplitude of the matter power spectrum relative to that measured at the time of decoupling by CMB data. This decrease is slightly redshift dependent: the smaller the redshift, the more the power spectrum gets suppressed. Besides that, the variation of \smnu{} induces a small shift in BAO scales, responsible for the oscillations visible in the right panel of \cref{fig:Pk_Mnu_linear_suppression}. 

The \Euclid weak-lensing probe traces total matter, while, to a very good approximation, the galaxy clustering probe traces only the fluctuations of cold and baryonic dark matter \citep{Castorina:2013wga}. The two panels in \cref{fig:Pk_Mnu_linear_suppression} also show the impact of varying \smnu{} on the cold-plus-baryonic matter power spectrum $P_{\rm cc}$ (dashed line). The effect of the neutrino mass on the amplitude of the linear small-scale spectrum $P_{\rm cc}$ is slightly reduced compared to the total matter case, with a suppression given roughly by $\Delta P_{\rm cc}(k)/P_{\rm cc}(k) \simeq -6f_{\nu}$. 

\subsection{Number of relativistic species\label{sec:neff}}

\begin{figure}
    \centering
    \includegraphics[width=0.48\textwidth]{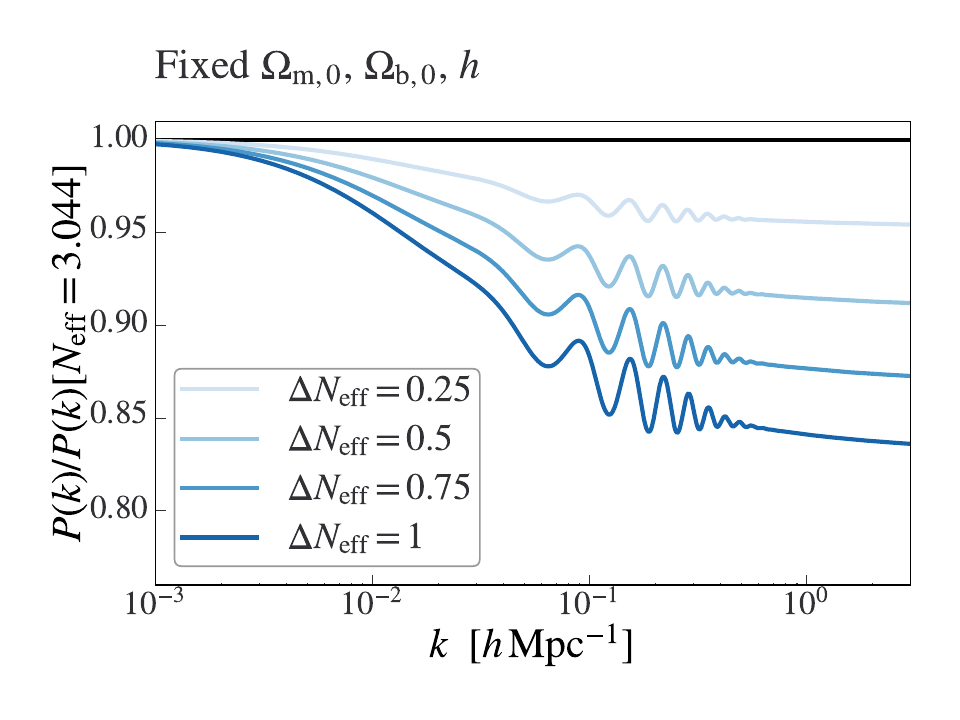}
    \includegraphics[width=0.48\textwidth]{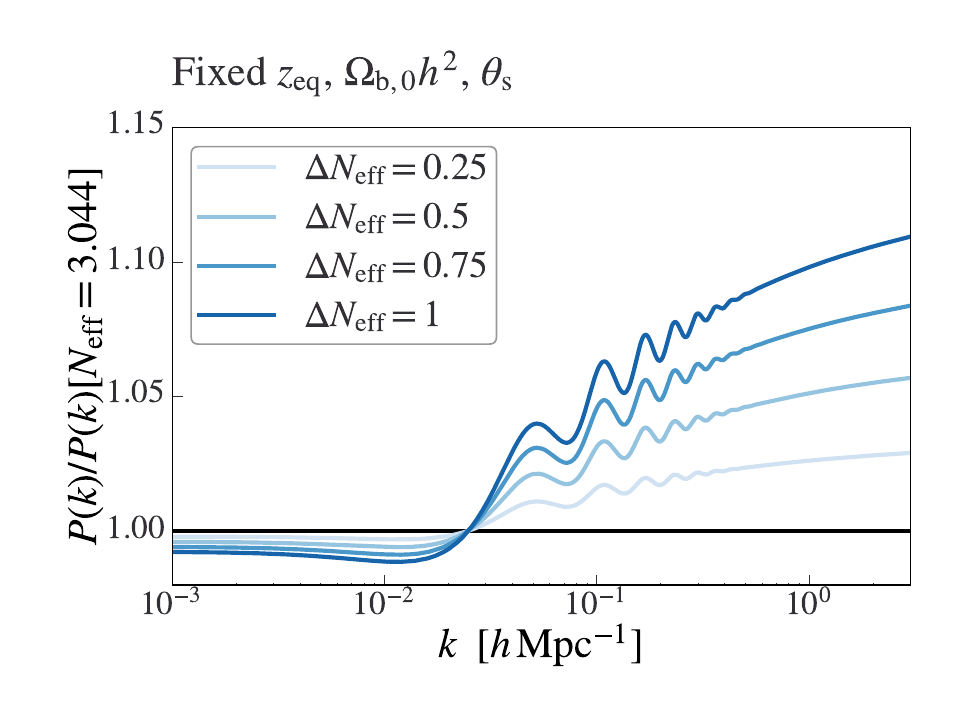}
    \caption{Effect of the effective number of relativistic degrees of freedom $\neff=3.044 + \dneff$ on the linear total matter power spectrum, presented as ratios with respect to the $\Lambda$CDM spectrum with $\neff=3.044$. \emph{Left}: We simply fix the density and Hubble parameters $\{\Omega_{{\rm m},0}, \Omega_{{\rm b},0}, h\}$. \emph{Right}: to show what is left for \Euclid to measure, we fix the quantities best constrained by CMB data, that is, $\{z_{\rm eq}, \Omega_{\mathrm{b},0}h^2, \theta_\mathrm{s}\}$. Here we assume only massless neutrinos.}
    \label{fig:Pk_Neff_linear_suppression}
\end{figure}

The contribution of ultra-relativistic species to the background density of radiation $\rho_{\rm r}$ (including that of neutrinos before their non-relativistic transition) can be parameterized in terms of an effective number of neutrinos $\neff$ as
\begin{equation}
 \rho_{\rm r} = \left[1 + \frac{7}{8}\left(\frac{4}{11}\right)^{4/3}\neff\right]\rho_{\gamma}\,,
\end{equation}
where $\rho_{\gamma}$ stands for the photon background density. In the above expression, the factor $7/8$ accounts for the Fermi--Dirac statistics of neutrinos, and $\left(4/11\right)^{4/3}$ accounts for the neutrino-to-photon temperature ratio in the instantaneous decoupling approximation~\citep{Lesgourgues:2006nd}. 
One expects $\neff = 3$ for three standard model neutrinos that thermalized in the early Universe and decoupled well before electron-positron annihilation. However, theoretical predictions set $\neff = 3.044$ in the standard cosmological model \citep{Mangano:2001iu,Froustey:2020mcq,Bennett:2020zkv,Drewes:2024wbw} because neutrinos decouple gradually with residual scatterings during this time.\footnote{During the completion of this work, there is a claim that this value shall be revised to $\neff = 3.043$ following a high-precision calculation of QED effects during electron-positron annihilation \citep{Cielo:2023bqp}. In this work we stick to $\neff = 3.044$ -- this difference has no impact on our sensitivity forecasts.} This value would change in the presence of non-standard neutrino features \citep[see for example][for the case of a large leptonic asymmetry, non-thermal distortions from low-temperature reheating, non-standard interactions, etc.]{Lesgourgues:2013sjj} or additional relativistic relics contributing to the energy budget \citep[see for example][]{Dvorkin:2022jyg}. 

At the level of background cosmology all these effects and models are captured by a single parameter $\neff$. However, at the level of perturbations some model-dependent features may arise, for instance if the additional relics carry small masses or feature (self-)interactions. In this work, for simplicity, we stick to the class of models where additional non-relativistic relics are decoupled, massless and free-streaming. However, we must simultaneously take the effect of neutrino masses into account -- in most of our forecasts, $\{ \smnu, \neff \}$ are two free model parameters. 
In this work, we will consider two ways to distribute the total mass $\smnu$ over the different species.
\begin{itemize}
    \item[(a)] By default, we consider three active neutrino species degenerate in mass ($m_\nu = \smnu/3$), thermally distributed, and with a temperature chosen in such a way that each species contributes to $\neff$ by $3.044/3$. Then, the additional free-streaming and massless dark radiation mentioned before enhances the total effective neutrino number as $\neff = 3.044 + \dneff$ with  $\dneff \geq 0$. In particular, this assumption is used throughout the results presented in \cref{sec:results_euclid}. Considering three degenerate massive neutrinos offers the advantage of predicting a matter power spectrum nearly indistinguishable from that of the realistic NO and IO scenarios with the same total mass, while models with only one or two massive species provide poorer approximations \citep{Lesgourgues:2004ps,Lesgourgues:2006nd,Archidiacono:2020dvx}.
    \item[(b)] In a single section, namely for the validation of \cref{sec:validation}, we adopt the same model as in most previous forecasts, both for the sake of comparison with earlier work and because the calculation of Fisher matrices is easier when parameters can vary symmetrically around their fiducial value. Then, like in the baseline analysis of \cite{Planck:2018vyg}, we stick to a single massive neutrino ($m_\nu = \smnu$), with the same thermal distribution and temperature as in the previous model. We consider additional free-streaming and massless species, contributing to the effective neutrino number by $\neff-3.044/3$, such that $\neff$ can be either bigger or smaller than $3.044$. 
\end{itemize}
We note that some of the models mentioned before (that feature non-standard physics in the neutrino sector) would require other model-dependent schemes for the splitting of the total mass across species and for the phase-space distribution of each massive species. However, the sensitivity of \Euclid to the parameters $\smnu$ and $\neff$ is expected to depend only weakly on each particular scheme, such that our sensitivity forecast assuming case (a) is representative of other cases.

The effect of varying $\neff$ on the linear matter power spectrum depends on which other parameters are kept fixed. For instance, the left panel of \cref{fig:Pk_Neff_linear_suppression} shows the impact of increasing $\neff$ with fixed $\Omega_{{\rm m},0}$, $\Omega_{{\rm b},0}$, and $h$. The leading effect is then a shift in the redshift of equality, inducing a suppression of the power spectrum.
However, like for neutrino masses, in order to understand what is left for \Euclid to measure, it is interesting to fix the cosmological parameters that are best constrained by CMB data, namely the redshift of radiation-to-matter equality $z_\mathrm{eq}$, the baryonic matter density, and the sound horizon angular scale \citep{Hou:2011ec,Lesgourgues:2013sjj,ParticleDataGroup:2022pth}. With such a choice, the effect of varying $\neff$ is displayed in the right panel of \cref{fig:Pk_Neff_linear_suppression}. Since $z_\mathrm{eq}$ is given by the matter-to-radiation density ratio $\Omega_{{\rm m},0}/\Omega_{{\rm r},0}$, increasing the radiation density with a fixed baryonic matter density implies an increase in the cold dark matter density, and thus, a decrease of the baryon-to-cold matter density ratio $\Omega_{{\rm b},0}/\Omega_{{\rm c},0}$. This has two well-known consequences: an enhancement of the amplitude of the matter power spectrum on scales $k>k_{\rm eq}$, where $k_{\rm eq} \sim {\cal O}(10^{-2})\, h \, {\rm Mpc}^{-1}$ is the wavenumber crossing the Hubble radius at equality; and a decrease in the amplitude of BAO oscillations. The scale of BAO peaks is also slightly shifted due to an enhanced neutrino drag effect \citep{Bashinsky:2003tk,Lesgourgues:2013sjj,Baumann:2019keh}. In the right panel of \cref{fig:Pk_Neff_linear_suppression}, one can clearly identify the power enhancement at $k>k_{\rm eq}$, as well as the oscillations produced by the shift in BAO phase and amplitude. The small power suppression on large scales comes from the fact that the matter fractional density $\Omega_{{\rm m},0}$, which controls the overall amplitude of the matter power spectrum on those scales, decreases slightly when one increases $\Omega_{{\rm m},0} h^2$ while fixing $\theta_{\rm s}$.

\section{Theoretical predictions for \texorpdfstring{\Euclid{}}{Euclid} observables\label{sec:euclid-observables}}

The modelling of \Euclid observables follows the recipes presented in \cite{Euclid:2019clj}, EC20 hereafter, and subsequently updated in \cite{EUCLID:2023uep}. However, the equations must be modified to account for the presence of massive neutrinos. Below we will briefly review the main equations, highlighting the relevant differences needed in massive neutrino cosmologies.

\subsection{Photometric survey\label{sec:photo}}

The information from the \Euclid\ imaging and photometric surveys can be embedded in three primary observables: the weak gravitational lensing (hereafter WL); the photometric reconstruction of galaxy clustering (hereafter $\GCph$); and their cross-correlation. For the purpose of our forecast, we decompose the observables into spherical harmonics leading to 2-dimensional angular power spectra ($C_\ell$). Assuming the Limber \citep[see also Ref.][]{Kilbinger2017} and the flat-sky approximations, the equation for the $C_\ell$ reads
\begin{equation}
 C^{XY}_{ij}(\ell) = c\int_{z_{\rm min}}^{z_{\rm max}}\de z\,{\frac{W_i^X(k_\ell, z)W_j^Y(k_\ell,z)}{H(z)r^2(z)}P^{XY}_{\delta \delta}(k_\ell,z)} + N^{XY}_{ij}(\ell)\,,\label{eq:photo_Cl}
\end{equation}
where $X$ and $Y$ each stand either for $\mathrm{L}$ (referring to WL) or $\mathrm{G}$ (referring to $\GCph$), $i$ and $j$ denote the redshift bins,\footnote{We assume 10 equally populated redshift bins in the range $z_{\rm min}=0.001$ to $z_{\rm max}=2.5$.} and $H(z)$ is the Hubble rate.
The nonlinear power spectrum of density fluctuations is denoted by $P^{XY}_{\delta\delta}(k,z)$ and is evaluated at $k_\ell=(\ell+1/2)/r(z)$ \citep{Kilbinger2017}, with $r(z)$ being the comoving distance.
Finally, $W_i^X(z)$ is the window function of the $i$-th redshift bin at $z$ for the $X$ observable,\footnote{In \cref{eq:photo_Cl}, $H(z)$ has units of $\kmsMpc$, $c$ of km\,s$^{-1}$, $r$ of Mpc, $P^{XY}_{\delta\delta}$ of Mpc$^3$, and $W_i^X(z)$ of Mpc$^{-1}$, while all other quantities are dimensionless.} which can be written
\begin{align}
 W_i^{\rm L}(k,z) = &\; \frac{3}{2} \Omega_{{\rm m},0} \frac{H_0^2}{c^2} (1+z)r(z)
\int_z^{z_{\rm max}}{\de z'\frac{n_i(z')}{\bar{n}_i}\frac{r(z'-z)}{r(z')}}
 + W^{\rm IA}_i(k,z)\, , \label{eq:window_wl} \\
 W_i^{\rm G}(k,z) = &\; \frac{H(z)}{c} b_i(k,z)\frac{n_i(z)}{\bar{n}_i}\,, \label{eq:window_gc} 
\end{align}
for WL and for $\GCph$, respectively. Here $n_i(z)/\bar{n}_i$ is the normalised galaxy distribution and $b_i(k,z)$ is the galaxy bias in the $i$-th redshift bin. The contribution of the intrinsic alignment is embedded in $W^{\rm IA}_i(k,z)$ and we adopt the modelling through nonlinear alignment method used in \citetalias{Euclid:2019clj}. The intrinsic alignment window function then reads
\begin{equation}
    W^{\rm IA}_i(k,z)=-\frac{\mathcal{A}_{\rm IA}\mathcal{C}_{\rm IA}\Omega_{\rm m,0}\mathcal{F}_{\rm IA}(z)}{D(z,k)}\frac{n_i(z)}{\bar{n}_i(z)}\frac{H(z)}{c}\,, \label{eq:window_ia}
\end{equation}
where $D(z,k)$ is the scale-dependent linear growth factor.
The function $\mathcal{F}_{\rm IA}(z)$ depends on the luminosity function and is given by
\begin{equation}
 \mathcal{F}_{\rm IA}(z)=(1+z)^{\eta_{\rm IA}}\left[\frac{\langle L\rangle(z)}{L_\star(z)}\right]^{\beta_{\rm IA}}\,,
\end{equation}
where $\langle L\rangle(z)$ and $L_\star(z)$ are the redshift-dependent mean and characteristic luminosities of source galaxies. The parameters $\mathcal{A}_{\rm IA}$ and $\eta_{\rm IA}$ are nuisance parameters, and are allowed to vary around the fiducial values $ \{ \mathcal{A}_{\rm IA}, \eta_{\rm IA} \} = \{ 1.72, -0.41 \}$; the parameters $\mathcal{C}_{\rm IA}=0.0134$ and $\beta_{\rm IA}= 2.17$ are kept fixed.

Finally, $N^{XY}_{ij}$ is the shot-noise term, which is zero for the cross-correlation between observables [$N_{ij}^{\rm GL}(\ell)=0$], while for the auto-correlation it can be written as
\begin{equation*}
   N^{\rm GG}_{ij}(\ell) = \frac{1}{\bar{n}_i}\delta_{ij}\,, \qquad  N^{\rm LL}_{ij}(\ell) = \frac{\sigma_\epsilon^2}{\bar{n}_i}\delta_{ij}\,,
\end{equation*}
where $\bar{n}_i$ is the average number of galaxies per redshift bin expressed in steradians and obtained by dividing the expected total number of observed galaxies, $n_{\rm gal}=30\,{\rm arcmin}^{-2}$, by the number of redshift bins, $N_{\rm bins}=10$, and $\sigma_\epsilon=0.30$ is the variance of the observed ellipticities. Following \citetalias{Euclid:2019clj}, we neglect any subdominant contributions from redshift-space distortions and lensing magnification.

The primary impact of the neutrinos can be summarised through their scale-dependent growth and their impact on the expansion history. Obviously, the neutrino mass enters as an additional component in the computation of $H(z)$. Besides this trivial difference, the first relevant difference in the modelling of \Euclid observables in massive neutrino cosmologies with respect to $\Lambda$CDM concerns $P^{XY}_{\delta\delta}(k_\ell,z)$.
Indeed, it has been shown \citep{Castorina:2013wga, DEMNUni1} that in the presence of massive neutrinos the tracer of galaxy clustering is given by the clustering of cold dark matter and baryons, neglecting the contribution of neutrinos. Defining the power spectrum in terms of cold dark matter only [$P_{\delta\delta}^{\rm GG}(k_\ell,z)=P_{\rm cc}(k_\ell,z)$ with ${\rm c} = {\rm CDM+baryons}$], makes the bias less scale dependent at linear and at mildly nonlinear scales. 
Therefore, we can approximate $b_i(k,z) \simeq b_i(z)$ in \cref{eq:window_gc} and we can model the bias with only one nuisance parameter for each redshift bin. We still take into account the scale-dependent growth in all relevant terms, such as for example in \cref{eq:window_ia}. 
A wrong definition of the bias in terms of total matter power spectrum in massive neutrino cosmologies leads to an overestimate of the sensitivity of future galaxy surveys to $\smnu$ \citep{Vagnozzi:2018pwo}.

On the other hand, massive neutrinos do contribute to the gravitational potential, which is the source of the weak lensing effects. Therefore, we set $P_{\delta\delta}^{\rm LL}(k_\ell,z) = P_{\rm mm}(k_\ell,z)$. For the cross-correlation of these two probes, we assume $P_{\delta\delta}^{\rm GL}(k,z) = \sqrt{P_{\delta\delta}^{\rm GG}(k,z)P_{\delta\delta}^{\rm LL}(k,z)}$. Finally, let us stress that the recipe used here to model the photometric observables implies that all the power spectra are nonlinear, thus the nonlinear corrections described in \cref{sec:nl} are applied to both $P_{\rm mm}(k_\ell,z)$ and $P_{\rm cc}(k_\ell,z)$.

The final log-likelihood is modelled as a Gaussian with respect to the angular power spectrum and can be written as 
\begin{align}
    \label{eq:chisq_photometric}
    \chi^2 = f_{\rm sky} \sum_\ell (2\ell+1) \left( \frac{d^{\rm mix}_\ell}{d^{\rm th}_\ell} + \ln{ \frac{d^{\rm th}_\ell}{d^{\rm fid}_\ell}}-N\right)\,,
\end{align}
with
\begin{align}
    d^{\rm th}_\ell &= \det \left[ C_{ij}(\ell)\right]\, , \nonumber\\
    d^{\rm fid}_\ell &= \det \left[ C_{ij}^{\rm fid}(\ell)\right]\, , \\
    d^{\rm mix}_\ell &= \sum_k \det \left[ \left\{ \begin{array}{ll}
        C_{ij}(\ell)&\text{ if }j\neq k \\
          C_{ij}^{\rm fid}(\ell)&\text{ if }j = k
    \end{array} \right.\right] \,.\nonumber
\end{align}
Here ``fid'' denotes the values of $C_{ij}(\ell)$ computed for the fiducial model parameters, and $f_\mathrm{sky}=0.3636$ is the sky fraction covered by the wide survey.\footnote{Here we use the same sky fraction, as well as other specifications, of Ref.~\citepalias{Euclid:2019clj}, however, some of them have changed since then \citep[see, for instance, the reduction of wide survey area reported in Ref.][]{Euclid:2021icp}, and keep changing continuously based on the performance of the satellite.}
Following \citetalias{Euclid:2019clj} and \cite{EUCLID:2023uep}, we assume two different sets of specifications for photometric probes, a pessimistic and an optimistic one, as listed in \cref{tab:spec_photo}. We will use both settings in our validation tests, but in \cref{sec:results_euclid} all our final forecast results rely solely on the pessimistic settings, for which multipoles are included in the likelihood only up to $\ell_{\rm max}=1500$. Note that at higher multipoles, it becomes important to include baryonic feedback effects in the modelling of the observable power spectrum. These effects could be potentially degenerate with those of the neutrino mass and number density. However, as hinted in previous works like, for instance, \cite{SpurioMancini:2023mpt}, and as explicitly shown in \cref{app:bf}, this is not the case when the data is cut at $\ell_{\rm max}=1500$. Thus, we neglect baryonic feedback in the rest of this work.

\begin{table}[ht]
\renewcommand{\arraystretch}{1.2}
    \centering
    \caption{Specifications assumed in the modelling of the photometric observables.}
    \begin{tabular}{c c c}
    \hline
    & Pessimistic & Optimistic \\
        \hline \vspace{-2mm}\\
        $\ell^\mathrm{WL}_\mathrm{max}$ & 1500 & 5000 \\ \vspace{-3mm}\\
        $\ell^\mathrm{\GCph}_\mathrm{max}$ & 750 & 3000 \\ \vspace{-3mm}\\
        $\ell^\mathrm{\XCph}_\mathrm{max}$ & 750 & 3000 \\
        \hline
    \end{tabular}
    \label{tab:spec_photo}
\end{table}

\subsection{Spectroscopic survey\label{sec:gc}}

The observable extracted from the \Euclid spectroscopic survey is a 3D galaxy power spectrum, which can be written as:
\begin{equation}
P_\text{obs}(k_\text{fid},\mu_\text{fid};z) = 
\frac{1}{q_\perp^2(z) q_\parallel(z)} 
\left\{\frac{\left[b\sigma_8(z)+f(k,z)\sigma_8(z)\mu^2\right]^2}{1+f(k,z)^2k^2\mu^2\sigma_{\rm p}^2(z)}\right\}
\, \frac{P_\text{dw}(k,\mu;z)}{\sigma_8^2(z)}  
F_z(k,\mu;z) 
+ P_\text{s}(z) \,, 
\label{eq:spectro_pk}
\end{equation}
where $\mu = \vec{k} \cdot \vec{r} / (kr)$ is the cosine of the line-of-sight angle with respect to the wavenumber $\vec{k}$ (with absolute value $k=|\vec{k}|$), and the subscript ``$\text{fid}$'' denotes the quantities computed with the fiducial cosmology (which is used to convert angles and redshifts to physical distances). 
In the following, we will briefly review the modelling of the main observational effects that are taken into account in \cref{eq:spectro_pk} to convert the cold dark matter and baryon power spectrum $P_{\rm cc}(k_\ell,z)$ into the observed galaxy power spectrum $P_\text{obs}(k_\text{fid},\mu_{\theta,\text{fid}};z)$. We note that, as already explained in \cref{sec:photo}, since the tracers of this observable are galaxies, we will always refer to $P_{\rm cc}(k_\ell,z)$, hence removing the contribution of neutrinos, rather than to the total matter power spectrum. 
Let us quickly summarise the impact of the neutrinos on the various terms:

The first term on the right-hand side is the \AP{} effect \citep{Alcock:1979mp}, arising from the assumption about the underlying cosmology applied in the conversion of redshifts and angles into parallel and perpendicular distances:
 \begin{equation}
   q_{\perp}(z) = \frac{D_{\rm A}(z)}{D_{\rm A,\, fid}(z)}\,, \qquad
   q_{\parallel}(z) = \frac{H_\text{fid}(z)}{H(z)}\,,
 \end{equation}
where $D_{\rm A}(z)$ is the angular diameter distance, and $H(z)$ is the Hubble rate. Both quantities are obviously affected by the impact of neutrinos on the expansion rate. We also note that as usual $\mu = \mu_{\rm fid} q_\perp/q_\parallel/G$ as well as $k = k_{\rm fid} \, G/q_\perp$, where $G^2= 1+\mu_{\rm fid}^2 (q_\perp^2/q_\parallel^2-1)$. Here ``fid'' denotes that the values are set to their fiducials, following the recipe of \citetalias{Euclid:2019clj} and \cite{EUCLID:2023uep}.

The term in the numerator in curly brackets, $\left[b(z)\sigma_8(z)+f(k,z)\sigma_8(z)\mu^2 \right]$, accounts for redshift-space distortions, which are anisotropic perturbations appearing in redshift-space, due to the Doppler effect being an additional source of redshift beyond the cosmological one. The effect is modelled according to the Kaiser formula \citep{Kaiser:1987qv}. Here $b(z)$ is the bias (see \cref{sec:photo} for a justification of dropping the scale dependence) and $f(k,z)$ is the scale-dependent growth factor of CDM+baryons, explicitly excluding neutrinos since we are interested in clustered objects (galaxies). The growth factor is computed with respect to the CDM+baryon component as $f(z,k) = \frac{1}{2}\frac{\de \ln\left[P_{cc}(z,k)\right]}{\de \ln a}$.\footnote{In this context, it is not obvious whether one should use $P_{\rm cc}$ or $P_{\rm mm}$. A comparison with simulations suggests, however, that using $P_{\rm cc}$ is a better approximation, see Eq.~(8) of \cite{Villaescusa-Navarro:2017mfx} or the discussion that follows Eq.~(6.13) of \cite{DEMNUni1}.} 
Finally, $\sigma_8(z)$ is defined also using CDM+baryons only. As such, the entire numerator is impacted by the mass of the neutrinos through the scale dependence of the growth as well as the reduction of the amplitude $\sigma_8(z)$.

The term in the denominator in curly brackets, $1+f(k,z)^2k^2\mu^2\sigma_{\rm p}^2(z),$ represents the fingers-of-God effect, due to the additional redshift coming from galaxy peculiar velocities. The effect is modelled as a Lorentzian factor, where $\sigma_{\rm p}(z)$ is the distance dispersion corresponding to the velocity dispersion $\sigma$, or explicitly $\sigma_{\rm p}(z)=\sigma/[H(z)a(z)]$. Given the uncertainty in the modelling and in the redshift dependence of $\sigma_{\rm p}(z)$, we treat it as four additional nuisance parameters, one for each redshift bin.

$P_\text{dw}(k,\mu_{\theta};z)$ is the partially de-wiggled power spectrum, computed starting from the linear $P^{\rm GG}_{\delta\delta}$,\footnote{Contrary to the recipe for photometric observables, the recipe for spectroscopic observables is built on the linear power spectrum, and the nonlinear effects are modelled through the Lorentzian factor and the damping of the wiggle-only power spectrum to which a de-wiggled contribution is added to restore the broadband shape of the power spectrum; the nonlinear corrections of \cref{sec:nl} are never applied here.} which is $P_{\rm cc}$, and accounting for the smearing of the baryon acoustic oscillations due to nonlinear effects; it can be written as \citep{Wang:2012bx}
\begin{equation}
P_\text{dw}(k,\mu;z) = P^{\rm GG}_{\delta\delta}(k;z)\,\text{e}^{-g_\mu k^2} + P_\text{nw}(k;z)\left(1-\text{e}^{-g_\mu k^2}\right) \,,
\label{eq:pk_dw}
\end{equation}
where $P_\text{nw}$ is the no-wiggle power spectrum obtained by removing the BAO from $P^{\rm GG}_{\delta\delta}$ \citep{Boyle:2017lzt}, and
\begin{equation}
    g_\mu(k,\mu,z)= [\sigma_{\rm v}(z)]^2 \left\{ 1-\mu^2+\mu^2 [1+f^{\rm fid}(z,k)]^2 \right\} \,,
\end{equation}
where $\sigma_{\rm v}(z)$, which has dimensions of length, reflects the galaxy velocity dispersion and is being treated as four additional nuisance parameters, one for each redshift bin, as for $\sigma_{\rm p}(z)$.

The term $F_z(k,\mu;z)=\exp{\left[-k^2\mu^2\sigma_{\rm r}^2(z) \right]}$ represents the damping due to the spectroscopic redshift errors along the line of sight, where $\sigma_{\rm r}(z)=\frac{c}{H(z)} \sigma_{z}$ with a redshift-independent error $\sigma_{z}=0.002$.

The last term in \cref{eq:spectro_pk}, $P_\text{s}(z)$, is the shot-noise caused by the Poissonian distribution of measured galaxies on the smallest scales due to the finite total number of observed galaxies. We model it like in \citetalias{Euclid:2019clj} and \cite{EUCLID:2023uep}, with one contribution given by the fiducial inverse number density in each bin, and a second contribution accounting for residual shot noise that we treat as a nuisance parameter in each bin.

The final likelihood is modelled as a Gaussian, comparing the observed power spectrum to a fiducial one. The $\chi^2$ is given by
\begin{align}
\label{eq:chisq_spectro}
    \chi^2 = \sum_i \int_{-1}^{1} \int_{10^{-3}}^{k_{\rm max}} k_{\rm fid}^2 \frac{V_i^{\rm fid}}{8 \pi^2} \left\{\frac{P_{\rm obs}\left[k(k_{\rm fid},\mu_{\rm fid},z_i),\mu(\mu_{\rm fid},z_i),z_i\right]-P_{\rm obs}^{\rm fid}\left(k_{\rm fid},\mu_{\rm fid},z_i\right)}{ P_{\rm obs}\left[k(k_{\rm fid},\mu_{\rm fid},z_i),\mu(\mu_{\rm fid},z_i),z_i\right]}\right\}^2 \de k_{\rm fid}\, \de \mu_{\rm fid}\,,
\end{align}
where $i$ denotes the index of the redshift bin and $V_i$ is the comoving volume of the spherical shell of the redshift bins probed by the experiment and $k_{\rm max}$ is in \cref{tab:spec_gc}. The partial sky coverage of the survey is approximately taken into account through multiplication of the volume by a sky fraction $f_\mathrm{sky}=0.3636$.

Following \citetalias{Euclid:2019clj} and \cite{EUCLID:2023uep} we assume two different sets of specifications for spectroscopic galaxy clustering, as reported in \cref{tab:spec_gc}.
\begin{table}[ht]
\renewcommand{\arraystretch}{1.2}
    \centering
    \caption{Specifications assumed in the modelling of the spectroscopic galaxy clustering.}
    \begin{tabular}{c c c}
    \hline
    & Pessimistic & Optimistic \\
        \hline \vspace{-2mm}\\
        \multicolumn{1}{c}{$k_\mathrm{max}\,\,\,\left[h^{\rm fid}{\rm Mpc}^{-1}\right]$} & \multicolumn{1}{c}{0.25} & \multicolumn{1}{c}{0.30}\\
        \hline
    \end{tabular}
    \label{tab:spec_gc}
\end{table}

\subsection{Nonlinear modelling\label{sec:nl}}

\begin{figure}
    \centering
    \includegraphics[width=.8\textwidth]{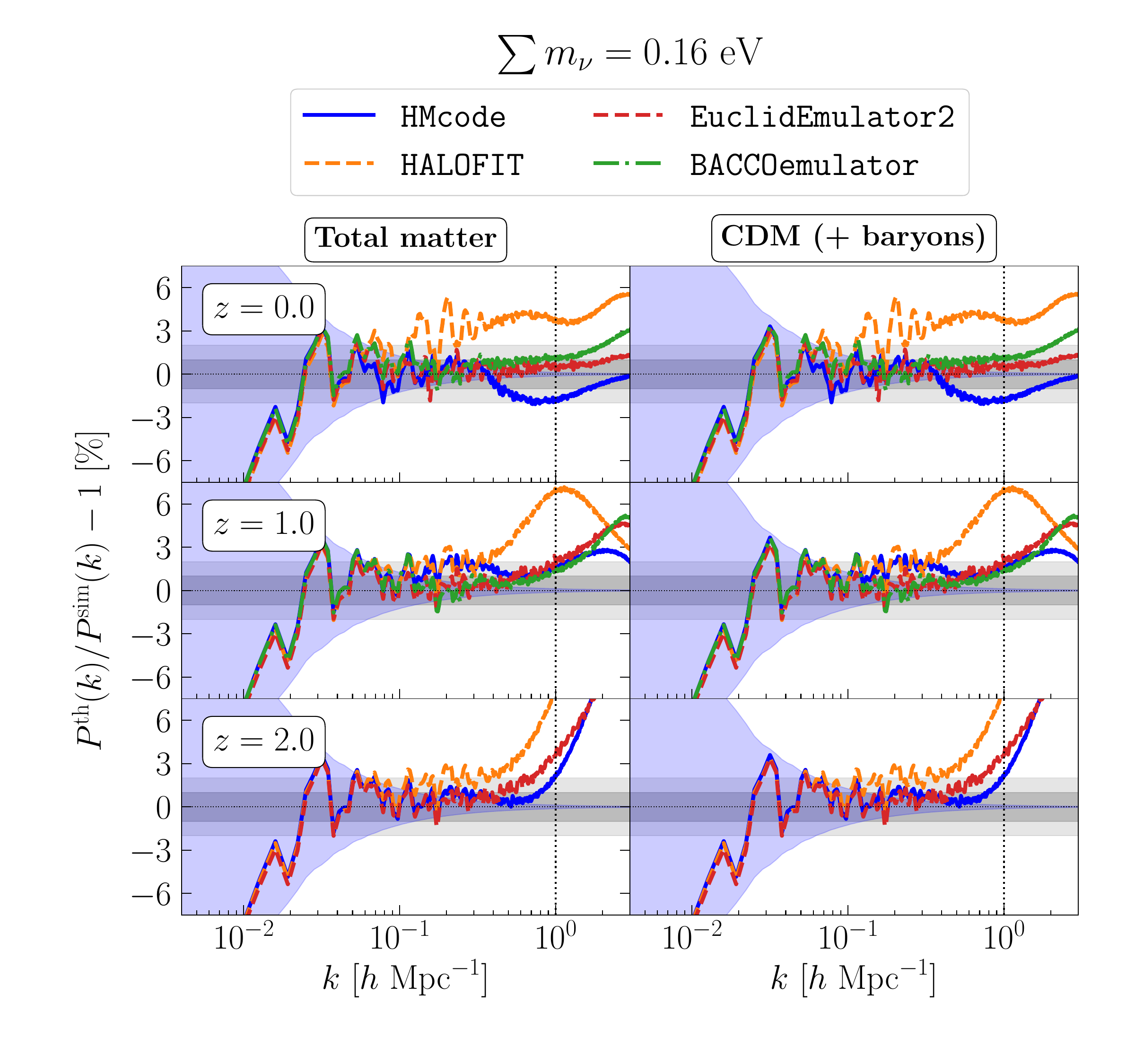}
    \caption{Relative percentage difference of the nonlinear power spectrum computed with various recipes with respect to the DEMNUni simulations for the case $\smnu=0.16\,{\rm eV}$. We note that here, contrary to the forecast analysis, and to be consistent with the DEMNUni simulations, we assume the total neutrino mass to be equally shared among the three neutrino species $m_1=m_2=m_3=\smnu/3$. The left column shows the total matter power spectrum, while the right column shows the cold dark matter power spectrum. The spectra are evaluated at $z=0,1,2$ (first, second and third rows, respectively). The theoretical predictions are provided by {\tt HALOFIT} (dashed orange line), {\tt HMcode} (solid blue line), {\tt EuclidEmulator2} (dashed red line), and {\tt BACCOemulator} (dot-dashed green line). The purple shaded area represents the shot noise of the simulations, and the vertical dotted line the maximum wavenumber. In the third row the predictions of {\tt BACCOemulator} are missing because the emulator is trained only up to $z=1.5$. The predictions for the cold dark matter power spectrum (neglecting the contribution of neutrino perturbations) is computed according to the approximate formula \cref{eq:Pk_sum_components} (see text for details).}
    \label{fig:Pk_Mnu_DEMNUni}
\end{figure}
\begin{figure}
    \centering
    \includegraphics[width=.9\textwidth]{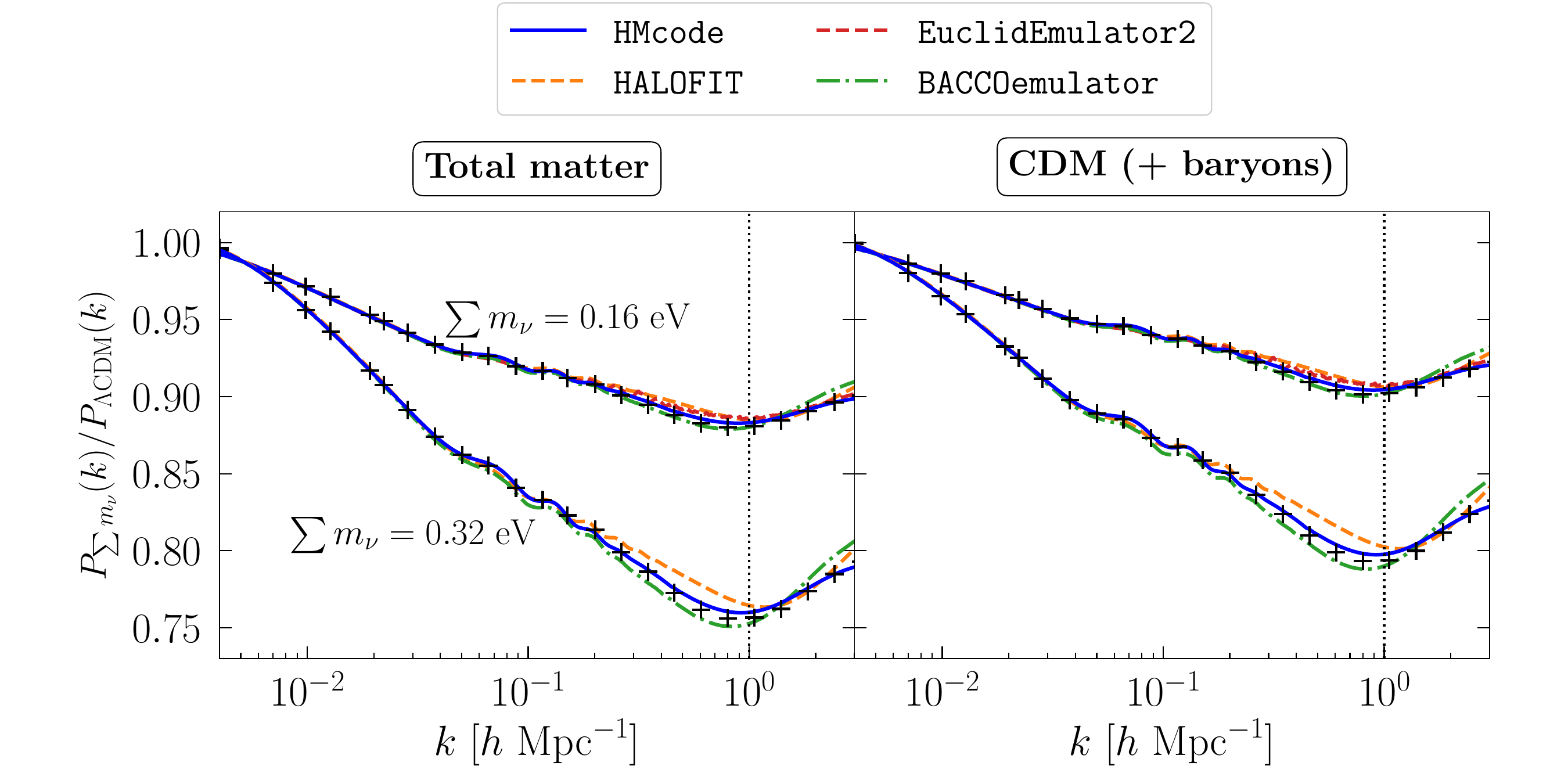}
    \caption{Theoretical predictions of the suppression of the total matter power spectrum (left) and of the cold dark matter power spectrum (right) in massive neutrino cosmologies ($\smnu=0.16\,{\rm eV}$ and $\smnu=0.32\,{\rm eV}$) with respect to the pure $\Lambda$CDM case with massless neutrinos at $z=0$. The colour coding of the theoretical predictions is the same as in \cref{fig:Pk_Mnu_DEMNUni}. In the case $\smnu=0.32\,{\rm eV}$, {\tt EuclidEmulator2} is not shown because the value of the neutrino mass is too far away from the range of validity of the emulator. The DEMNUni simulations are depicted as black pluses.}
    \label{fig:Pk_Mnu_supp_DEMNUni}
\end{figure}
Modelling nonlinear clustering in massive neutrino cosmologies is essential to achieve robust constraints on their mass. A thorough comparison of the performance of several $N$-body codes and emulators is shown in \cite{Euclid:2022qde}. 
Their findings show that, in cosmologies where only the total neutrino mass is varied, the most up-to-date emulators ({\tt HMcode}, {\tt EuclidEmulator2}, {\tt BACCOemulator}) agree with simulations at the $1-2\%$ level for the matter power spectrum. However, none of the aforementioned emulators has been explicitly trained on or built for models with a varying number of neutrino-like species. We present below a comparison of these emulators to $N$-body simulations in order, firstly, to confirm that they accurately capture the effect of neutrino mass, and secondly, to check whether they can also account for the impact of varying $N_{\rm eff}$.

In \cref{fig:Pk_Mnu_DEMNUni} (left column) we show the accuracy in the prediction of the total matter power spectrum of {\tt HALOFIT} \citep{Takahashi:2012em} with the neutrino corrections of \cite{Bird:2011rb}, {\tt HMcode} \citep{Mead:2020vgs}, {\tt EuclidEmulator2} \citep{Euclid:2020rfv},\footnote{We note that {\tt EuclidEmulator2} was trained in the range $\smnu \in [0.0,0.15]\,{\rm eV}$. However, the minimum value in the DEMNUni simulations is $\smnu=0.16\,{\rm eV}$. Therefore, we had to slightly extend the range of validity of the emulator, assuming it would not cause any dramatic effect on the predictions.} and {\tt BACCOemulator} \citep{Angulo:2020vky}. The accuracy is computed with respect to the Dark Energy and Massive Neutrino Universe (DEMNUni) $N$-body simulations \citep{Carbone:2016nzj,Parimbelli:2022pmr} at redshifts $z=0,1,2$. We find that {\tt HALOFIT} is the least accurate in reproducing the results from the simulations, both at intermediate (BAO) scales and at smaller scales for any redshift. Instead, {\tt HMcode} and the two emulators show a very similar performance, apart from a few exceptions.
At $z=0$ {\tt HMcode} underestimates the power at small scales with respect to the two emulators, which describe the neutrino mass suppression with $1-2\%$ accuracy. At $z=1$ {\tt HMcode} slightly overestimates power at BAO scales, while deeply in the nonlinear regime it seems to reproduce the results of the simulations more accurately. Finally, at $z=2$ {\tt HMcode} performs slightly better than {\tt EuclidEmulator2}, while {\tt BACCOemulator} is not yet trained up to this redshift. Overall, both {\tt HMcode} and the emulators are within $2\%$ accuracy at any redshift and at any scale where the simulations can be trusted ($k<1 \, h\,{\rm Mpc}^{-1}$).

In our usage case, it is not only important to accurately model the clustering of the total power spectrum but also of the CDM+baryon power spectrum (see \cref{sec:euclid-observables}).
In the right column of \cref{fig:Pk_Mnu_DEMNUni} we show the accuracy of the same nonlinear recipes in predicting the cold dark matter and baryons power spectrum $P_{\rm cc}(k,z)$ extracted from the DEMNUni simulations.
In order to convert the nonlinear $P_{\rm mm}(k)$ into the nonlinear $P_{\rm cc}(k)$ we use the formula
\begin{equation}
    P_{\rm mm}(k) = f_{\rm c}^2P_{\rm cc}(k) + 2f_\nu f_{\rm c} P_{{\rm c}\nu}(k) + f_\nu^2 P_{\nu\nu}(k)\,,
    \label{eq:Pk_sum_components}
\end{equation}
where $P_{\rm mm}(k)$ is the total matter auto-correlation power spectrum, $P_{\nu\nu}(k)$ the neutrino auto-correlation power spectrum, $P_{{\rm c}\nu}(k)$ the cross-correlation cold dark matter--neutrino power spectrum, $f_{\rm c}$ the cold fraction of dark matter, and $f_\nu=(1-f_{\rm c})$ the hot one. The formula is correct as long as all the power spectra appearing both in the left-hand side and in the right-hand side are either linear or nonlinear. An approximation arises when mixing linear and nonlinear power spectra. Here we consider nonlinear $P_{\rm mm}(k)$ and $P_{\rm cc}(k)$, while $P_{{\rm c}\nu}(k)$ and $P_{\nu\nu}(k)$ are assumed to be linear. This assumption is accurate for $P_{\nu\nu}(k)$, because for $\smnu \lesssim 0.6\,{\rm eV}$ the neutrino free-streaming length is larger than the nonlinear scale today (and even more so at higher redshift). On the other hand, while we expect the cold dark matter--neutrino cross-power spectrum $P_{{\rm c}\nu}(k)$ to exhibit some nonlinearities, these are found to be negligible.

The right panel of \cref{fig:Pk_Mnu_DEMNUni} shows precisely the accuracy of this approximation: we compare results from the DEMNUni simulations against the nonlinear $P_{\rm cc}(k)$ of the various emulators obtained by inverting \cref{eq:Pk_sum_components}, where only $P_{\rm mm}(k)$ is assumed to be fully nonlinear. 
Concerning the accuracy on $P_{\rm cc}(k)$ of the different prescriptions adopted to compute the theoretical $P_{\rm mm}(k)$ [from which we derive $P_{\rm cc}(k)$] the same considerations drawn for the total matter power spectrum hold true here. While {\tt HALOFIT} fails to accurately reproduce the results of the simulations, the accuracy of the emulators is better than the one of {\tt HMcode}, especially at low redshift. However, both {\tt HMcode} and the emulators remain within $2\%$ accuracy with respect to the DEMNUni simulations at any redshift and scale.

In \cref{fig:Pk_Mnu_supp_DEMNUni} we show the massive neutrino ($\smnu=0.16\,{\rm eV}$ and $\smnu=0.32\,{\rm eV}$) induced suppression in the total matter power spectrum (left plot) and in the CDM+baryon power spectrum (right plot), with respect to a $\Lambda$CDM cosmology for the DEMNUni simulations and for the nonlinear predictions described above. For larger neutrino masses, {\tt HMcode} outperforms not only {\tt HALOFIT} but also the emulators in the precision of modelling this suppression. We note that a neutrino mass of about $0.3\,{\rm eV}$, although it is excluded in the minimal $\Lambda$CDM + $\smnu$ scenario by the most stringent cosmological constraints to date \citep[e.g.][]{eBOSS:2020yzd, Palanque-Delabrouille:2019iyz}, remains within reach in extended models \citep{Lambiase:2018ows}.

\begin{figure}
 \centering
 \includegraphics[width=.8\textwidth]{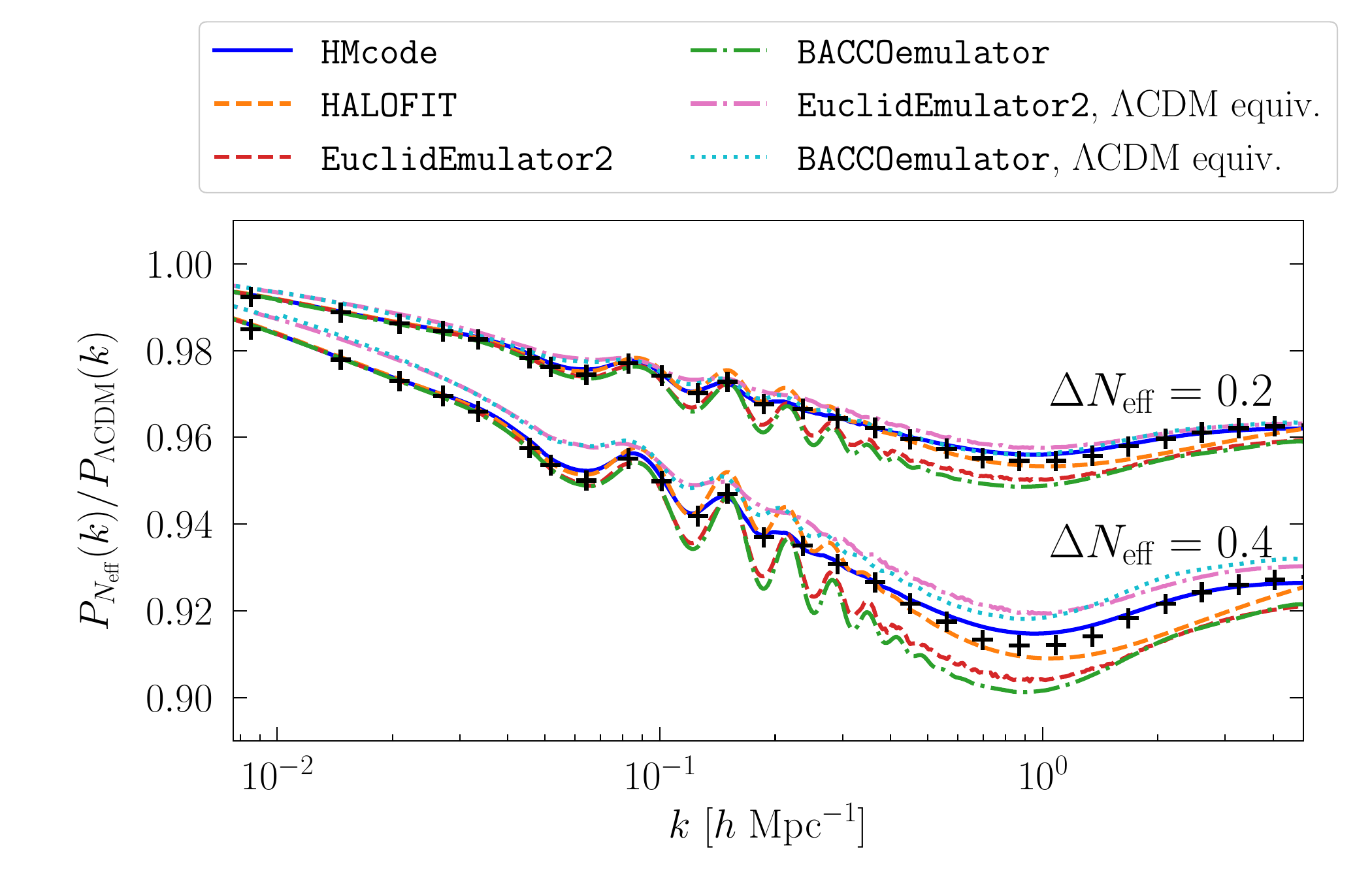}
 \caption{Theoretical predictions of the suppression of the total matter power spectrum in non-standard $\neff$ cosmologies with respect to the standard one, at $z=0$, and assuming fixed values of $\{\Omega_{{\rm m},0}, \Omega_{{\rm b},0}, h\}$, like in the left panel of \cref{fig:Pk_Neff_linear_suppression}. The colour coding of the theoretical predictions is the same as in \cref{fig:Pk_Mnu_DEMNUni}. We add here two more prescriptions, obtained by rescaling $\Omega_{\mathrm{c},0}$, $\Omega_{\mathrm{b},0}$ and $h$ to mimic changes in \neff. We refer to it as ``$\Lambda$CDM equivalent'': the dot-dashed pink line represents results for {\tt EuclidEmulator2}; and the dotted cyan line represents the {\tt BACCOemulator} prediction. Black crosses represent results from $N$-body simulations.}
    \label{fig:Pk_Neff_supp_DEMNUni}
\end{figure}

Finally we are also going to vary the effective number of neutrino-like species (\neff) in our main analysis.
In order to understand the nonlinear clustering in cosmologies in this case, we performed simulations with $1024^3$ particles in a box with size $L=1024\,h^{-1}\,{\rm Mpc}$ varying $\neff$ between $0.2$ and $0.4$. In \cref{fig:Pk_Neff_supp_DEMNUni} we checked the accuracy of {\tt HALOFIT}, {\tt HMcode}, {\tt EuclidEmulator2}, and {\tt BACCOemulator} with respect to these $\neff$ simulations. The two emulators fail in reproducing the variations of the number of neutrino species (green and red lines, {\tt BACCOemulator} and {\tt EuclidEmulator2}, respectively); this was expected since they are not trained on cosmologies with non-standard $\neff$. In order to overcome the parameter extension with the emulators, we tried to remap the variation of $\neff$ on large scales to a variation of the total matter and cold dark matter density parameters $\{ \Omega_{{\rm m},0}, \Omega_{{\rm c},0} \}$ and of the reduced Hubble constant $h$ using the relationship derived in \cite{Rossi:2014nea}. However, tweaking the parameters does not improve the accuracy of the reconstruction of the $\neff$ variations by means of emulators to a level competitive with {\tt HMcode}. On the other hand, {\tt HMcode} provides a good fit of the relative difference in the nonlinear clustering induced by variations of the number of neutrino-like particles, both for the phase-shift in the BAO scale, and for the overall suppression at small scales. Indeed, even though the {\tt HMcode} parameters are not fit to simulations with varying $\neff$, the model is based on the convolution with the linear power spectrum, which naturally embeds the effect of $\neff$. 

To summarise, the emulators reach $1\%$ accuracy in reproducing the CDM+baryon power spectrum in massive neutrino cosmologies out to $k=1\,h \, {\rm Mpc}^{-1}$, thus performing slightly better than {\tt HMcode}, as already noted in \cite{Euclid:2022qde}. However, given their range of validity in terms of redshift ({\tt BACCOemulator}) and neutrino mass ({\tt EuclidEmulator2}) and especially given that they are not trained in varying $\neff$ cosmologies, we opted for {\tt HMcode} as our recipe to compute nonlinear corrections.

\section{Additional probes}\label{sec:add_prob}

Through the history of the Universe neutrinos evolve from behaving like radiation to contributing to the total matter density. In order to capture neutrino effects at different epochs and remove parameter degeneracies, it is crucial to combine primary probes with additional secondary probes on larger scales from the \Euclid survey, such as cluster number counts, and with external data probing the early Universe, such as from CMB anisotropies.

\subsection{Cluster number counts from the \texorpdfstring{\Euclid\ survey}{Euclid survey}}\label{sec:clusters}

Clusters of galaxies are potentially a strong cosmological probe \citep{1997A&A...317....1O,White:1993wm,Bahcall:1998ur,Reiprich:2001zv,Mantz:2007qh,Vikhlinin:2008ym}. Measurements of their abundance and evolution allow to constrain both the geometry of our Universe and the growth of its density perturbations \citep{2005ASPC..339..140M,2009ApJ...692.1060V,Allen:2011zs,Kravtsov:2012zs}. This could be achieved in particular from galaxy cluster number-counts experiments where, assuming that a halo number density mass function is predicted in a given cosmological model, one can confront the number of observed clusters computed in a given survey volume with its theoretical prediction and, from it, constrain cosmological parameters \citep{SPT:2018njh,Kirby:2019mrb,DES:2020cbm,Sakr:2021jya,Lesci:2020qpk}.
\Euclid will study galaxy clusters abundance, an independent and complementary probe to the two primary ones, allowing the optical detection of clusters previously unattainable in terms of the depth and area covered.
Under optimistic assumptions about the calibration of the mass-observable relation, \cite{Carbone2012JCAP,Basse:2013zua,Cerbolini:2013uya,2016MNRAS.459.1764S} find that the promising increase in the number of detected clusters will provide tight constraints on models of dark energy or non-minimal massive neutrinos.

To forecast constraints from cluster number counts, we have followed a similar approach to \cite{2016MNRAS.459.1764S} on the modelling of the mass-observable scaling relation and selection function. The cluster's mass is derived from a scaling relation with an observable quantity, such as the cluster richness, luminosity, velocity dispersion or shear from gravitational lensing. We call $M_{\rm obs}$ this mass. The estimated number counts of these clusters for a redshift bin $l$ and mass bin $m$, corresponding to $z_l$ and $M_{{\rm obs},m}$, can then be expressed as
\begin{equation}\label{eq:nlm}
     N_{l, m} = \int_{\Omega_{\rm tot}} \de \Omega \int^{z_{l+1}}_{z_l}\de z \;\frac{\de V}{\de z\,\de \Omega}\, \int^{+\infty}_0 \de M \; \frac{\de n(M,z)}{\de M} \frac{1}{2}\left[ \text{erfc}  (x_{m,l}) - \text{erfc}(x_{m+1,l})\right]\,,
\end{equation}
under the assumption of a log-normal observed mass distribution. Here $\de n(M,z)/ \de M$ is the cluster mass function of \cite{Euclid:2022dbc} defined below, $\de \Omega$ is the solid angle element in steradians, $\de V/(\de z \,\de \Omega)$ is the derivative of the comoving volume with respect to the redshift and solid angle element, 
\begin{equation}
 \frac{\de V}{\de z\,\de \Omega}= \, \frac{c\,(1+z)^2\, D_{\rm A}^2(z)}{H(z)}\,, 
 \label{eq:volume_element}
\end{equation}
with $D_{\rm A}(z)$ being the angular diameter distance and $c$ the speed of light, while erfc$(x)$ is the complementary error function, with the argument $x$ being the (biased) logarithm of the mass (see below). We can explicitly write this as $x_{m,l}=x\,(M_{{\rm obs}, m, l})$ defined in each mass bin $m$ and redshift bin $l$ as 
\begin{equation}
    x_{m,l} = x\,(M_{{\rm obs},m,l}) = \frac{\ln (M_{{\rm obs}, m,l}/M) - \mathcal{B}_{\mathrm{bias},l}}{ \sqrt{2 \, \sigma^2_{ M_{{\rm obs}, m,l}}}}\,.
\end{equation}
We define the bias and the variance to be
\begin{align} \label{eq:scaling_euclid}
    \mathcal{B}_{\mathrm{bias},l} = \, B_{M,0} + b_\sfont{E} \, \ln{\Bigg(\frac{M}{M_\mathrm{pivot}} \Bigg)} + \alpha_\sfont{E} \, \ln(1+z_l)\,, \quad \mathrm{and} \quad \sigma^2_{M_{{\rm obs}, m,l}} =  \sigma^2_{M_{\rm obs},0} -1 +(1+z_l)^{2\,\beta_\sfont{E}}\,,  
\end{align}
where $B_{M,0}$, $b_\sfont{E}$, and $\alpha_\sfont{E}$ are free parameters to quantify, respectively, the change in calibration, slope, and redshift dependence in the mass-observable-biased log-normal mass distribution, with $M_\mathrm{pivot} = 3 \times 10^{14} \,h^{-1} M_\odot$, while $\sigma^2_{M_{\rm obs},0}$ and $\beta_\sfont{E}$ serve also to respectively calibrate and account for the redshift dependence in the variance.
The predicted number density of halos of mass $M$ at redshift $z$ (the mass function) is given by
\begin{equation}
    \frac{\de n(M,z)}{\de M} = \frac{\rho_{{\rm bc},0}}{M}\,\mathcal{F}\big[\sigma(R,z)
\big] \,  \frac{\de \ln \big( \, [\sigma(R,z) \big ]^{-1} \, \big)}{\de M} \,,
\label{eq:mf}
\end{equation}
where $\sigma(R,z)$ is the variance of the density field within a sphere of radius $R$ at redshift $z$, and it is computed following \cite{Costanzi2013} neglecting the massive neutrino component, from the linear CDM+baryon power spectrum $P_{\rm cc}(k,z)$ as
\begin{equation}
\sigma^2(R,z) = \frac{1}{2\pi^2} \int_0^\infty \diff k \, k^2 P_{\rm cc}(k,z) W^2(kR) \,,
\end{equation}
where $W(kR)$ is the top-hat filter in $k$-space,
\begin{equation}
   W(kR) = 3\frac{\left[\sin(kR)-kR\cos(kR)\right]}{(kR)^3} \,,
\end{equation}
$R=R(M)= \left(3M/4 \pi \rho_{\rm bc,0}\right)^{1/3} $ is the radius of a sphere enclosing a mass $M$, and $\rho_{\rm bc,0}= \rho_{\rm crit,0}\,\Omega_{\rm bc,0}$ is the mean CDM+baryon energy density at $z=0$. Here we have used the critical density $\rho_{\rm crit,0}$ and the CDM+baryon density fraction $\Omega_{\rm bc,0} = \Omega_{\rm m,0} - \Omega_{\nu,0}$.
Finally, the multiplicity function reads, according to the modified Press--Schechter formalism of \cite{Euclid:2022dbc}:
\begin{equation}
\label{eqn:mult}
    \mathcal{F}(\nu,z)=\nu \, A(p,q) \sqrt{\frac{2a}{\pi}} \mathrm{e}^{-a\nu^2/2} \left(1+ \frac{1}{(a\nu^2)^p} \right) (\nu\sqrt{a})^{q-1} \,,
\end{equation}
where $\nu(M,z)=\delta_c(z)/\sigma(R,z)$, and $\delta_c(z)$ is the linear density contrast for spherical collapse. This can be computed following the prescription of \cite{Weinberg:2002rd}:
\begin{equation}
   \delta_{c, {\rm lin}}(z) \simeq  \frac{3}{20} (12\pi)^{2/3}[1+0.0123 \logten \Omega_{\rm m}(z)]\,.
\end{equation}
Note that this formula does not take into account any massive neutrino, except for $\Omega_{\rm m}$.
Here the matter density fraction $\Omega_{\rm m}(z)$ at redshift $z$ is computed from the present day total matter density $\Omega_{\rm m,0}$ as
\begin{equation}
    \Omega_{\rm m}(z) = \Omega_{\rm m,0} (1+z)^3 \frac{H_0^2}{H^2(z)} \, .
\end{equation}
The redshift and scale-dependence of the parameters of \cref{eqn:mult} can be written as
\begin{align}
    A(p,q) = & \left\{ \frac{2^{-0.5-p+q/2}}{\sqrt{\pi}} \left[ 2^p \Gamma(q/2)+ \Gamma(-p+q/2)  \right] \right\}^{-1} \,,\\
    q = & \, q_{\rm R} \, [\Omega_{\rm bc}(z)]^{q_z} \,, \\
    q_{\rm R} = & \, q_1 + q_2 \left[ \frac{\de \ln \sigma(R,z)}{\de \ln R} + 0.5 \right]\,, \\
    p = & \, p_1 + p_2 \left[ \frac{\de \ln \sigma(R,z)}{\de \ln R} + 0.5 \right]\,, \\
    a = & \, a_{\rm R} \, [\Omega_{\rm bc}(z)]^{a_z}\,, \\
    a_{\rm R} = & \, a_1 + a_2 \left[ \frac{\de \ln \sigma(R,z)}{\de \ln R} + 0.6125 \right]^2 \, ,
\label{eq:parevol}
\end{align}
where the parameters $a_i$, $p_i$, and $q_i$ are calibrated to the simulation. The adopted values of the seven fitted parameters are listed in \cref{tab:hmf_param}. 

\begin{table}
    \centering
    \footnotesize
    \caption{Parameters of the multiplicity function $\mathcal{F}(\nu,z)$ (Eq. \ref{eqn:mult}).}
    \label{tab:hmf_param}
   \begin{tabular}{lr}
   
	Parameter			&  Value~~	\\
    \hline \vspace{-2.0mm} \\
	$a_1$		& $0.7962$ 		\vspace{0.5mm} \\
	$a_2$		& $0.1449$ 		\vspace{0.5mm} \\
        $a_z$		& $-0.0658$ 	\vspace{0.5mm} \\
        $p_1$		& $-0.5612$ 	\vspace{0.5mm} \\
        $p_2$		& $-0.4743$ 	\vspace{0.5mm} \\
        $q_1$		& $0.3688$ 	    \vspace{0.5mm} \\
        $q_2$		& $-0.2804$ 	\vspace{0.5mm} \\
        $q_z$		& $0.0251$ 	    \vspace{0.5mm} \\
    \hline \vspace{-3mm}\\
    \end{tabular}
\end{table}
The corresponding likelihood function is based on Poisson statistics \citep{1979ApJ...228..939C,Holder:2001db, Bonamente:2019efn}:
\begin{equation}
\ln \mathcal{L}_{\rm CC} = \ln \mathcal{P} (n_{l,m}|N_{l,m}) = \sum_{l=1}^{N_l}\sum_{m=1}^{N_m} [n_{l,m} \ln(N_{l,m}) -
  N_{l,m} - \ln(n_{l,m}!)]\,,
\end{equation}
where $\mathcal{P}(n_{l,m}|N_{l,m})$ is the Poisson distribution probability of finding $n_{l,m}$ clusters given an expected number of $N_{l,m}$ in each bin in redshift and mass. The $N_{l,m}$ are computed from \cref{eq:nlm}, whereas the $n_{l,m}$ are the fiducial values of the $N_{l,m}$.
To estimate the number counts, we have considered equally spaced redshift bins in $z \in [0.2, 1.8]$, with a width of $\Delta z=0.1$. As for the limiting mass, we have followed the selection function used in \cite{2016MNRAS.459.1764S} in the pessimistic case where the lower mass for clusters is defined as the one corresponding to the significance of detection threshold, or the ratio between the cluster galaxy number count and the field RMS, $N_{\rm c}/\sigma_{\rm field}$ is above $5$. 
In our analysis, we also conservatively vary the nuisance parameters. In order to compute the fiducial mocks the latter are set to $(B_{M,0},\, b_\sfont{E}, \, \alpha_\sfont{E},\,  \sigma_{\ln M,0},\, \beta_\sfont{E})=(0.0,\, 1.0, \,0.0, \, 0.2, \, 0.125)$.

\subsection{Cosmic microwave background}\label{sec:cmb}

In \cref{sec:results_euclid} we forecast the sensitivity to neutrino parameters of the \Euclid probes combined with CMB data from the \Planck satellite. In the context of a forecast, it is easier to describe the \Planck constraining power not through the actual data and likelihood, but through mock data and a synthetic likelihood mimicking the sensitivity of \Planck. This allows us to assume the same underlying cosmology for our \Euclid and \Planck mock data, and perform our forecast in ideal conditions -- of course, one should keep in mind that this is a very different exercise from a real data analysis, which can bring surprises, especially if different data sets are in tension due to statistical flukes, incomplete systematic modelling, or unaccounted physical effects.

As in many previous works, we use for this purpose the {\tt fake\_planck\_realistic} likelihood of the public \texttt{MontePython} package, which accounts for the measurement of CMB temperature, polarisation, and lensing by \Planck, with a sky coverage of $57\%$ and noise spectra very close to the actual ones. This Gaussian likelihood accounts for three correlated observables: the CMB temperature map; the E-mode polarisation map, and the reconstructed CMB lensing potential map. In principle, these observables are slightly correlated with \Euclid observables because the same clusters can shear high-redshift \Euclid galaxies and distort patterns on the last scattering surface. This correlation will be taken into account in the analysis of real \Euclid data, but, for simplicity, we neglect it in the present forecasts.

Next, we estimate the potential of \Euclid data in combination with future CMB data from the LiteBIRD satellite, optimised for large angular scales, and from the CMB Stage-IV survey (CMB-S4), optimised for smaller angular scales. As in \cite{Brinckmann:2018owf} we model this combination with two mock likelihoods in the public \texttt{MontePython} package: the {\tt litebird\_lowl} likelihood accounts for LiteBIRD temperature and polarisation data in the multipole range where the survey is the most constraining, $2 \leq \ell \leq 50$; and the {\tt cmb\_s4\_highl} likelihood for temperature, polarisation, and lensing data from CMB-S4 at $\ell>50$. Details on the assumed sensitivity can be found directly in the numerical package or in \cite{Brinckmann:2018owf}. In that case, we follow the same methodology as for \Planck: the CMB mock data account for the same fiducial model as \Euclid and we neglect correlations between the CMB and large-scale structure data, although this correlation will be more important for a survey very sensitive to CMB lensing like CMB-S4.  

For CMB experiments, the predicted error on \neff{} depends mainly on the sensitivity of data to small angular scales. This means that in our modelling of LiteBIRD+CMB-S4 the assumptions regarding the CMB-S4 sensitivity at large $\ell$ are crucial. As in the rest of this forecast paper, we stick to conservative assumptions. Similarly to \cite{Brinckmann:2018owf}, we assume that foregrounds can be removed from CMB temperature and polarisation data only up to
$\ell_\mathrm{max}=3000$,
that the Galactic cut imposes a sky coverage
$f_\mathrm{sky}=0.40$,
and that the CMB-S4 instrument can be modelled with a beam width of
$\theta=3.0\,$arcmin
and a sensitivity to temperature of
$\sigma_\mathrm{T}=1.0\,\mu$K\,arcmin.
With such assumptions, the combination of LiteBIRD+CMB-S4 predicts
$\sigma(\neff)=0.038$
in the $\Lambda$CDM+\smnu+\neff{} case -- see Table~5 of \cite{Brinckmann:2018owf}. Since in this case the posterior is nearly Gaussian, this implies a 95\% CL upper bound
$\Delta \neff< 0.076$.
This is consistent with the forecasts of the CMB-S4 white paper \cite{CMB-S4:2016ple}, 
but that paper explores many other assumptions on CMB-S4 data, with a maximum multipole for the temperature data in the range
$\ell_\mathrm{max}^\mathrm{TT} \in [3000, 5000]$,
a sky fraction
$f_\mathrm{sky} \in [0.1,0.8]$,
a beam width of
$\theta \in [1.0, 3.0]\,$arcmin,
and a sensitivity of
$\sigma_\mathrm{T} \in [1.0, 3.0]\,\mu$K\,arcmin
-- see figures~22--24 of \cite{CMB-S4:2016ple}. We should stress that in that regard the summary plot of figure~23 (right panel) -- which can be compared with our~\cref{fig:neff} -- CMB-S4 forecasts are performed under extremely optimistic assumptions:
$\ell_\mathrm{max}^\mathrm{TT} = 5000$, $f_\mathrm{sky} = 0.5$ (respectively 0.7), $\theta = 1.0\,$arcmin, and $\sigma_\mathrm{T}= 1.0\,\mu$K\,arcmin,
which gives a 95\% CL upper bound $\Delta \neff< 0.054$ (respectively 0.046).
While these numbers provide an example of what CMB Stage-IV may ideally achieve, we want to give more conservative estimates of what will be possible with the combined surveys.

\section{Forecast method\label{sec:method}}

Having described our modelling of observables and likelihood functions in \cref{sec:euclid-observables,sec:add_prob}, we now explain how such likelihoods can be used for the purpose of \Euclid forecasts. Our forecast methods are based on \cite{EUCLID:2023uep}, adjusted to account for the presence of massive neutrinos. The four different pipelines described in this section are validated against each other in \cref{sec:validation}, and one of them -- the MCMC pipeline -- is used for deriving our main results in \cref{sec:results_euclid}.

For well constrained parameters we can approximate the posteriors with a multivariate Gaussian. Since such a distribution can be described by a mean vector and a covariance matrix, with such an approximation all the information concerning the sensitivity of the experiment to the parameters of the model is contained in the Hessian matrix $F_{\alpha \beta}$ of the log-likelihood at the best fit, also called the Fisher Information (FI) matrix:
\begin{align}
    F_{\alpha\beta} \equiv -\langle \partial_\alpha \partial_\beta \left. \ln \mathcal{L} \right|_\mathrm{best\,fit}\rangle \,,
    \label{eq:fi}
\end{align}
where the indices $\{\alpha, \beta\}$ run over model (cosmological or nuisance) parameters. Our version of the \montepython{} package implements mock likelihood functions ${\cal L}(\theta_\alpha)$ for each of the \Euclid probes described in \cref{sec:euclid-observables}. The pipeline referred to later as {\tt MP/Fisher} (the \montepython{} package used in Fisher mode) directly evaluates the FI matrix from \cref{eq:fi}. First, the code calls the Einstein--Boltzmann solver (EBS) \class{} to evaluate the cosmological observables required by the likelihood calculation (power spectra, growth rate, etc.). Then, it performs a numerical calculation of two-sided second-order derivatives, based on the evaluation of the likelihood at the best fit and in $2N^2$ neighbouring points, where $N$ is the number of model parameters.\footnote{To compute the second derivatives using a finite difference scheme, after computing the likelihood at the best fit, the code needs to perform two likelihood evaluations for each of the $N$ second derivatives with $\alpha=\beta$, and four evaluations for each of the $N(N-1)/2$ cross-derivatives with $\alpha \neq \beta$, leading to a total of $2N^2$ evaluations.} 

Due to the computational cost and complexity of evaluating second-order derivatives, one can use the fact that the likelihoods have a relatively simple Gaussian dependence on the power spectra (but not on the parameters) to express the Fisher matrix in terms of first derivatives only. For the photometric probe, the FI matrix can be expressed as 
\begin{equation}
F_{\alpha \beta} = \frac{1}{2} f_\mathrm{sky} \sum_\ell (2\ell+1) \mathrm{Tr} \left\{ \left[\tens{C}^\mathrm{fid}(\ell)\right]^{-1} \left[ \left. \partial_\alpha \tens{C}^\mathrm{th}(\ell)\right|_\mathrm{fid}\right] \left[{\tens{C}^\mathrm{fid}(\ell)}\right]^{-1} \left[\left. \partial_\beta \tens{C}^\mathrm{th}(\ell)\right|_\mathrm{fid} \right]
\right\}\,,
\label{eq:fisherWL}
\end{equation}
where for each multipole $\ell$ the matrices $\tens{C}^\mathrm{fid}$ and $\tens{C}^\mathrm{th}$ are built out of the angular power spectra $C_{ij}^{XY}$ defined in \cref{eq:photo_Cl}, evaluated, respectively, at fiducial parameter values in the case of $\tens{C}^\mathrm{fid}$, or at two points (per dimension) in parameter space to compute numerically the first-order derivative $\partial_\alpha \tens{C}^\mathrm{th}|_\mathrm{fid}$. Since with this method only the first-order derivatives are evaluated, the total number of points needed reduces to $2N$ (using a first-order double-sided finite difference). We refer the reader to \cite{EUCLID:2023uep} for more details on the definition of the matrices $\tens{C}^\mathrm{fid}$ and $\tens{C}^\mathrm{th}$. For the spectroscopic probe, the FI matrix becomes instead
\begin{equation}
    F_{\alpha \beta} = \sum_i \frac{1}{8\pi^2} \int_{-1}^{1} \mathrm{d}\mu \int_{10^{-3}}^{k_{\rm max}} \mathrm{d}k\, k^2\, \partial_\alpha \mathrm{ln} (P_\mathrm{obs})\big|_\mathrm{fid} \,\, \partial_\beta \mathrm{ln} (P_\mathrm{obs})\big|_\mathrm{fid}  V_i^\mathrm{fid}\,,
    \label{eq:fisherGC}
\end{equation}
where $P_\mathrm{obs}$ is the observable power spectrum defined in \cref{eq:spectro_pk}, $k_{\rm max}$ is given in \cref{tab:spec_gc} and the index $i$ runs over four redshift bins with effective volume $V_i^\mathrm{fid}$. Details on the assumed survey characteristics and binning strategy are specified in \cite{EUCLID:2023uep}.

The pipeline that we refer to as {\tt CF}, that is, the \cosmicfish{} code, uses \cref{eq:fisherWL,eq:fisherGC} to compute the FI matrix. It calculates the matrices $\tens{C}(\ell)$ and the spectra $P_\mathrm{obs}$ at the fiducial point and at a given number of neighbouring points in parameter space, and numerically evaluates the first derivatives of these observables with respect to model parameters. By default, \cosmicfish{} uses two-sided first-order derivatives (two neighboring points). To check the numerical stability of this method, we have checked that we get similar result when using a 4-point forward stencil derivative, see \cref{app:der}. The version of \cosmicfish{} used in this work internally calls one of the EBSs (\camb{} or \class{}) to evaluate the cosmological observables, providing us with two forecast pipelines {\tt CF/CAMB} and {\tt CF/CLASS}.\footnote{These pipelines correspond to those called `\cosmicfish{} internal' in \cite{EUCLID:2023uep}, denoted as {\tt CF/int/CAMB} and {\tt CF/int/CLASS}. In this paper, we omit the `{\tt int}' since we do not use external precomputed spectra.}

Our fourth pipeline called {\tt MP/MCMC} is the \montepython{} package used in its default mode, which is running Markov chain Monte Carlo (MCMCs) with a Metropolis-Hasting algorithm to explore the parameter space and infer the posterior from the likelihood according to Bayes theorem. We always assume top-hat priors on model parameters, with prior edges chosen far in the exponential tails of the posteriors, such that the confidence intervals inferred from the MCMCs at the 95\% CL are unaffected by the choice of prior edge. The only exception is the summed neutrino mass $\smnu$, for which we impose a theoretical prior $\smnu>0$. The two main advantages of MCMC forecasts is that they rely neither on a Gaussian approximation to the likelihood, nor on the calculation of numerical derivatives. We note that the {\tt MP/Fisher} and {\tt MP/MCMC} pipelines call the very same \Euclid mock likelihoods, computing the same functions $\cal L(\theta_\alpha)$, which are the only parts of the pipeline depending on the cosmology; thus, it is impossible by construction to introduce a mistake in the modelling of cosmology (or of the \Euclid survey) when passing from one pipeline to the other. This means that the validation of one pipeline automatically implies the validation of the other pipeline (as long as the FI matrix is well conditioned and the MCMC has converged).

This statement could be questioned in the case in which one of the two approaches (FI matrix or MCMC) requires a more accurate calculation of the cosmological observables or of the likelihood functions. In the case at hand, we know that the highest precision is required by the FI matrix approach, because a robust evaluation of numerical derivatives with small step sizes requires high accuracy and low numerical noise. However, our approach consists of validating the {\tt MP/Fisher} pipeline first, which is more demanding in terms of accuracy; this in turn validates the {\tt MP/MCMC} pipeline, which is less demanding. We also explicitly compare the results from the two pipelines in \cref{sec:validation} to make sure that they agree.

\section{Code validation\label{sec:validation}}

In order to validate the four forecast pipelines described in \cref{sec:method} against each other, we need to choose a framework (fiducial model and set of parameters) in which the FI formalism can be applied in a robust way. In the results section, we will be interested in extensions of the minimal 5-dimensional flat $\Lambda$CDM model with up to four additional free parameters $\{ \smnu, \neff, w_0, w_a \}$.\footnote{We recall that $\tau_\mathrm{reio}$ is relevant for CMB observables but not for cosmic shear or galaxy clustering, making the model effectively 5- instead of 6-dimensional.} However, floating these nine parameters simultaneously opens up strong degeneracies in the parameter space and leads to very non-Gaussian posteriors. In that case, the Gaussian approximation breaks down. Then, the FI matrix calculation becomes unstable, since it depends heavily on the choice of numerical derivative step sizes. Besides, the FI matrix no longer provides a good estimate of the experimental sensitivity.

To overcome this issue, we decided to validate our forecasting pipelines by looking separately at three 7-parameter models for which the posterior remains close to Gaussian (as confirmed by contour plots from MCMC runs displayed in \cref{sec:val_mcmc}). This strategy does not limit the range of validity of our final results, since at the end, after validating the pipelines against each other, we will use the MCMC pipeline for our main results. This pipeline does not rely on any Gaussian approximation and can be used even for full 9-parameter forecasts.

The three 7-parameter models used for validation are defined as follows. Firstly, we stick to a $w_0w_a$CDM cosmology, with a time-varying dark energy equation of state parameter $w(a) = w_a (1-a) + w_0$ \citep{Chevallier:2000qy,Linder:2002et} but fixed $\{ \sum m_\nu, \neff \}$. This model was already used in the forecasts of  \cite{EUCLID:2023uep} and allows us to cross-check that the previous validation still holds.  Secondly, we investigate cosmologies where we vary the neutrino mass and the effective number of relativistic degrees of freedom, $\Lambda$CDM + $\sum m_\nu$+$\neff$, but stick to a cosmological constant. This framework allows us to cross-check our implementation on neutrino effects on the observables. Thirdly, we consider models with varying neutrino mass and dark energy equation of state, $w_0$CDM+$\sum m_\nu$, with fixed $\neff=3.044$ and $w_a=0$. This last set up will prove that our modelling of nonlinear corrections is consistent across the different pipelines even when neutrino and DE parameters are varied simultaneously.

The fiducial values of cosmological parameters used in \cref{sec:validation,sec:results_euclid} are summarised in \cref{tab:fiducial_validation}. Despite identical fiducial values, the models used for validation in this section and for forecasts in the next section feature two subtle differences with respect to each other. First, as already explained in \cref{sec:neff} (case b), we assume here a single massive neutrino species with mass $m_
\nu=\sum m_\nu$, which allows for easier comparison with earlier work and for simpler FI matrix calculations (given that the $\neff$ posterior remains symmetric around the fiducial value). Second, we fix here the nonlinear modelling parameters $\sigma_{\rm p}$ and $\sigma_{\rm v}$ to their fiducial values in order to keep the contours of the spectroscopic probe more Gaussian. In \cref{sec:results_euclid}, we will instead consider three degenerate massive neutrino species, each with mass $m_
\nu=\sum m_\nu/3$ (see \cref{sec:neff}, case a) and we will marginalise over $\sigma_{\rm p}$ and $\sigma_{\rm v}$.

\begin{table}[ht]
\renewcommand{\arraystretch}{1.2}
    \caption{Fiducial values of the cosmological parameters used in the validation and result sections. The optical depth to reionization is fixed to the \Planck best-fit value $\tau=0.0543$ when only \Euclid observables are included in the analysis, while we allow it to vary in the joint analysis with CMB probes.}
    \centering
    \begin{tabular}{ccccc|cccc}
    \hline
    \multicolumn{9} {c}{{\bf{Fiducial}}}\\ 
    \hline
     \multicolumn{5} {c}{$\Lambda$CDM} & \multicolumn{4} {|c}{Extensions} \\
     \hline
 \multicolumn{1}{c}{$\Omega_{\rm m,0}$} & \multicolumn{1}{c}{$100\,\Omega_{\rm b,0}$} & \multicolumn{1}{c}{$h$} & \multicolumn{1}{c}{$n_{\rm s}$} & \multicolumn{1}{c}{$\sigma_{8}$} & \multicolumn{1}{|c}{$\sum m_\nu$[${\rm meV}$]} & \multicolumn{1}{c}{$\neff$} & \multicolumn{1}{c}{$w_0$} & \multicolumn{1}{c}{$w_a$}\\
  \hline
        0.314571 & 4.92 & $0.6737$ & $0.9661$ & $0.81$ & $60$ & $3.044$ & $-1$ & $0$ \\
    \end{tabular}
    \label{tab:fiducial_validation}
\end{table}

We perform validation tests for each of the three 7-parameter models, each probe (photometric or spectroscopic), and each case (pessimistic or optimistic). Thus, the number of validation tests amounts to 12. In each of these 12 situations, we proceed as follows:
\begin{itemize}
    \item[1.] validate the predictions and accuracy settings of the EBSs by comparing the {\tt CF/CAMB} and {\tt CF/CLASS} pipeline results;
    \item[2.] validate our \montepython{} likelihood and our FI matrix implementation by comparing the results from the {\tt CF/CLASS} and {\tt MP/Fisher} pipelines;
    \item[3.] check the validity and stability of our FI forecasts (choice of step size in numerical derivatives and validity of Gaussian approximation) by comparing the previous two pipelines with {\tt MP/MCMC} runs.
\end{itemize}
In practice, we perform steps 1 and 2 simultaneously, by comparing all three derived FI matrices. For steps 1 and 2, our test consists of comparing the marginalised and unmarginalised errors on each cosmological and nuisance parameter.\footnote{Marginalised errors are obtained by an inversion of the Fisher matrix. They represent the error on one parameter when the posterior is integrated over all possible values of the other parameters, effectively erasing or averaging over their information.} Validation is achieved when all errors are within 10\% of the median. This subjective criterion\footnote{As detailed in Sect. 3.1.6 of \citetalias{Euclid:2019clj}, a fractional error of 10\% on the marginalized $1\,\sigma$ forecasted errors corresponds roughly to a  requirement at the $10^{-4}$ level on the numerical accuracy of the elements of the Fisher matrix and its inverse.} was already adopted in previous \Euclid validation papers like \citetalias{Euclid:2019clj} and \cite{EUCLID:2023uep}. The results of these tests are shown in \cref{sec:val_fi}. For step 3 we compare 1-dimensional posteriors and 2-dimensional confidence contours in triangle plots in \cref{sec:val_mcmc}.

\subsection{Validation of Fisher pipelines \label{sec:val_fi}}

\begin{figure}[htp]
    \centering
    \includegraphics[width=0.49\textwidth]{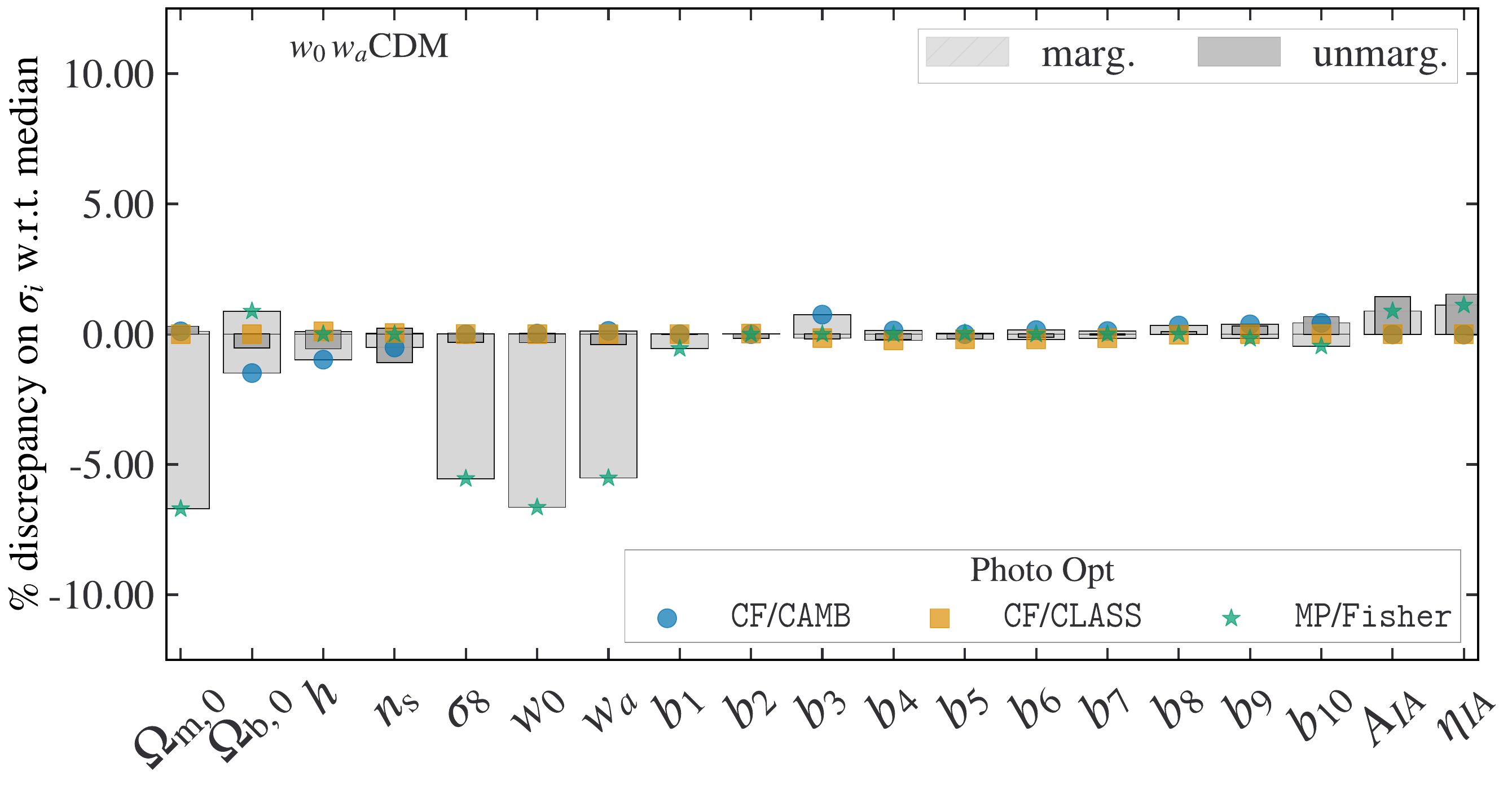}
    \includegraphics[width=0.49\textwidth]{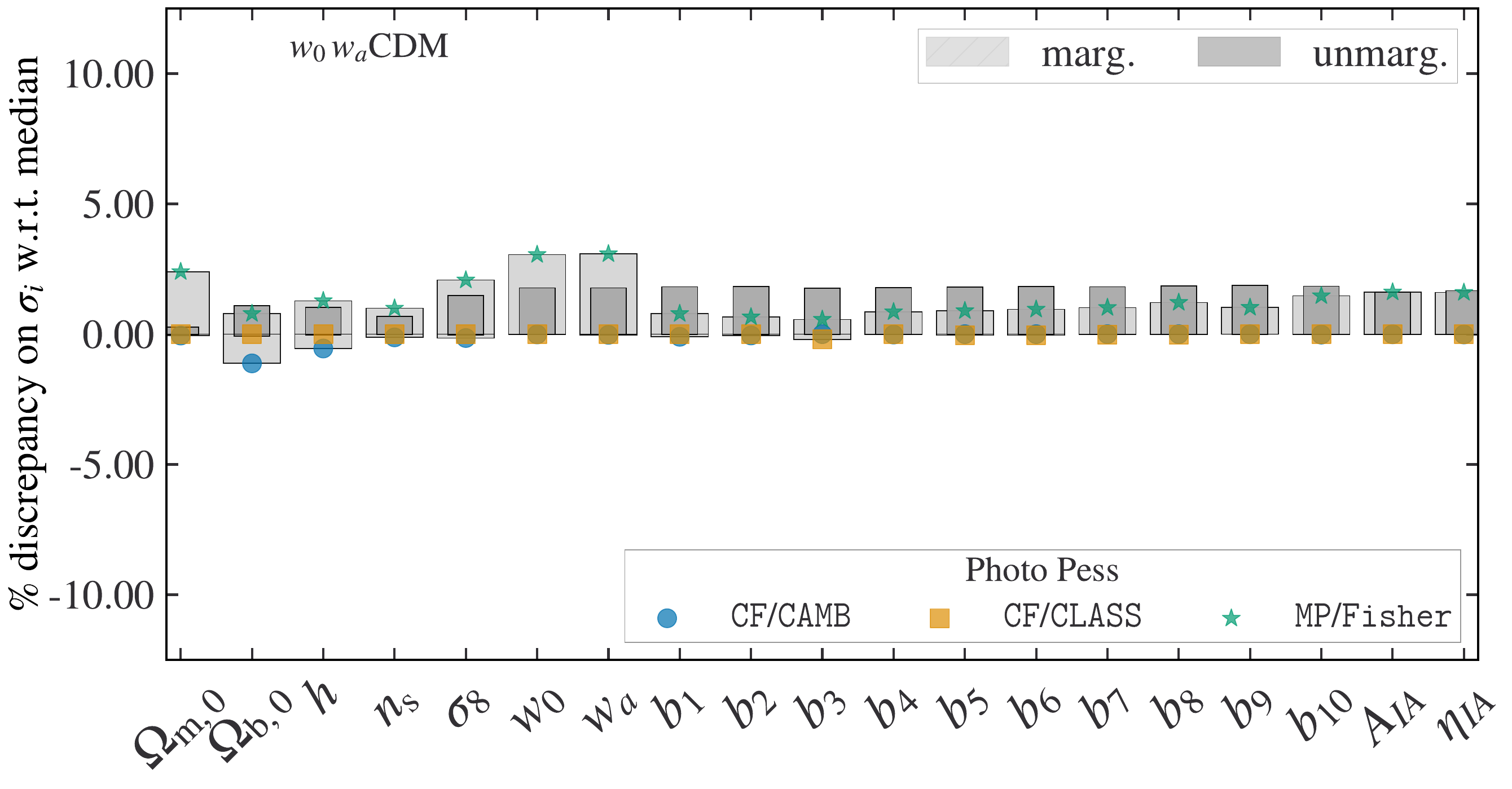}\\
    \includegraphics[width=0.49\textwidth]{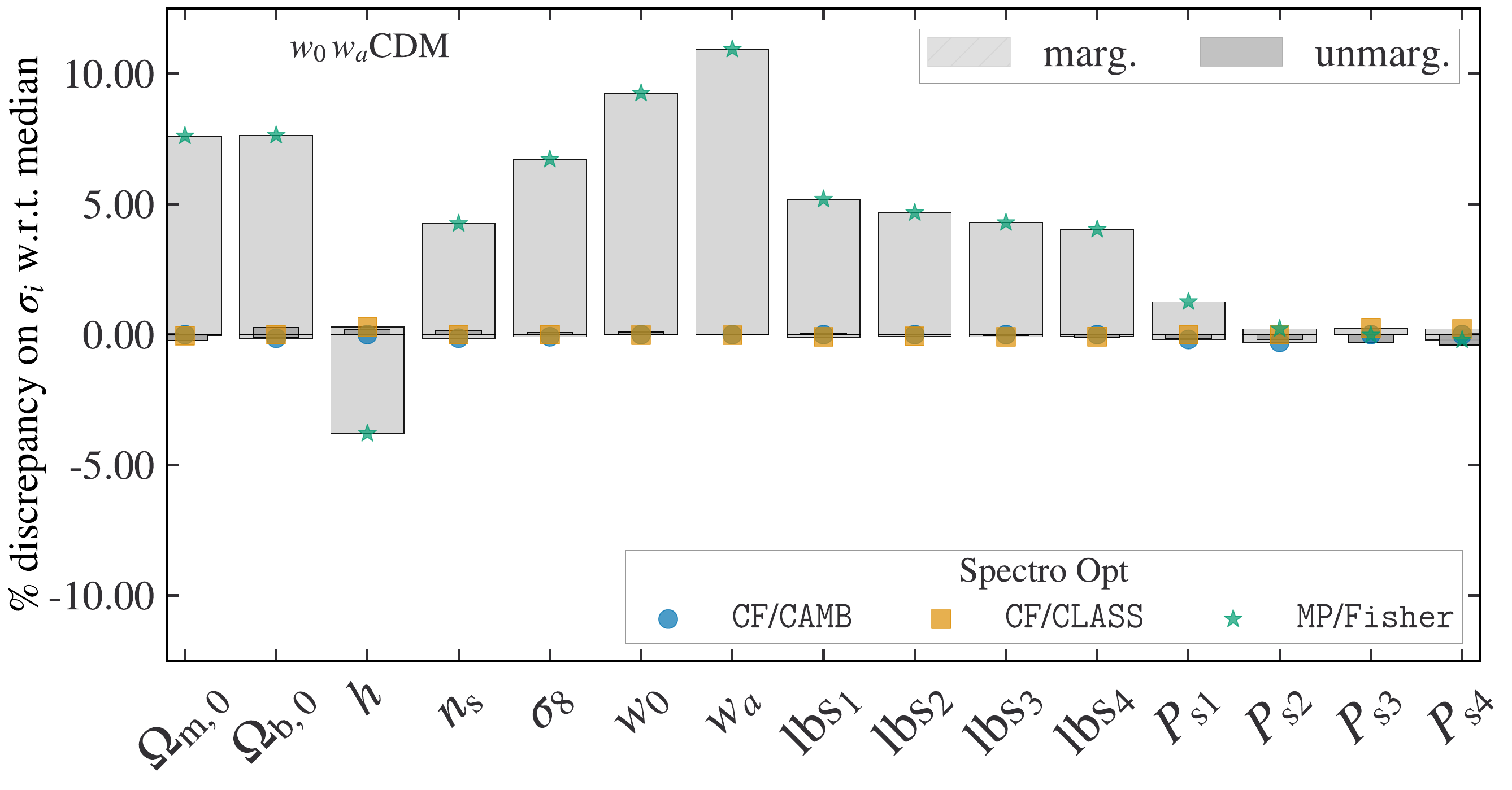}
    \includegraphics[width=0.49\textwidth]{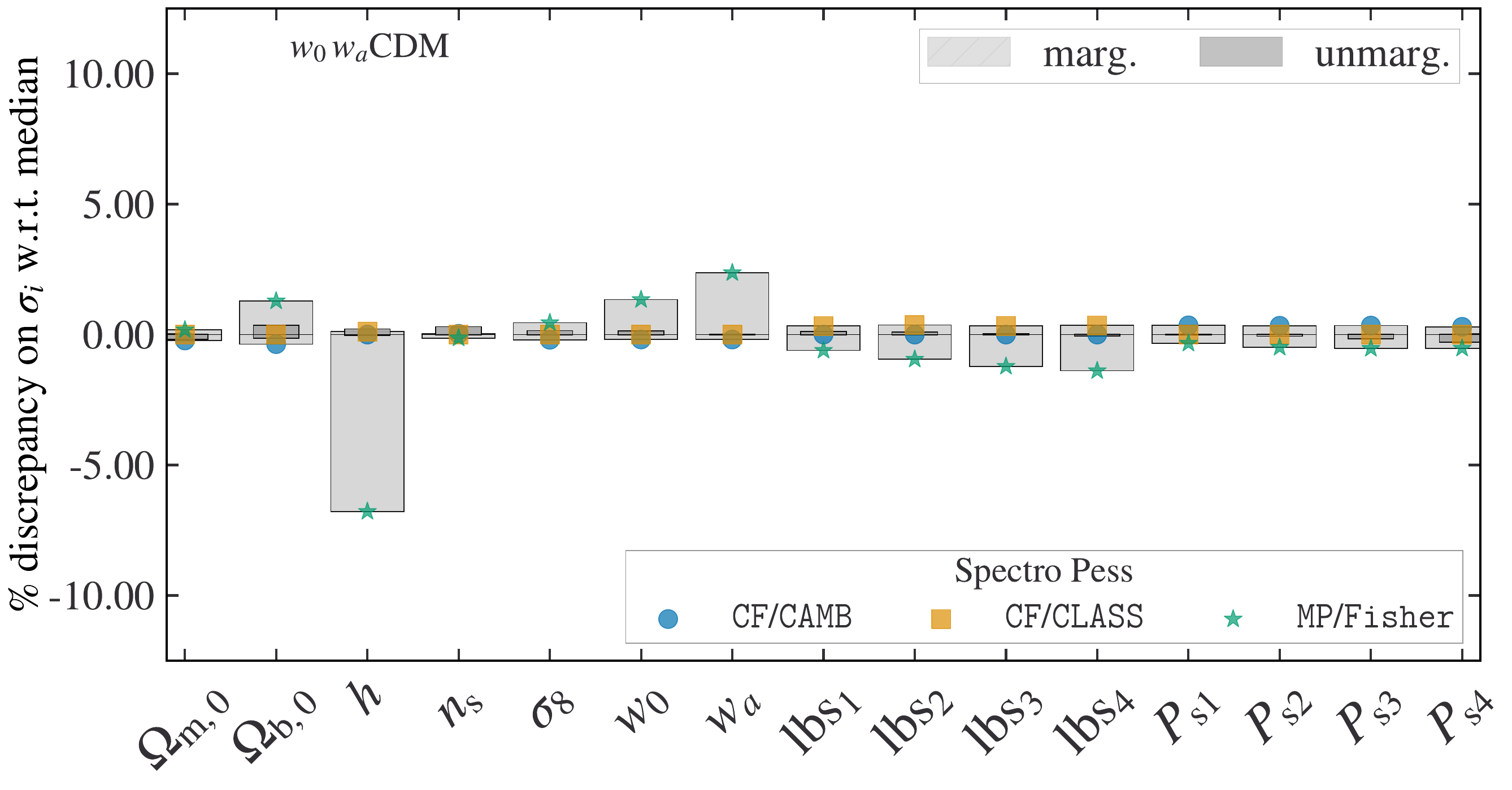}
    \caption{For the $w_0w_a$CDM model, comparison of each of the Fisher marginalised (light grey) and unmarginalised (dark grey) errors on each cosmological and nuisance parameter for: \emph{top left}, photometric optimistic probe; \emph{top right}, photometric pessimistic probe; \emph{bottom left}, spectroscopic optimistic probe; and \emph{bottom right}, spectroscopic pessimistic probe. For the spectroscopic probe the labels $lbs_i$ are short for the parameters $\ln(b  \,\sigma_8)_i$.
    \label{fig:1}
    }
\end{figure}

\begin{figure}[htp]
    \centering
    \includegraphics[width=0.49\textwidth]{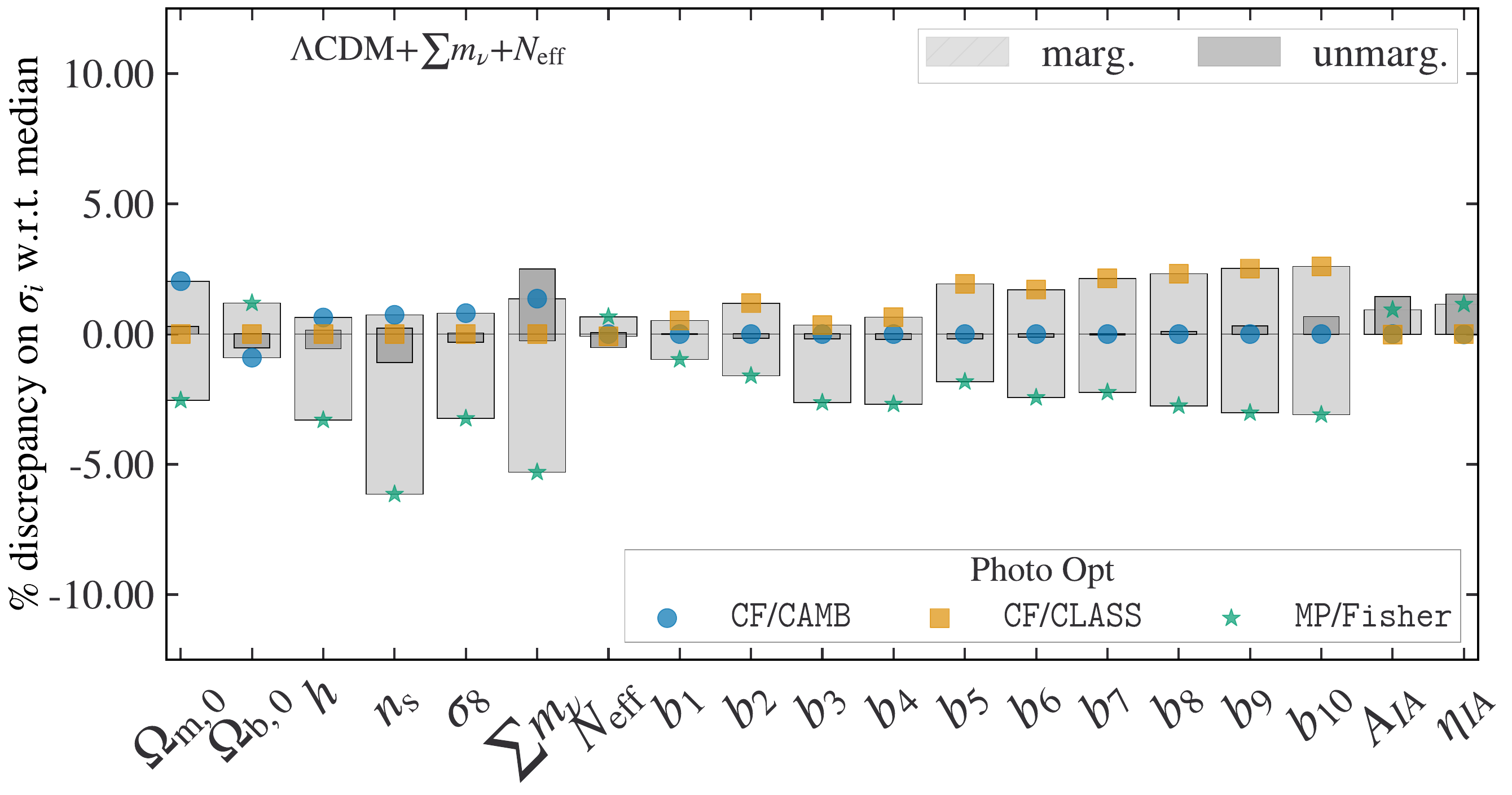}
    \includegraphics[width=0.49\textwidth]{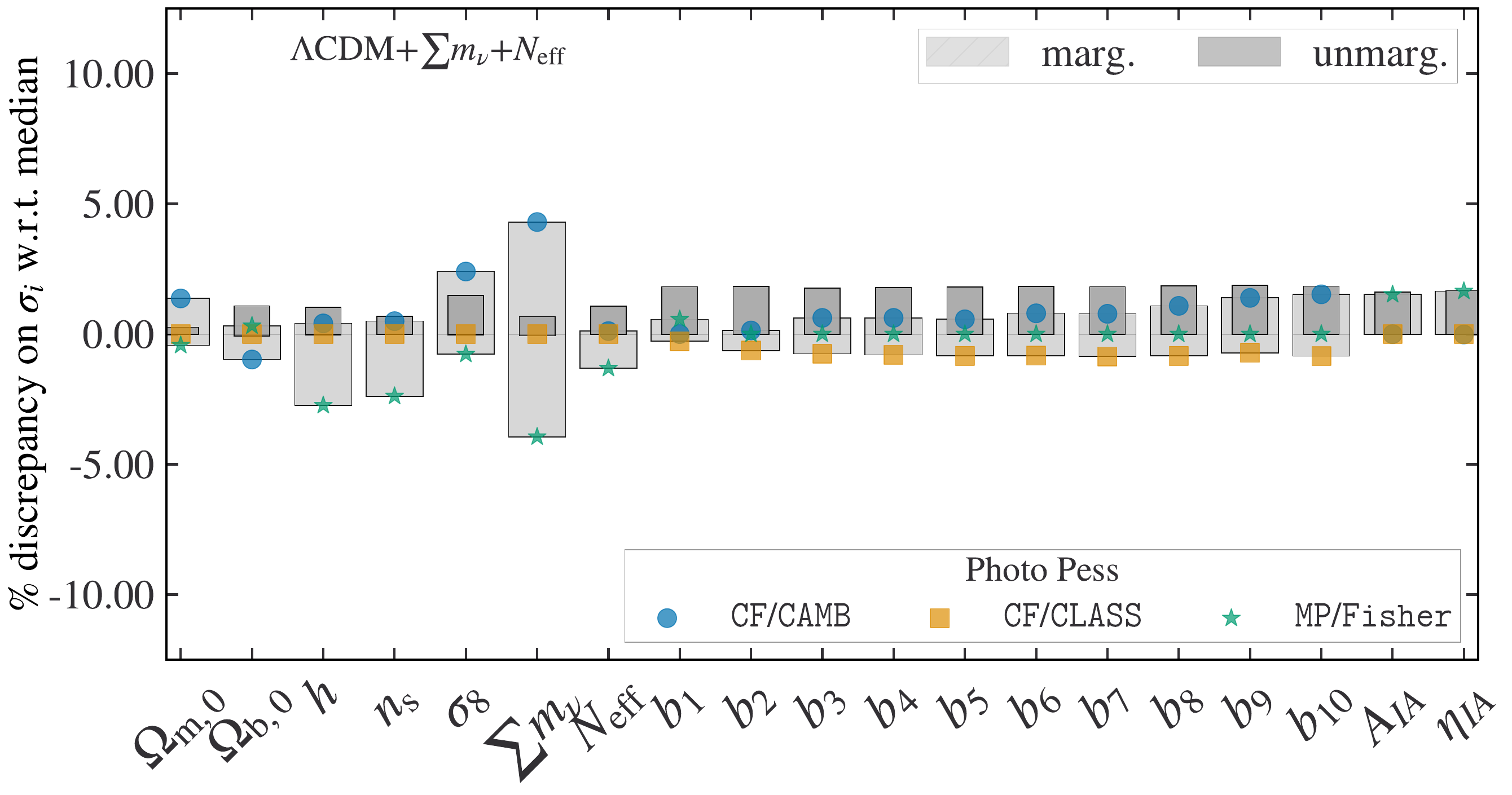}\\
    \includegraphics[width=0.49\textwidth]{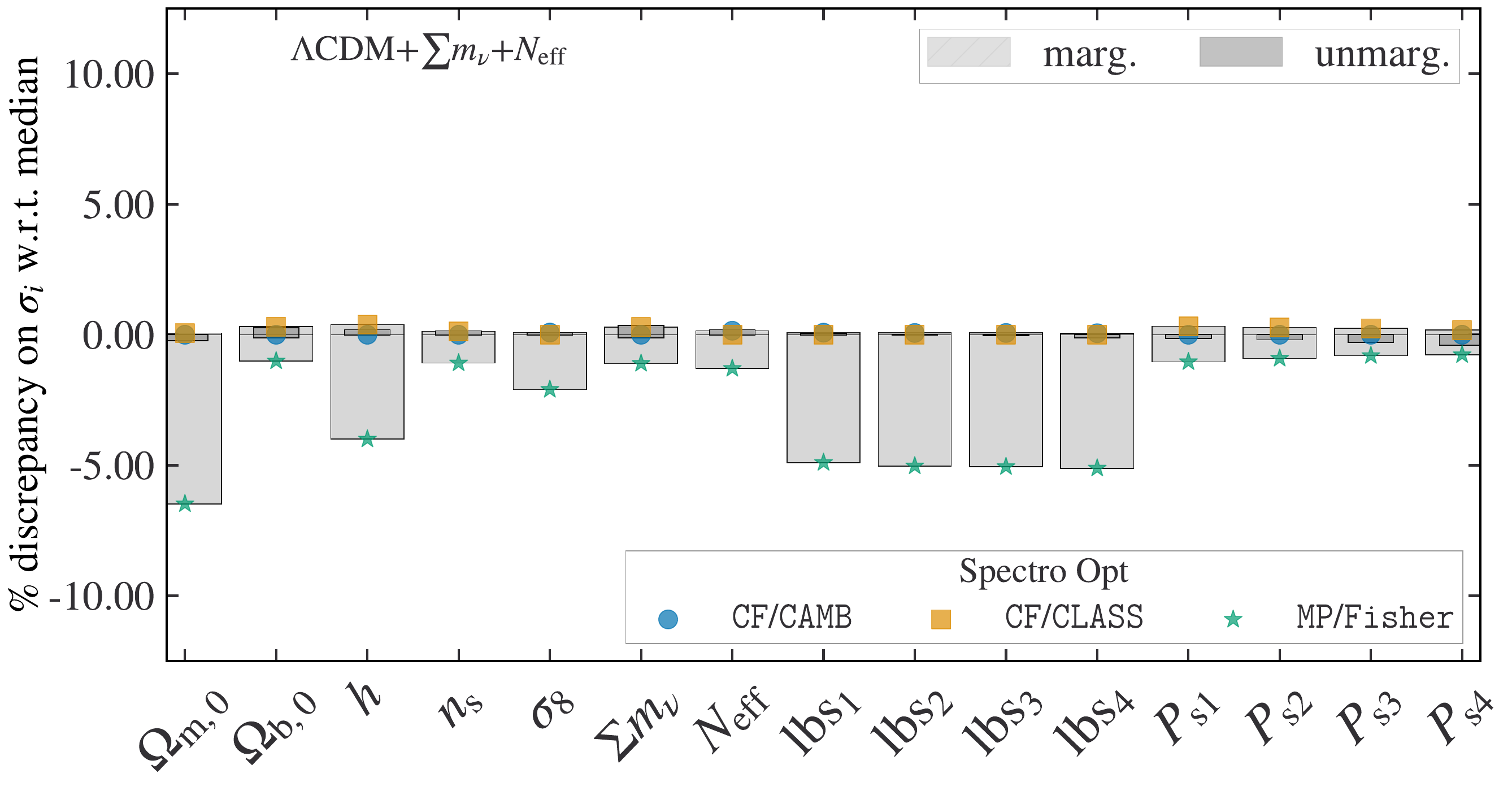}
    \includegraphics[width=0.49\textwidth]{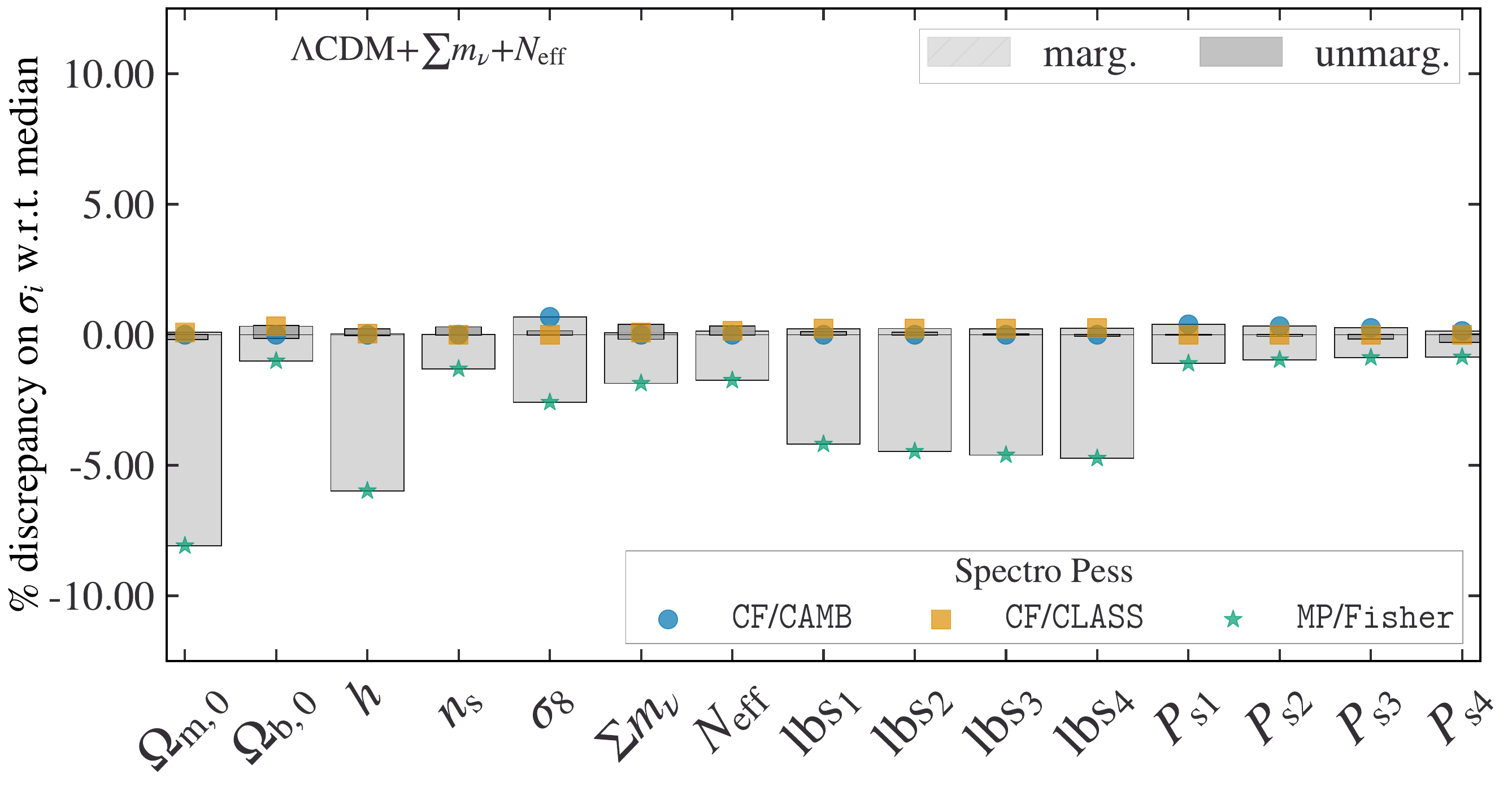}
    \caption{Same as \cref{fig:1} for the model $\Lambda$CDM+$\sum m_\nu$+$\neff$.
    \label{fig:2}
    }
\end{figure}

\begin{figure}[htp]
    \centering
    \includegraphics[width=0.49\textwidth]{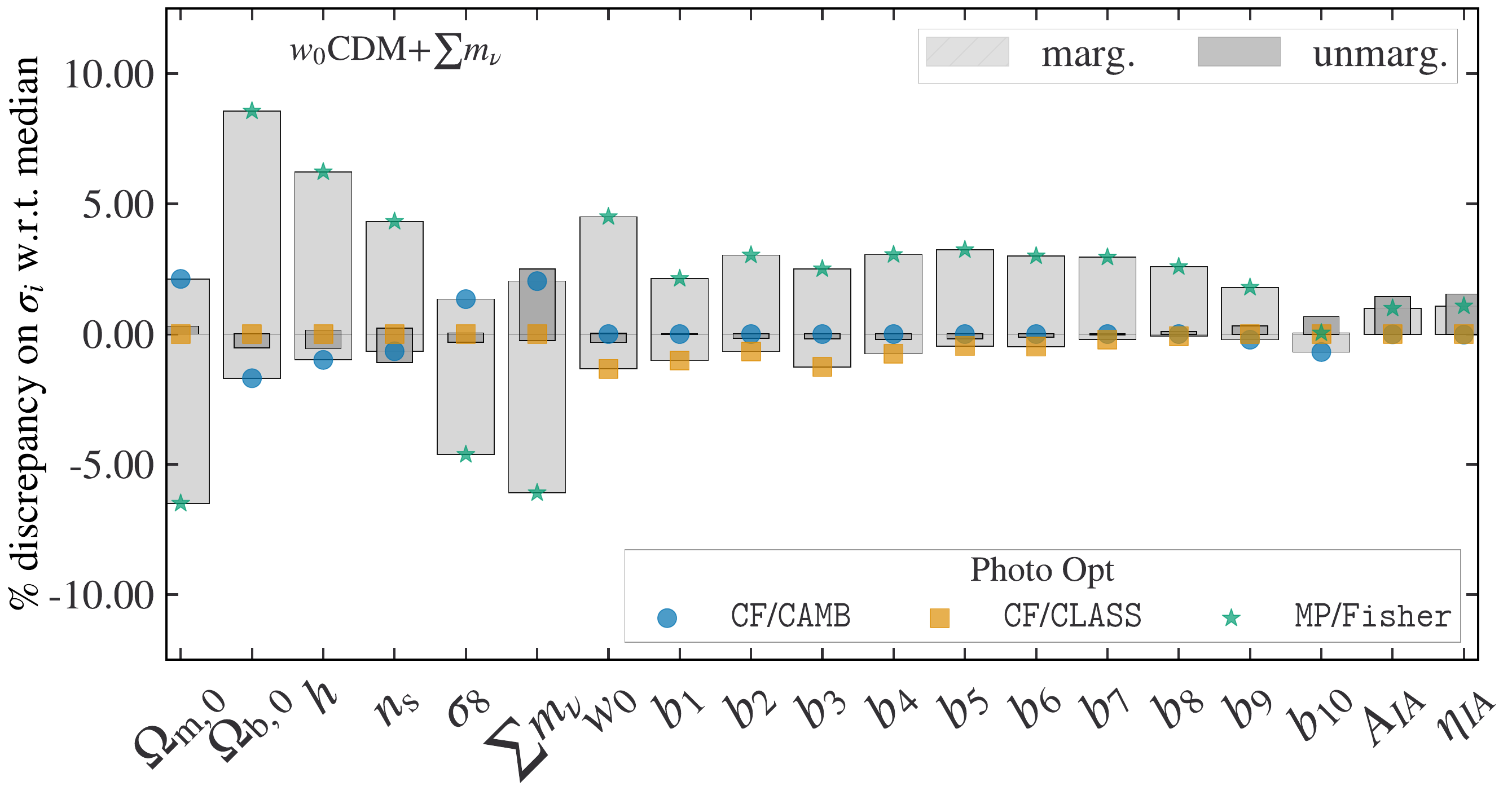}
    \includegraphics[width=0.49\textwidth]{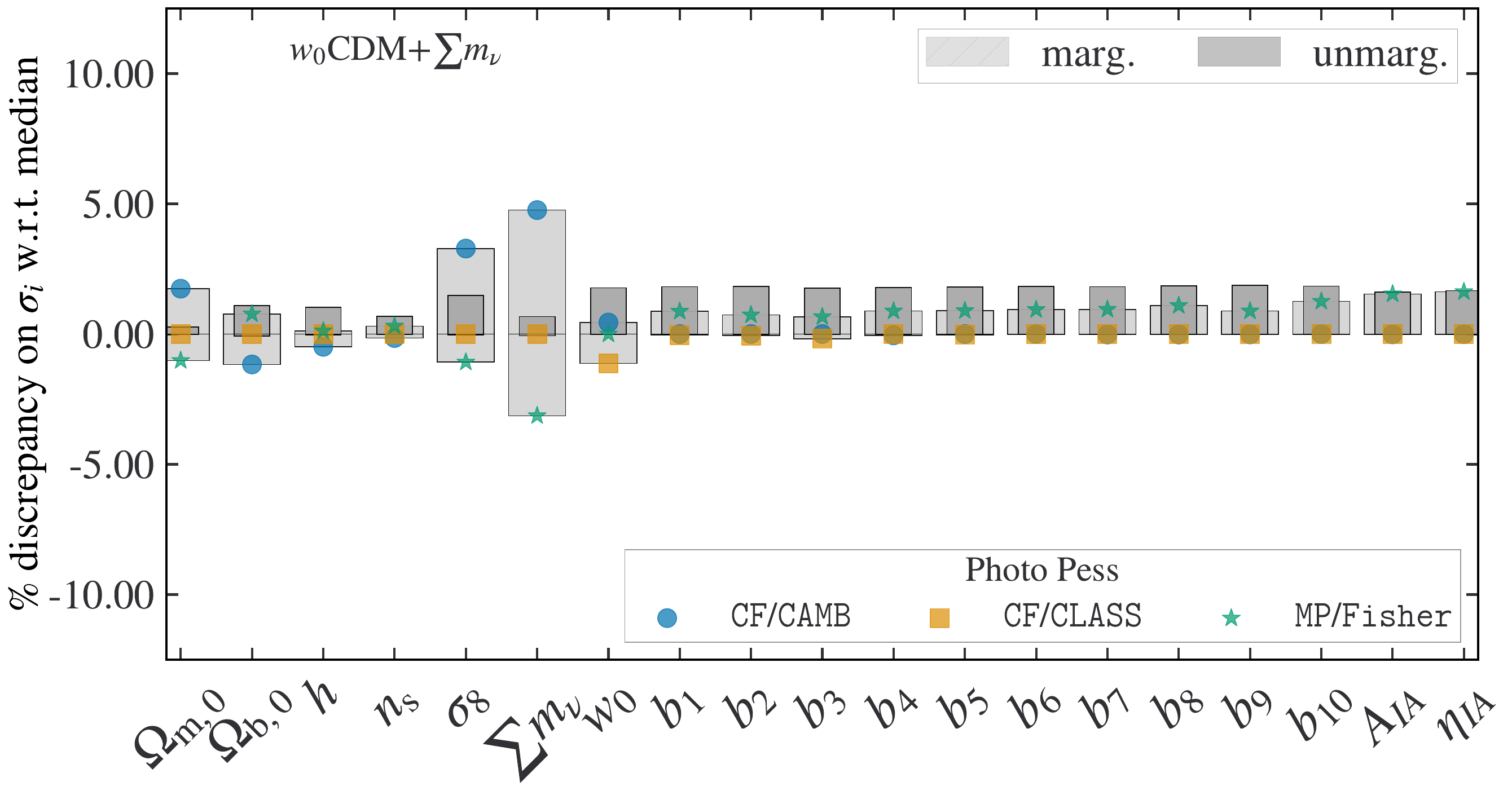}\\
    \includegraphics[width=0.49\textwidth]{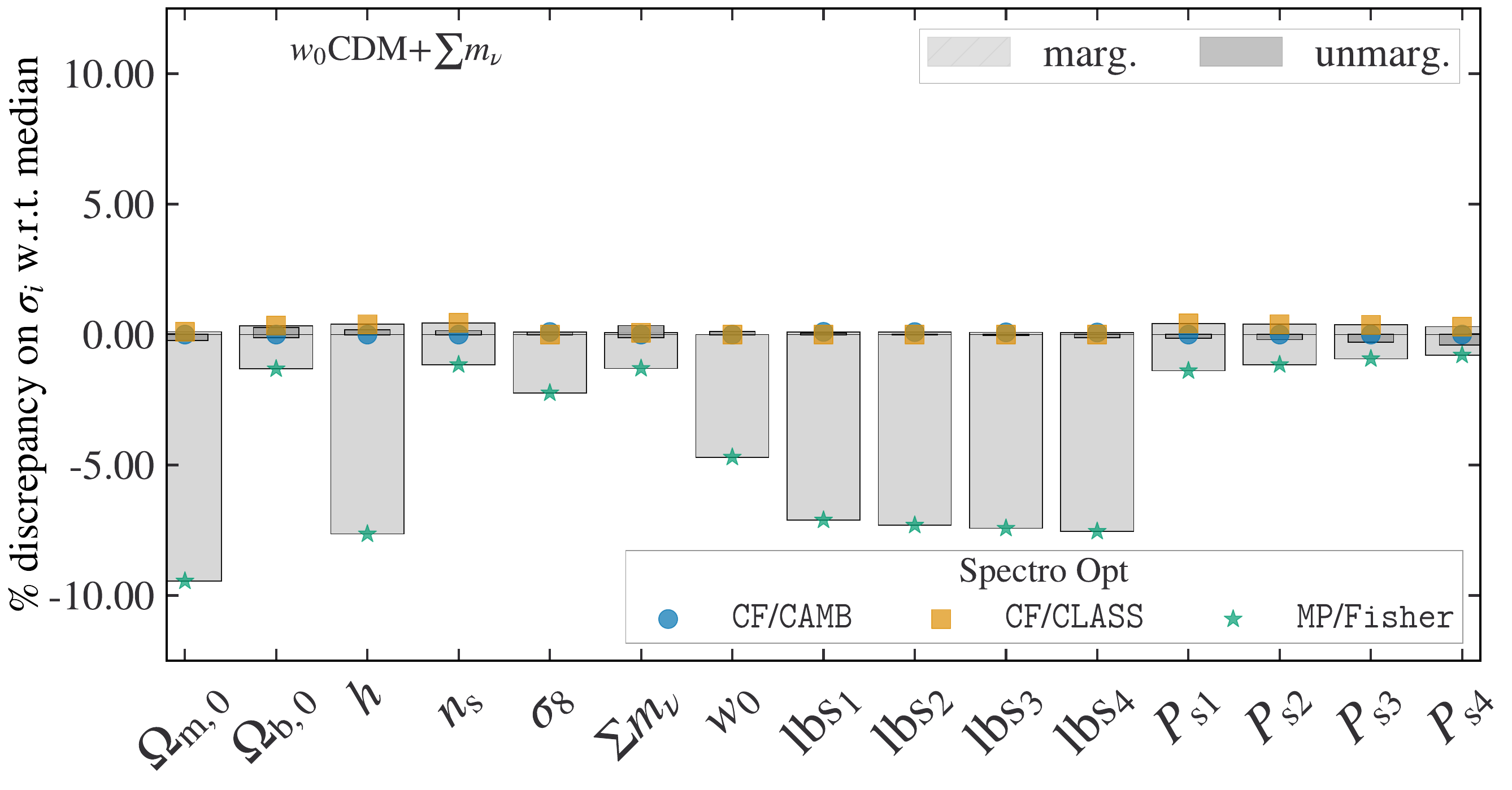}
    \includegraphics[width=0.49\textwidth]{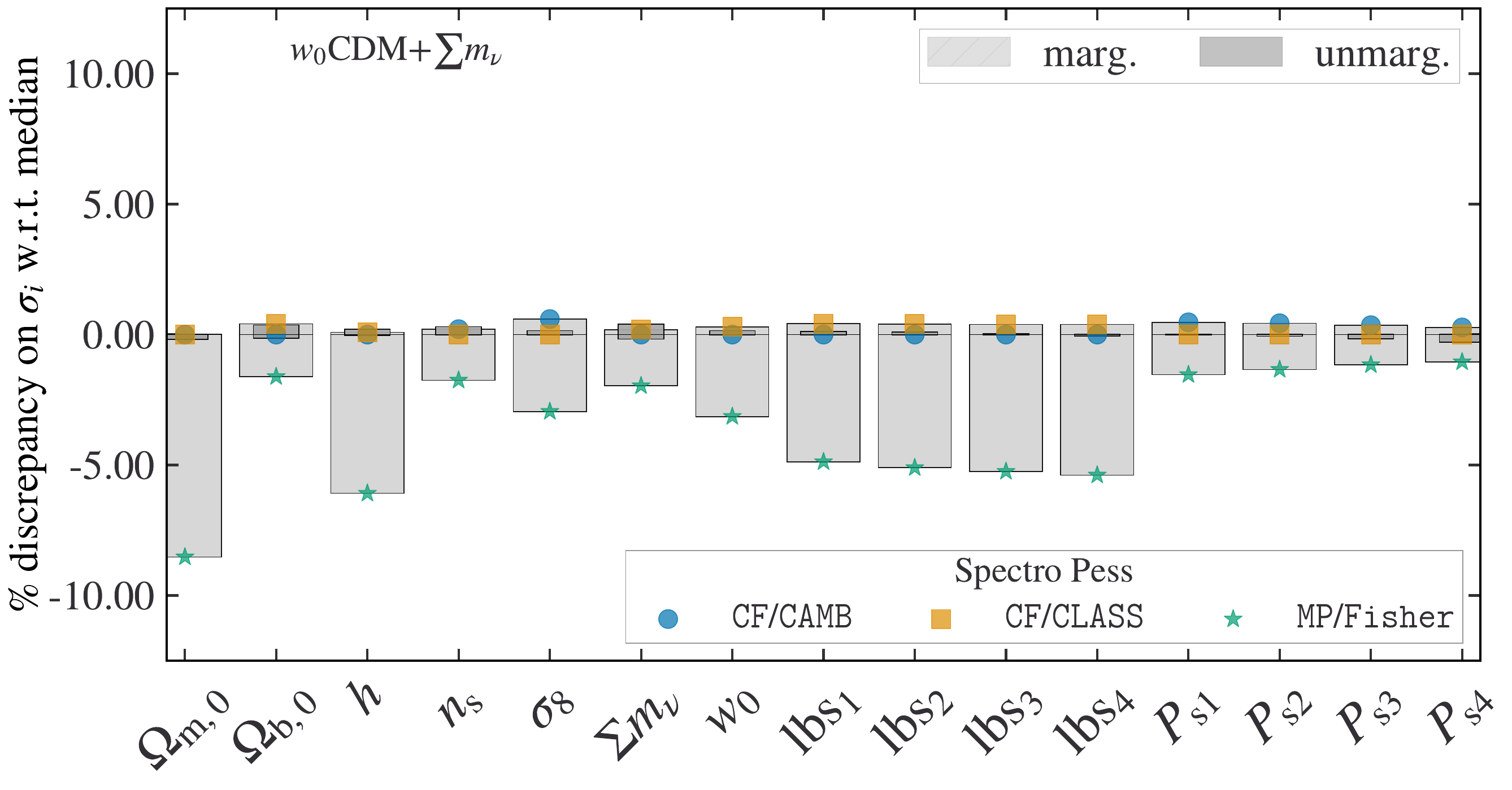}
    \caption{Same as \cref{fig:1} for the model $w_0$CDM+$\smnu$ model.
    \label{fig:3}
    }
\end{figure}

We show the marginalised and unmarginalised errors predicted by each  pipeline ({\tt CF/CAMB}, {\tt CF/CLASS}, {\tt MP/Fisher}) used to compute the FI matrix, each probe, and each case, respectively, in \cref{fig:1} for the $w_0w_a$CDM model, in \cref{fig:2} for the $\Lambda$CDM+$ \sum m_\nu$+$\neff$ model, and in \cref{fig:3} for the $w_0$CDM+$ \sum m_\nu$ model. In all cases, the errors are within 10\% of the median, leading to the validation of the pipelines.

Uncertainties from {\tt CF/CAMB} and {\tt CF/CLASS} are usually within 2\% of each other, which shows the excellent agreement between the two EBSs. Such a level of agreement was achieved after a careful setting of the parameters describing underlying assumptions on the cosmological models (especially those controlling the modelling of the neutrino sector) and of accuracy settings. Our choice of input and accuracy settings for \camb{} and \class{} is described in \cref{app:EBS}. The errors from {\tt CF/CAMB} and {\tt CF/CLASS} differ by more than 2\% only in the photometric case, in which the $\sigma_8$ and $\sum m_\nu$ errors are 3\% to 5\% away from each other. We believe this is due to the tiny but irreducible difference in the neutrino treatment between \class{} and \camb. As shown in \cref{fig:comparasion_fiducial_spectra}, the suppression of the power spectrum on small scales due to neutrinos seems to start earlier in \class{} than \camb, inducing differences of the order of 0.1\% in the linear matter power spectrum. Still, the final error bars achieve a level of agreement that is remarkable and sufficient for \Euclid{} purposes.

Errors from {\tt CF/CLASS} and {\tt MP/Fisher} are within 10\% of each other. The precise level of agreement is sensitive to the choice of step sizes, especially in the {\tt MP/Fisher} pipeline. For this choice, we adopted the following strategy
\begin{itemize}
    \item For {\tt CF/CAMB} and {\tt CF/CLASS}, we adopt by default relative steps of 1\% in each parameter (or 0.01 for $w_a$). However, we observed that this choice is inappropriate for $ \sum m_\nu$, since this parameter has a smaller impact on the matter power spectrum. The effect of a 1\% variation of the neutrino mass on both $P_\mathrm{mm}$ and $P_{\rm cc}$ is so small that numerical derivatives would be dominated by numerical noise from the EBSs. Thus, for $\sum  m_\nu$, we adopt a step size of 10\%, that is, $\Delta (\sum m_\nu)=6\,{\rm meV}$. We checked that the results are relatively stable against small variations in the step sizes (see \cref{app:step}) and against a change in the numerical derivative method (see \cref{app:der}).
    \item In {\tt MP/Fisher}, we stick to the recommendation of \cite{EUCLID:2023uep} and choose the step size in relation to the marginalised 1-dimensional parameter uncertainty for each individual parameter for a given model. To get the uncertainties, we could either run an MCMC or use the error obtained from \cosmicfish. In this case we use the uncertainties obtained from \cosmicfish{}. In the case of the photometric probe, we choose a step size of 10\% of the posterior error, while for the spectroscopic probe we use a step size of 5\% of the error.
\end{itemize}

These results are very sensitive to the modelling of nonlinear corrections. In \cref{sec:nl}, we argued that the most recent version of {\tt HMcode} \citep{Mead:2020vgs} reproduces simulations featuring massive neutrinos significantly better than {\tt HALOFIT} \citep{Takahashi:2012em,Bird:2011rb}. We checked that using {\tt HALOFIT} instead of {\tt HMcode} does affect the marginalised forecasted errors by a large factor of 1.5 (see \cref{app:halofit}). The assumption that galaxies trace cold-plus-baryonic matter instead of total matter is also important for \Euclid, but to a lesser extent. When replacing $P_{\rm cc}$ by $P_\mathrm{mm}$ everywhere in the spectroscopic likelihood, the uncertainties vary by at most 5\% on $\sum m_\nu$ (for our fiducial value $\smnu=60\,{\rm meV}$), as shown in \cref{app:cb}. 

\subsection{Comparison of Fisher ellipses and MCMC contours \label{sec:val_mcmc}}

\begin{figure}[htp]
    \centering
    \includegraphics[width=0.49\textwidth]{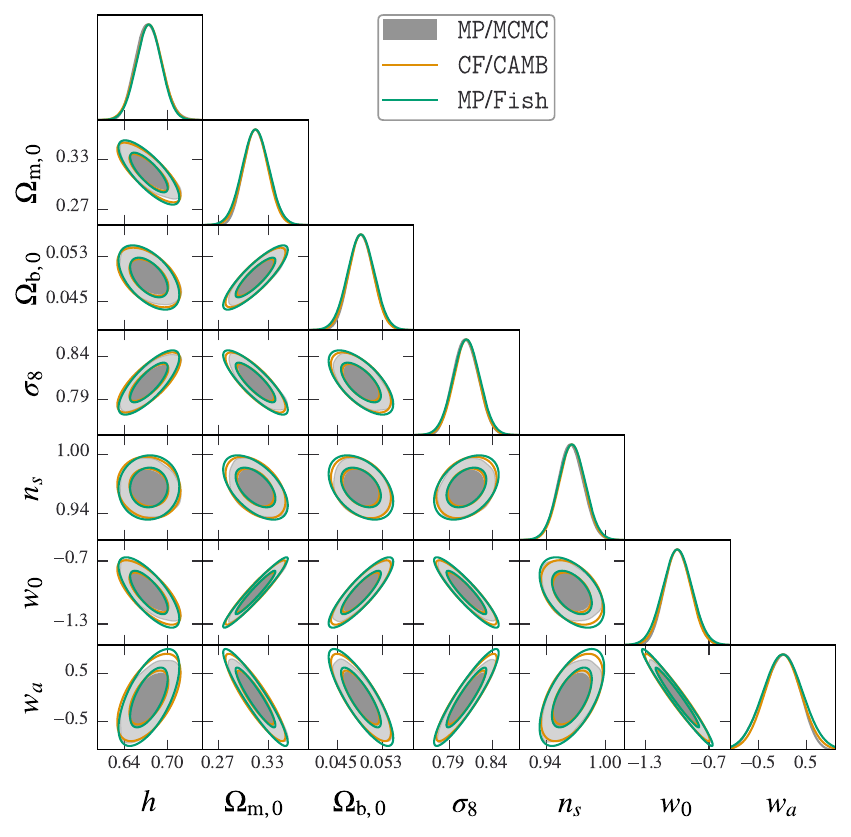}
    \includegraphics[width=0.49\textwidth]{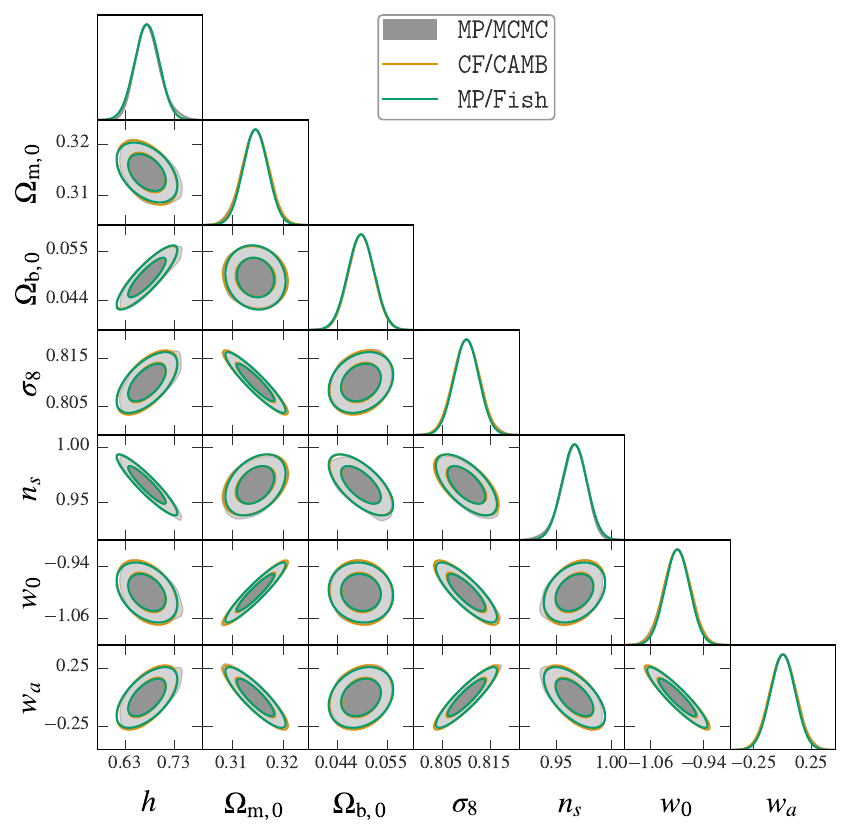}
    \caption{For the $w_0w_a$CDM model, comparison of the 1D marginalized posteriors and 2D (95\% C.L. and 68\% C.L.) marginalized contours obtained with the {\tt MP/MCMC} pipeline to the Gaussian posteriors and elliptical contours from the FI pipelines {\tt CF/CAMB} and {\tt MP/Fisher}: \emph{Left}: spectroscopic optimistic probe, \emph{Right}: photometric optimistic probe.
    }
    \label{fig:ValidationTriangle_w0waCDM}
\end{figure}

\begin{figure}[htp]
    \centering
    \includegraphics[width=0.49\textwidth]{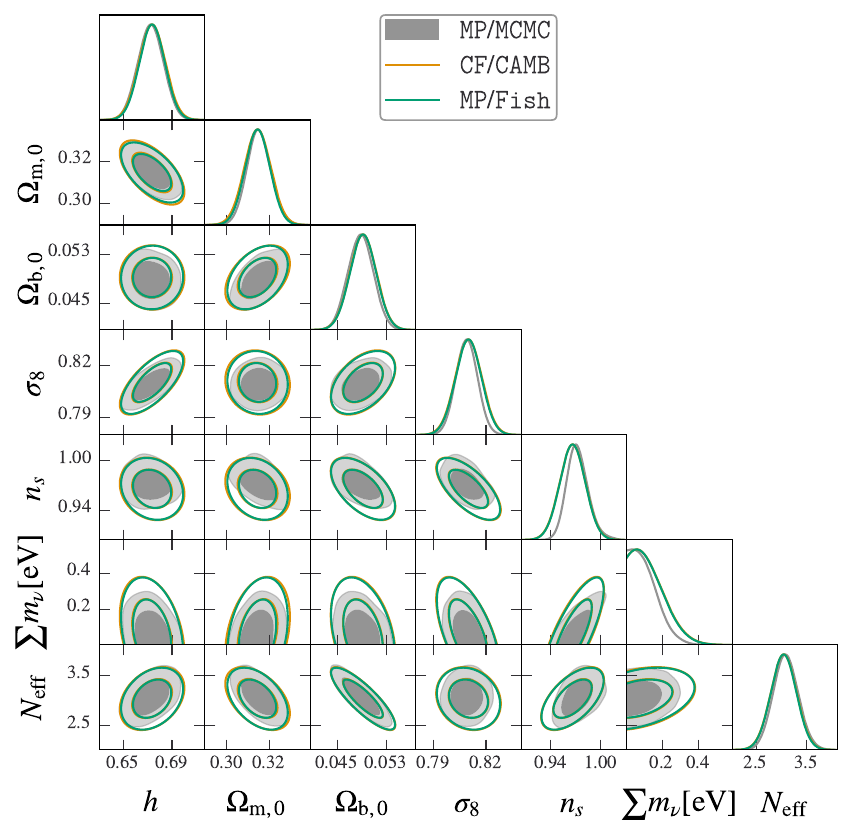}
    \includegraphics[width=0.49\textwidth]{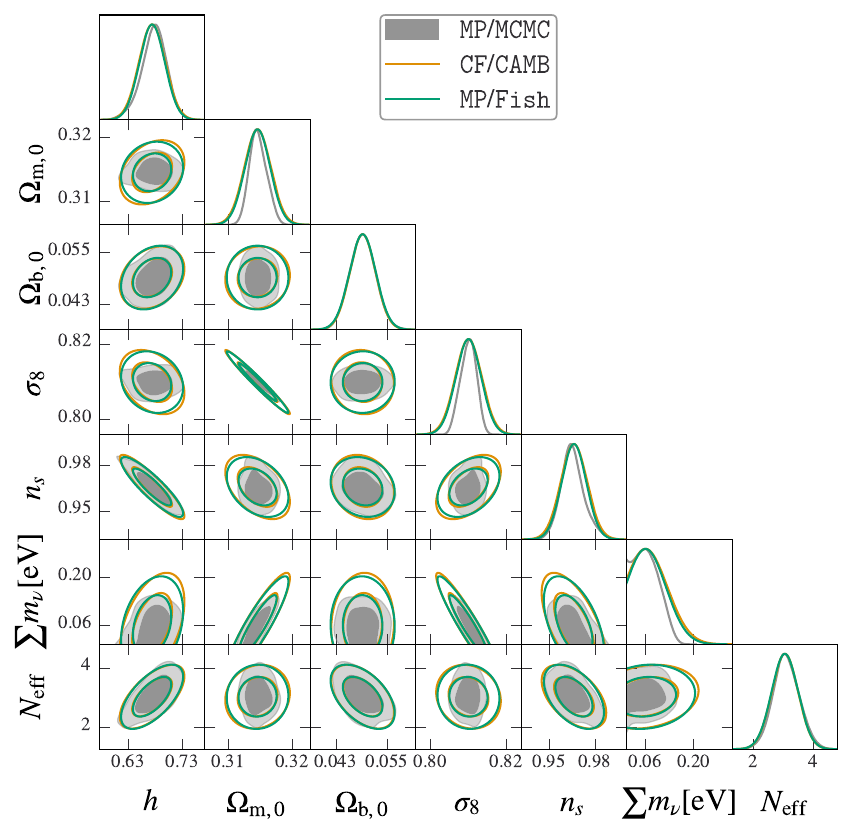}
    \caption{Same as \cref{fig:ValidationTriangle_w0waCDM} for the model $\Lambda$CDM+$\sum m_\nu$+$\neff$.
    \label{fig:ValidationTriangle_mnu+Neff}
    }
\end{figure}

\begin{figure}[htp]
    \centering
    \includegraphics[width=0.49\textwidth]{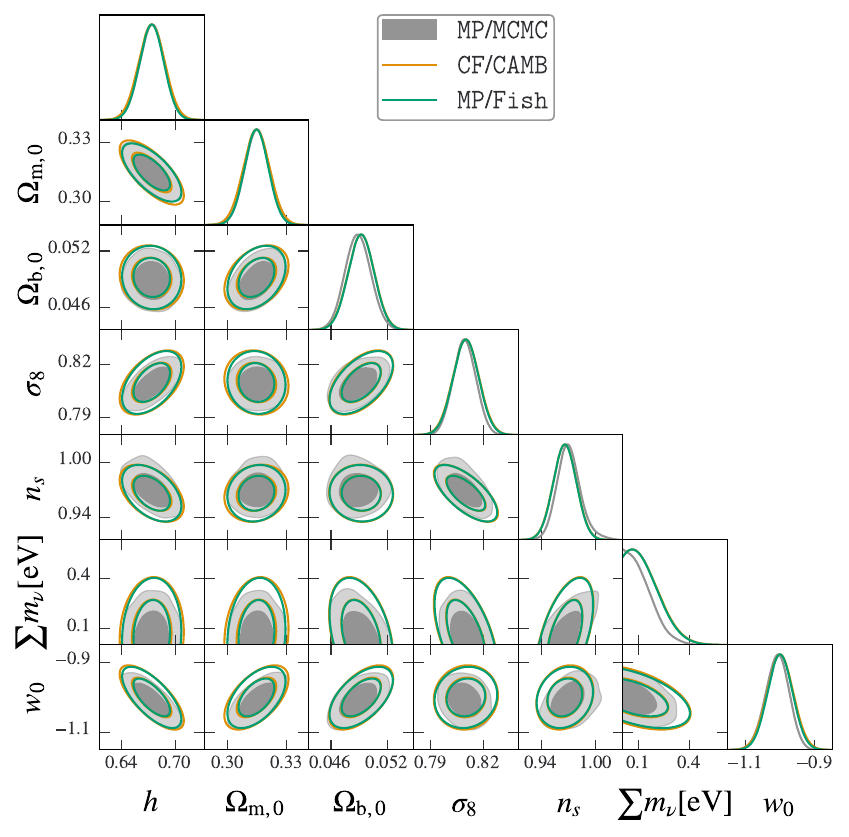}
    \includegraphics[width=0.49\textwidth]{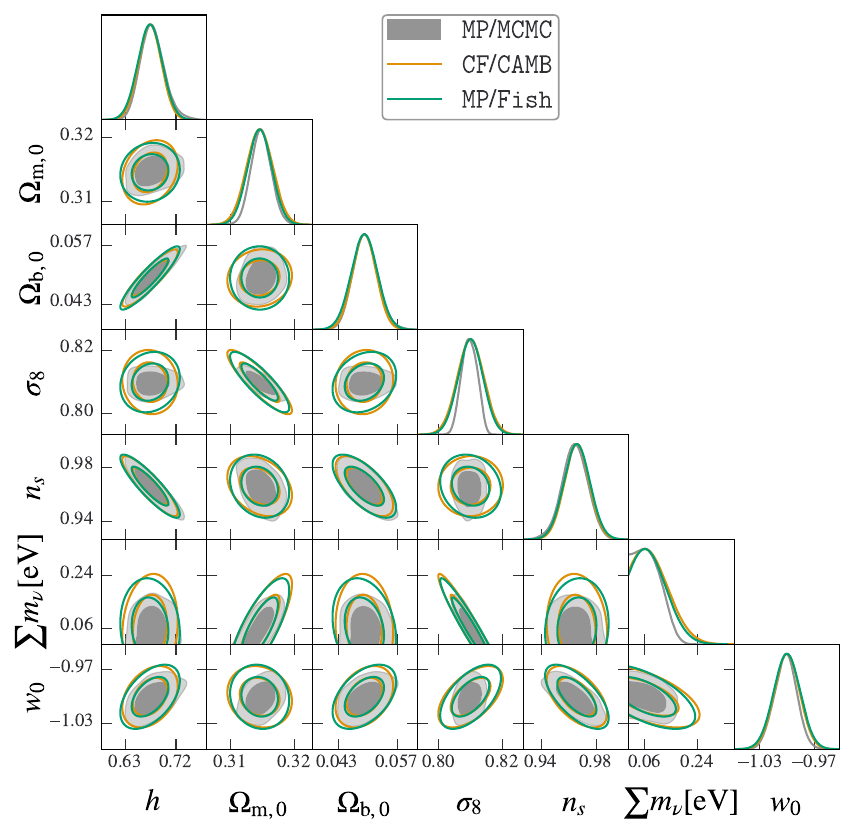}
    \caption{Same as \cref{fig:ValidationTriangle_w0waCDM} for the model $w_0$CDM model+$\sum m_\nu$.
    \label{fig:ValidationTriangle_w0+Mnu}
    }
\end{figure}

For each of the three validation models and each of the two probes, we show the 1-dimensional marginalized posteriors and the 2-dimensional marginalized confidence contours of our {\tt MP/MCMC} runs in the triangle plots of \cref{fig:ValidationTriangle_w0waCDM,fig:ValidationTriangle_mnu+Neff,fig:ValidationTriangle_w0+Mnu}. For this test, we only consider the optimistic case, which is the most sensitive to details in the modelling of the observables, and the most likely to reveal possible discrepancies in the numerical implementation of the likelihoods. This case is also the most constraining, and thus closer to the Gaussian approximation and the most suited for comparison with FI matrix predictions. In \cref{fig:ValidationTriangle_w0waCDM,fig:ValidationTriangle_mnu+Neff,fig:ValidationTriangle_w0+Mnu}, we actually plot the Gaussian posteriors and elliptical contours of the {\tt CF/CAMB} and {\tt MP/Fish} pipelines on top of the {\tt MP/MCMC} results.

First and foremost, we can see in \cref{fig:ValidationTriangle_w0waCDM} that the previous validation of photometric probes presented in \cite{EUCLID:2023uep} for the $w_0w_a$CDM model is reproducible, despite the changes in the likelihood code and nonlinear recipe triggered by the assumption of massive neutrinos. Concerning the spectroscopic probes, here we have corrected for an inconsistency in the unit conversion from ${\rm Mpc}^{-1}$ to $h\,{\rm Mpc}^{-1}$ that affected previous forecasts; thus, the sensitivity to $h$ degrades with respect to \cite{EUCLID:2023uep}, while the sensitivity to the other cosmological parameters is only mildly altered due to their correlations with $h$. Nevertheless, the conclusions drawn in \cite{EUCLID:2023uep} about the validation of {\tt MP/MCMC} for this case still hold: the posteriors of the photometric and spectroscopic probes are very close to multivariate Gaussians. We get another confirmation of this, since the MCMC and FI posteriors or contours are hardly distinguishable by eye and agree up to typical MCMC convergence errors.

In the case of the $\Lambda$CDM+$\smnu$+$\neff$ model, we see in \cref{fig:ValidationTriangle_mnu+Neff} that non-Gaussian effects come into play. This is particularly obvious when looking at the marginalised neutrino posteriors, which are truncated at $\smnu=0$ by the physical prior, and whose tails fall quicker than a Gaussian distribution for large masses. This non-Gaussianity propagates to the parameters that are most correlated with $\smnu$. Indeed, the orientation of the 2D contours shows that $h$, $\Omega_{{\rm m},0}$, $n_{\rm s}$, and $\sigma_8$ are significantly correlated with $\smnu$, while $\Omega_{{\rm b},0}$ and $\neff$ are not. The former parameters are also the ones whose posteriors deviate the most from a Gaussian shape. The fact that the MCMC contours are overall tighter than the FI ellipses is also a consequence of the non-Gaussianity with respect to $\smnu>0$, which cuts off a region of the parameter space (with either negative or too large neutrino mass) that would otherwise be reachable by FI contours. Deviations from Gaussianity are found to be even stronger within the photometric probe, both because the neutrino mass posterior is more suppressed for large masses and due to the very strong correlation between $\smnu$, $\Omega_{{\rm m},0}$, and $\sigma_8$.

Finally, in the case of the $w_0$CDM+$\smnu$ model, we see some qualitatively similar trends in \cref{fig:ValidationTriangle_w0+Mnu}. For the spectroscopic probe, we observe a `banana'-shaped contour in the $n_{\rm s}$-$\sum m_\nu$ plane and slightly non-elliptic contours in other planes. Once again, the non-Gaussianity is even more pronounced with the photometric probe, due to the cut-off of negative or too large neutrino masses.

As in \cref{sec:method} we stress that the {\tt MP/Fisher} and {\tt MP/MCMC} pipelines call the very same \Euclid mock likelihoods functions $\mathcal{L}(\theta_\alpha)$. Given that these functions have been validated by the tests presented in the previous subsection, and that the MCMC runs presented here have reached a high convergence level, the difference between the MCMC and FI contours can only be attributed to the intrinsic non-Gaussianity of the problem. This shows that for extended cosmological models, and particularly in the case of a free neutrino mass parameter, robust  \Euclid forecasts -- and even more so analyses of future real data -- require a full MCMC analysis. Having validated the \montepython{} likelihoods against \cosmicfish{} forecasts, we have proved that our {\tt MP/MCMC} pipeline offers a robust and accurate implementation of the physical assumptions performed in this work. Thus, we can use it in the next section in order to derive our main forecast results.

\section{Results\label{sec:results_euclid}}

Here we present the results of our forecast analysis.
We use the previously validated \texttt{MP/MCMC} pipeline based on the \montepython{} sampler, interfaced with the Boltzmann solver \class.
The underlying cosmology is either flat $\Lambda$CDM with the standard cosmological constant, or $w_0w_a$CDM, where dark energy is a fluid with a time-varying equation-of-state parameter according to \cite{Chevallier:2000qy} and \cite{Linder:2002et}.


As already explained in \cref{sec:neff} (case a), from the many possible ways of distributing the total mass $\smnu$ among the different species contributing to the effective number of neutrinos $\neff$ (which could lead to potentially different results), we make a choice that is representative of a large class of well-motivated scenarios. We assume that the three active neutrino species are degenerate in mass with a fiducial value $\smnu = 60\,{\rm meV}$ close to the minimum allowed by normal
ordering. We further assume that massive neutrinos contribute to $\neff$ through the standard value $3.044$ and that additional free-streaming massless particles account for $\dneff=\neff-3.044$, with $\dneff>0$ (see \cref{app:EBS_cosmo} for details on the specific settings in \class and \camb). Sticking to these assumptions, we consider four extended cosmological scenario with an increasing number of free parameters:
\begin{itemize}
\setlength\itemsep{1em}
    \item $\Lambda$CDM + $\smnu$ (baseline);
    \item $\Lambda$CDM + $\smnu$ + $\dneff$;
    \item $w_0w_a$CDM + $\smnu$;
    \item $w_0w_a$CDM + $\smnu$ + $\dneff$.
\end{itemize}
In \cref{tab:fiducial_validation} we list the fiducial values for the parameters of our models.

Here -- contrary to the validation of the MCMC pipeline -- we use pessimistic specifications, additionally also varying the nonlinear nuisance parameters $\sigma_{\rm v}$ and $\sigma_{\rm p}$ of the $\GCsp$ recipe. The reasoning behind this choice is that additional theoretical errors and systematic effects might have been overlooked in the modelling of the observables adopted in \citetalias{Euclid:2019clj} and \cite{EUCLID:2023uep}. Therefore, applying optimistic specifications might overestimate the sensitivity to the neutrino mass, while our aim is to provide a conservative forecast of the \Euclid discovery potential.
Moreover, the cut-off at $\ell_{\rm max}=1500$, implied by pessimistic specifications, prevents our results from being biased due to the uncertainties on the modelling of baryonic feedback, as we discuss in \cref{app:bf}.

We fit both \Euclid and CMB synthetic data (for the latter, see \cref{sec:cmb}). For \Euclid, the observables are the $3\times2$pt angular power spectra from photometric data (photometric galaxy clustering, weak lensing, and their cross-correlation, see \cref{sec:photo}), the 3D galaxy power spectrum from spectroscopic data (\cref{sec:gc}), and cluster number counts (CC, \cref{sec:clusters}).
    
The cosmological parameter fiducial values were already specified in \cref{tab:fiducial_validation}. The parameters $w_0$ and $w_a$ are kept fixed to the fiducial value when the underlying cosmological model is $\Lambda$CDM. The parameters of the baseline model are always varied, while the parameters of the extended models (such as the number of relativistic degrees of freedom and the dark energy equation of state parameters) are kept fixed, unless specified. The optical depth to reionization is fixed to the \Planck best fit value $\tau=0.0543$ when only \Euclid observables are included in the analysis, while we allow it to vary in the joint analysis with CMB probes.

\begin{figure}[htp]
    \centering
    \includegraphics[width=0.49\textwidth]{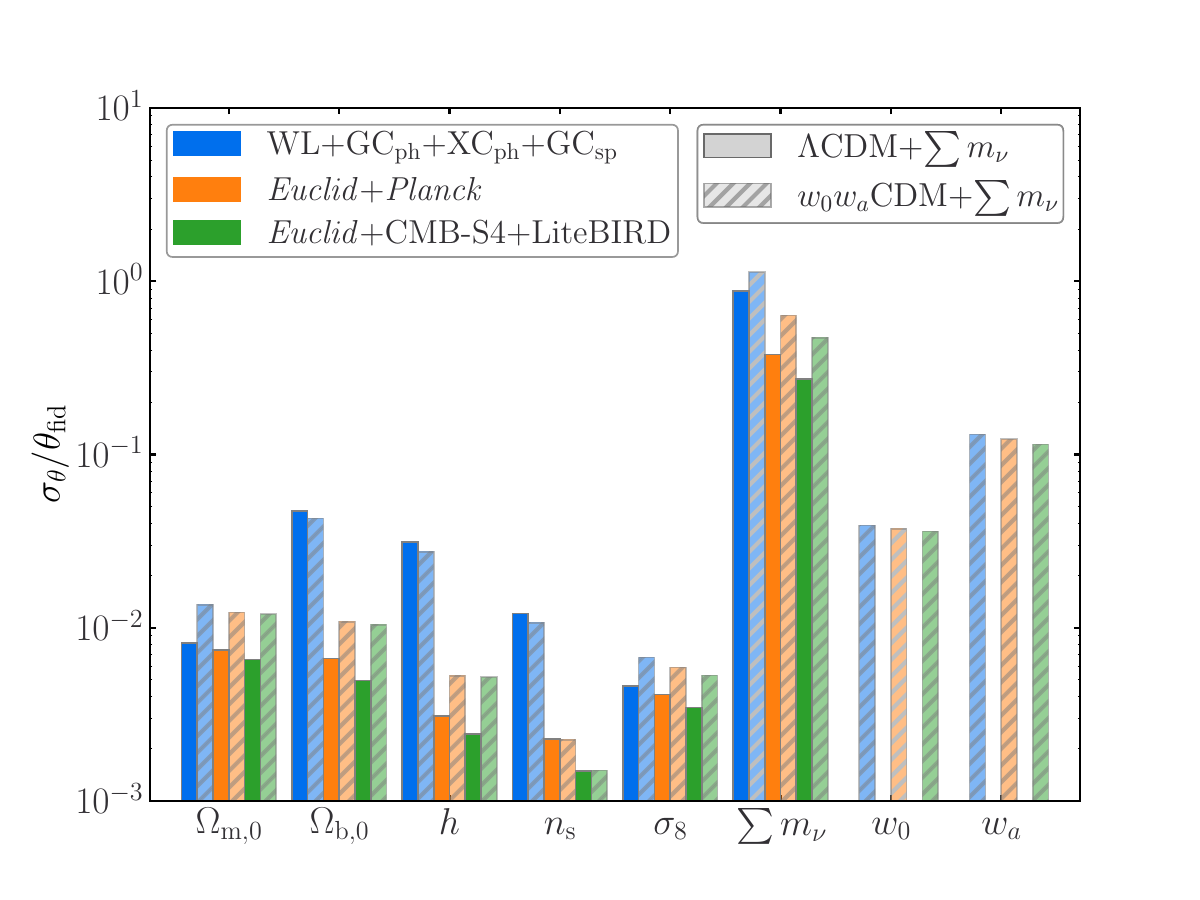}
    \includegraphics[width=0.49\textwidth]{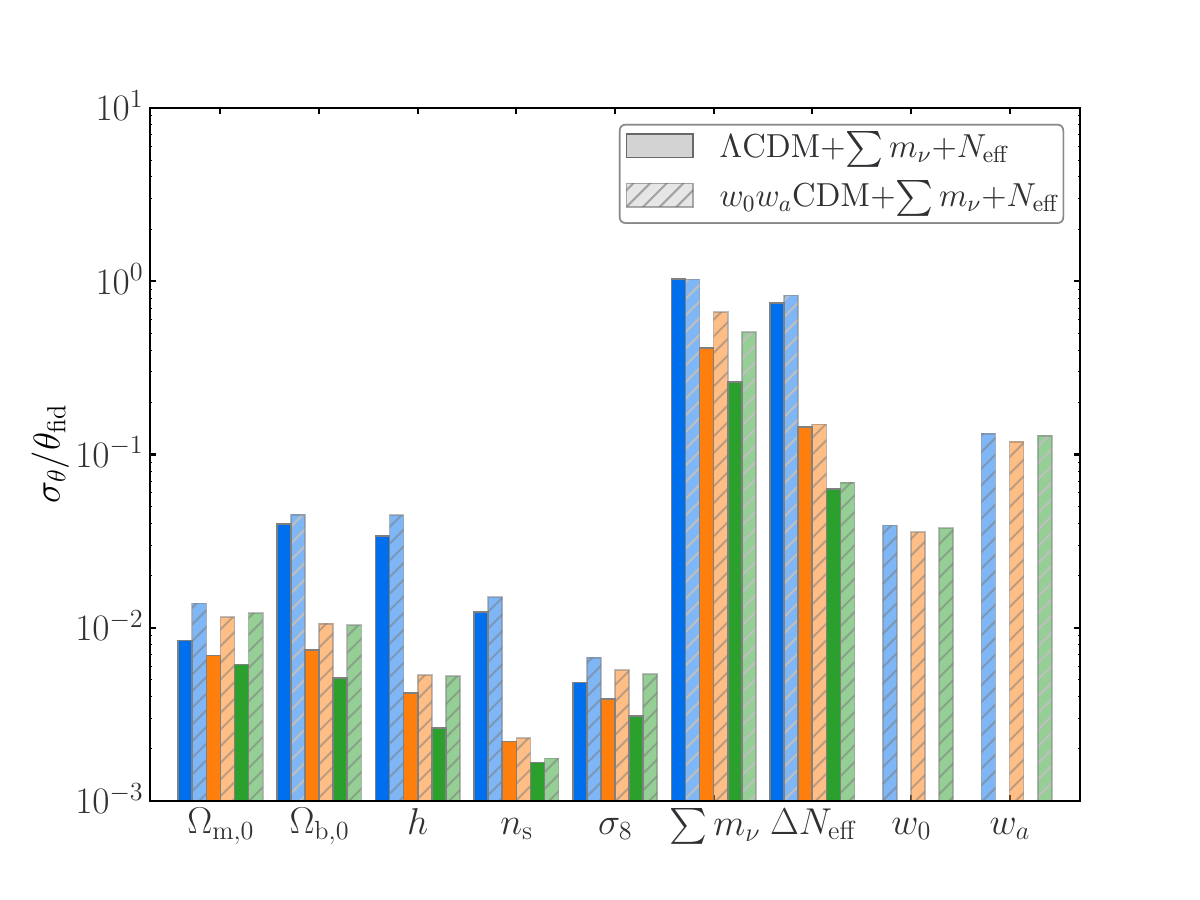}
    \caption{Sensitivity (relative errors with respect to the fiducial values) to the cosmological parameters, comparing $\Lambda$CDM cosmologies (solid colour bars) and $w_0w_a$CDM cosmologies (hatched colour bars), for models with (right panel) and without (left panel) varying $\neff$. The different colours correspond to the sensitivity of different data combinations: WL+$\GCph$+$\XCph$+$\GCsp$ (blue); \Euclid{}+\Planck (orange); and \Euclid{}+CMB-S4+LiteBIRD (green). We note that for $w_a$ we show the absolute errors, while for $\dneff$ we show the 95\% upper limits. For the neutrino mass we always show the relative $1\,\sigma$ uncertainty as computed from the posterior variance in order to facilitate the comparison between different data sets.}
    \label{fig:res_sens}
\end{figure}

In \cref{tab:results} we show the $1\,\sigma$ uncertainty (or the $95\%$ CL upper limit) on each cosmological parameter inferred from our MCMC forecasts, assuming different cosmological models and varying the combination of synthetic data.\footnote{In the table, we switch from the $1\,\sigma$ uncertainty to the 95\% CL upper limit whenever the posterior is compatible with the lower prior edge at the $1\,\sigma$ CL (otherwise, our estimates of the $1\,\sigma$ uncertainty could be biased).}
In \cref{fig:res_sens}, we show the same results expressed in terms of a dimensionless sensitivity, which is the $1\,\sigma$ uncertainty relative to the fiducial value.\footnote{In the case of the parameter $w_a$, whose fiducial value is zero, we quote the $1\,\sigma$ uncertainty in place of the sensitivity; while for $\dneff$ we quote the 95\% CL upper limit.} In each panel we compare the two underlying cosmologies, which are $\Lambda$CDM + $\smnu$ (solid colour bars) and $w_0w_a$CDM + $\smnu$ (hatched colour bars). The effective number of neutrino species is additionally varied in the right panel. The different colours show the sensitivity of three different data combinations: WL+$\GCph$+$\XCph$+$\GCsp$ (blue); \Euclid{}+\Planck (orange); and \Euclid{}+CMB-S4+LiteBIRD (green).

\begin{table}[ht]
\renewcommand{\arraystretch}{1.2}
\caption{Marginalised $1\,\sigma$ uncertainties on cosmological parameters in $\Lambda$CDM + $\smnu$, in $\Lambda$CDM + $\smnu$ + $\neff$, in $w_0w_a$CDM + $\smnu$, and in $w_0w_a$CDM + $\smnu$ + $\dneff$. We show results from \Euclid-only observables and in combination with current (\Planck) and future (CMB-S4+LiteBIRD) CMB surveys. The label \Euclid stands for $\GCsp$+WL+$\GCph$+$\XCph$ (all the primary probes from both photometric and spectroscopic data). In the baseline model ($\Lambda$CDM + $\smnu$) and in the extended model ($w_0w_a$CDM + $\smnu$ + $\dneff$) we also tested the combination $\GCsp$+WL+$\GCph$+$\XCph$+CC that includes, besides \Euclid primary probes, also \Euclid cluster number counts. For the $\dneff$ parameter, we always quote the upper bounds at 95\% CL. For $\smnu$, we quote the 95\% upper bounds when the posterior is one-sided.}
  \centering
  \resizebox{\textwidth}{!}{

\begin{tabular}{lccccccccc}
\hline
\rowcolor{cyan} \multicolumn{10} {c}{{\bf{$\boldsymbol \Lambda$CDM + $\boldsymbol \smnu$ }}}\\

 & \multicolumn{1}{c}{$\Omega_{\rm m,0}$} & \multicolumn{1}{c}{$100\,\Omega_{\rm b,0}$} & \multicolumn{1}{c}{$h$} & \multicolumn{1}{c}{$n_{\rm s}$} & \multicolumn{1}{c}{$\sigma_{8}$} & \multicolumn{1}{c}{$\smnu$[${\rm meV}$]} \\
   \hline
\rowcolor{gray}     \multicolumn{10} {l}{{{\Euclid-only}}} \\  
$\GCsp$	&$0.0068$	&$0.37$	&$0.033$	&$0.029$	&$0.0077$	&$<320$	\\ 
WL+$\GCph$+$\XCph$	&$0.0032$	&$0.36$	&$0.035$	&$0.017$	&$0.0047$	&$<260$	\\
 
WL+$\GCph$+$\XCph$+$\GCsp$	&$0.0026$	&$0.24$	&$0.022$	&$0.013$	&$0.0039$	&$56$	\\ 
WL+$\GCph$+$\XCph$+$\GCsp$+CC	&$0.0025$	&$0.24$	&$0.022$	&$0.012$	&$0.0037$	&$53$	\\ 
  \hline 
\rowcolor{gray}    \multicolumn{10} {l}{{{\Euclid{}+CMB}}} \\
   
\Euclid{}+\Planck	&$0.0023$	&$0.033$	&$0.0021$	&$0.0022$	&$0.0033$	&$23$	\\ 
\Euclid{}+CMB-S4+LiteBIRD	&$0.0021$	&$0.024$	&$0.0016$	&$0.0014$	&$0.0028$	&$16$	\\

\hline
\rowcolor{cyan} \multicolumn{10} {c}{{\bf{$\boldsymbol \Lambda$CDM + $\boldsymbol \smnu$ + $\boldsymbol \dneff$}}}\\

 & \multicolumn{1}{c}{$\Omega_{\rm m,0}$} & \multicolumn{1}{c}{$100\,\Omega_{\rm b,0}$} & \multicolumn{1}{c}{$h$} & \multicolumn{1}{c}{$n_{\rm s}$} & \multicolumn{1}{c}{$\sigma_{8}$} & \multicolumn{1}{c}{$\smnu$[${\rm meV}$]}  & \multicolumn{1}{c}{$\dneff$} \\
  \hline
\rowcolor{gray}   \multicolumn{10} {l}{{{\Euclid-only}}} \\ 
  
WL+$\GCph$+$\XCph$+$\GCsp$	&$0.0026$	&$0.19$	&$0.023$	&$0.012$	&$0.0039$	&$<220$	&$<0.746$	\\ 
   \hline  
\rowcolor{gray}     \multicolumn{10} {l}{{{\Euclid{}+CMB}}} \\
    
\Euclid{}+\Planck	&$0.0022$	&$0.037$	&$0.0028$	&$0.0021$	&$0.0031$	&$25$	&$<0.144$	\\ 
\Euclid{}+CMB-S4+LiteBIRD	&$0.0019$	&$0.025$	&$0.0018$	&$0.0016$	&$0.0025$	&$16$	&$<0.063$	\\

\hline
\rowcolor{cyan} \multicolumn{10} {c}{{\bf{$\boldsymbol{w_0w_a}$CDM + $\boldsymbol \smnu$ }}}\\ 

 & \multicolumn{1}{c}{$\Omega_{\rm m,0}$} & \multicolumn{1}{c}{$100\,\Omega_{\rm b,0}$} & \multicolumn{1}{c}{$h$} & \multicolumn{1}{c}{$n_{\rm s}$} & \multicolumn{1}{c}{$\sigma_{8}$} & \multicolumn{1}{c}{$\smnu$[${\rm meV}$]}  & \multicolumn{1}{c}{$w_0$} & \multicolumn{1}{c}{$w_a$}\\
   \hline
\rowcolor{gray}     \multicolumn{10} {l}{{{\Euclid-only}}} \\  

 WL+$\GCph$+$\XCph$+$\GCsp$	&$0.0043$	&$0.21$	&$0.019$	&$0.010$	&$0.0055$	&$<220$	&$0.04$	&$0.13$	\\ 

\hline
\rowcolor{gray} \multicolumn{10} {l}{{{\Euclid{}+CMB}}} \\

\Euclid{}+\Planck	&$0.0038$	&$0.053$	&$0.0036$	&$0.0022$	&$0.0048$	&$38$	&$0.04$	&$0.12$	\\ 
\Euclid{}+CMB-S4+LiteBIRD	&$0.0038$	&$0.051$	&$0.0035$	&$0.0015$	&$0.0043$	&$28$	&$0.04$	&$0.11$	\\

\hline
\rowcolor{cyan} \multicolumn{10} {c}{{\bf{$\boldsymbol{w_0 w_a}$CDM + $\boldsymbol \smnu$ + $\boldsymbol \dneff$}}}\\ 

 &\multicolumn{1}{c}{$\Omega_{\rm m,0}$} & \multicolumn{1}{c}{$100\,\Omega_{\rm b,0}$} & \multicolumn{1}{c}{$h$} & \multicolumn{1}{c}{$n_{\rm s}$} & \multicolumn{1}{c}{$\sigma_{8}$} & \multicolumn{1}{c}{$\smnu$[${\rm meV}$]}  & \multicolumn{1}{c}{$\dneff$} & \multicolumn{1}{c}{$w_0$} & \multicolumn{1}{c}{$w_a$}\\
  \hline
\rowcolor{gray}   \multicolumn{10} {l}{{{\Euclid-only}}} \\ 

  $\GCsp$	&$0.024$	&$0.52$	&$0.056$	&$0.043$	&$0.022$	&$<440$	&$<1.352$	&$0.21$	&$0.59$	\\ 
WL+$\GCph$+$\XCph$	&$0.0049$	&$0.38$	&$0.065$	&$0.029$	&$0.0065$	&$<260$	&$<1.705$	&$0.05$	&$0.18$	\\ 
WL+$\GCph$+$\XCph$+$\GCsp$	&$0.0043$	&$0.25$	&$0.036$	&$0.016$	&$0.0054$	&$<220$	&$<0.935$	&$0.04$	&$0.14$	\\
WL+$\GCph$+$\XCph$+$\GCsp$+CC	&$0.0037$	&$0.21$	&$0.029$	&$0.014$	&$0.0049$	&$<220$	&$<0.745$	&$0.03$	&$0.12$	\\ 
   \hline  
 \rowcolor{gray}   \multicolumn{10} {l}{{{\Euclid{}+CMB}}} \\
 
\Euclid{}+\Planck	&$0.0036$	&$0.052$	&$0.0036$	&$0.0022$	&$0.0046$	&$40$	&$<0.149$	&$0.04$	&$0.12$	\\ 
\Euclid{}+CMB-S4+LiteBIRD	&$0.0038$	&$0.051$	&$0.0035$	&$0.0017$	&$0.0044$	&$31$	&$<0.069$	&$0.04$	&$0.13$	\\

\end{tabular}
}
\label{tab:results}
\end{table}

{\bf Comparison with previous forecast.}
When comparing the results of the present forecast to those of \citetalias{Euclid:2019clj}, one has to consider that we assume the same specifications labelled as pessimistic there. However, our baseline case includes the variation of the neutrino mass, which was not done in \citetalias{Euclid:2019clj}.
In light of these considerations, we observe that the sensitivity of the photometric probes (WL+$\GCph$+\XCph) to $w_0$ and $w_a$ is consistent with the previous \Euclid forecast from Table 11 of \citetalias{Euclid:2019clj} (note that they cite relative sensitivities, while we cite absolute sensitivities). On the other hand, the sensitivity to the other cosmological parameters degrades because of the correlation with the neutrino parameters; for example, the sensitivity to $\sigma_8$ degrades by a factor of 1.6 in $w_0w_a$CDM+$\smnu$+$\dneff$ due to the free neutrino mass, while varying $\dneff$ does not have any impact.
A direct comparison of the results that include \GCsp with those of \citetalias{Euclid:2019clj} is not feasible because, as already mentioned in \cref{sec:val_mcmc}, here we have corrected an inconsistency in the unit conversion from $h\,{\rm Mpc}^{-1}$ to ${\rm Mpc}^{-1}$ that was present in the pipeline of \citetalias{Euclid:2019clj}. This issue produced a fake signal on $H_0$, and, thus, a spurious enhancement of the sensitivity to $h$, which affected also the sensitivity to other cosmological parameters that are correlated with $h$. Note that we checked that reintroducing the bug would make our results consistent with those of \citetalias{Euclid:2019clj}.
Our results are also consistent with the older \Euclid{}-like sensitivity forecast of \cite{Audren:2012vy}, which predicted a $1\,\sigma$ sensitivity to $\smnu$ of $32\,{\rm meV}$ for \Euclid ${\rm WL} + \Planck$ or $25\,{\rm meV}$ for \Euclid \GCsp + \Planck, while we obtain $23\,{\rm meV}$ for the combined ${\rm WL} + \GCph + \XCph + \GCph + \Planck$ data set. Interestingly, to account for the imperfect modelling of nonlinear scales, that forecast included a marginalisation over a theoretical error smoothly growing towards smaller scales, instead of cutting the data abruptly at some $k_{\rm max}$ or $\ell_{\rm max}$ like in the present work. \cite{Audren:2012vy} adopted a theoretical error amplitude suggested by the covariance of the nonlinear power spectra predicted by different $N$-body codes for the same cosmology. The fact that our results are comparable suggests that their theoretical error and our pessimistic settings are equally conservative. A slightly different modelling of the theoretical error in \cite{Sprenger:2018tdb} led to a sensitivity to $\smnu$ of $24\,{\rm meV}$ for \Euclid $\GCsp + {\rm WL} + \Planck$ with pessimistic specifications (there labelled as conservative), consistent with our results.

\begin{figure}[htp]
    \centering
    \includegraphics[width=0.9\textwidth]{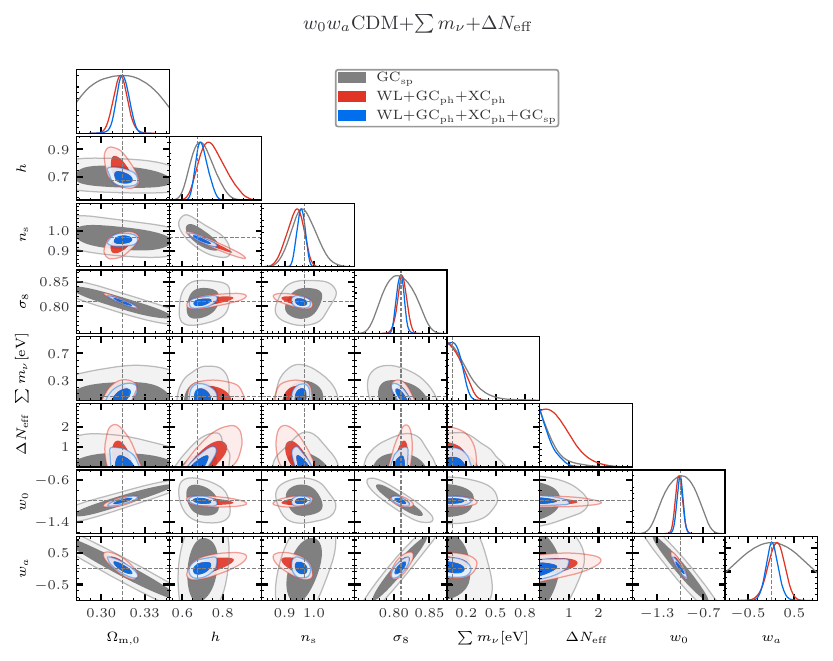}
    \caption{Marginalised 1D distribution (diagonal) and 2D (off-diagonal) 68\% and 95\% contours for a selection of parameters of the $w_0w_a$CDM+$\smnu$+$\neff$ model obtained by fitting $\GCsp$ (grey), WL+$\GCph$+$\XCph$ (red), and their combination (blue). The horizontal and vertical dashed lines mark the fiducial values.}
    \label{fig:res_w0waCDM}
\end{figure}

{\bf Complementarity of \Euclid probes.}
In \cref{fig:res_w0waCDM} we show the constraints on the extended $w_0w_a$CDM+$\smnu$+$\dneff$ model expected from \Euclid-only probes. The same figure for the baseline model $\Lambda$CDM + $\smnu$ can be found in \cref{app:res}. We note that in the extended $w_0w_a$CDM+$\smnu$+$\dneff$ model the constraints on $h$ and $\dneff$ are dominated by the spectroscopic probes and those on $\sigma_8$, $\smnu$, $w_0$, and $w_a$ by the photometric probes.
The complementarity of the photometric and spectroscopic probes is impressive: the $\GCsp$ contours exhibit parameter degeneracies in different directions than the WL+$\GCph$+\XCph ones in several planes of the parameter space. In particular, the correlations between the dimensionless Hubble constant $h$ and some of the other cosmological parameters that are present in \GCsp are broken by the photometric information. This leads to a remarkable sensitivity increase in the joint analysis of the two primary \Euclid probes. 

As far as the total neutrino mass is concerned, photometric probes are potentially more powerful than spectroscopic ones in constraining $\smnu$. As long as we are not including CMB information, the observable effect of the total neutrino mass is best described by the left panel of \cref{fig:Pk_Mnu_linear_suppression}, that is, a steplike suppression of the matter and cold-plus-baryonic power spectra. The longer lever arm of WL data thus provides the best constraint. However, the \Euclid probes alone cannot provide significant evidence for a non-zero neutrino mass, neither in the $w_0w_a$CDM cosmology ($\smnu \lesssim 0.22 \, {\rm eV}$, 95\% CL, see \cref{tab:results}) nor in $\Lambda$CDM [$\sigma(\smnu) = 56 \, {\rm meV}$, see \cref{tab:results}]. 

When using CMB data, there is a well-known correlation between $\smnu$ and $h$, due to their opposite effects on the angular diameter distance to last scattering. This strong correlation does not show up when fitting \Euclid probes alone, mainly thanks to the sensitivity of galaxy clustering data to the unique scale dependence of the growth factor induced by massive neutrinos. Note that this conclusion does not depend on the specific underlying cosmological model, whether it is $\Lambda$CDM or $w_0w_a$CDM, and whether $\dneff$ is varying or fixed.
We anticipate that the different  directions of degeneracy between $\smnu$ and $h$ in \Euclid and CMB probes will induce a significant improvement on the $\smnu$ constraints in the \Euclid{}+CMB joint analysis.
Finally, note that we do not find any specific correlation between the neutrino mass and the bias parameters, neither in \GCph nor in \GCsp, as we show in \cref{app:nuisance}.

Concerning dark radiation, as already mentioned, a variation in \dneff{} impacts the matter and cold-plus-baryonic matter power spectra both through a change in the overall shape (caused by a shift of the time of equality) and in the phase and amplitude of BAO peaks (due to neutrino drag), as can be checked in the left panel of \cref{fig:Pk_Neff_linear_suppression}. The first effect induces a correlation between \dneff{} and $h$, since both parameters can be increased while maintaining a fixed redshift of equality. This correlation, combined with the prior volume effect induced by the physical prior $\dneff>0$ \citep{Hadzhiyska:2023wae}, leads to a small bias in the mean value of $h$ and of $n_{\rm s}$ reconstructed from photometric probes. We have explicitly checked that this bias disappears in the models where $\dneff$ is fixed.
The \Euclid constraints on $\dneff$ are dominated by the spectroscopic probe, because \GCsp data partially break the correlation between $\dneff$ and $h$ thanks to a better measurement of BAO peaks. Nevertheless, the photometric probes contribute to the $\dneff$ constraints by breaking the degeneracy in the $\smnu$ versus $\dneff$ plane. Normally, one would expect a negative correlation between these parameters, due to their opposite effects on the BAO peak scale.\footnote{We note that the effects of $\smnu$ and $\dneff$ on the BAO peaks are intrinsically different. Indeed a variation of the number of relativistic species leads to an irreducible phase-shift of the BAO peaks due to the gravitational boost of CDM perturbations at horizon entry sourced by the anisotropic stress of free-streaming species. On the other hand, a variation of $\smnu$ changes the distance of the BAO peaks by means of background effects, which can be also compensated by a variation of other cosmological parameters (such as $h$ and $\Omega_{{\rm c},0}$).} This negative correlation is clearly visible on the \GCsp-alone contour of \cref{fig:res_w0waCDM}. Thanks to the large lever arm of WL data, the photometric probe allows us to distinguish the difference in overall shape and growth rate induced by variations of $\smnu$ or \dneff{} in the matter and cold-plus-baryonic power spectra, and to lift the degeneracy.
As a consequence, a combination of the probes improves the sensitivity to both $\dneff$ and $\smnu$.

On the other hand, under our assumptions concerning the modelling of the \Euclid{} cluster count data and likelihood, we find no significant improvement on cosmological parameter sensitivity when adding clusters. \Cref{tab:results} shows the result for WL+\GCph{}+\XCph{}+\GCsp{}+CC, where CC stands for cluster count, in the $\Lambda$CDM+$\smnu$ and $w_0w_a$CDM+$\smnu$+$\neff$ cases. The difference between the results with and without CC lies within MCMC convergence errors. Additionally, in the $w_0w_a$CDM+$\smnu$+$\neff$ case, the upper bound on $\dneff$ gets a tiny enhancement from a Bayesian prior volume effect related to small parameter degeneracies between $\dneff$ and the nuisance parameters of the CC likelihood. We recall that we performed a very conservative modelling of CC data with wide priors on several nuisance parameters. With a more optimistic modelling, the same data may lead to tighter constraints.


\begin{figure}[ht]
    \centering
    \includegraphics[width=0.9\textwidth]{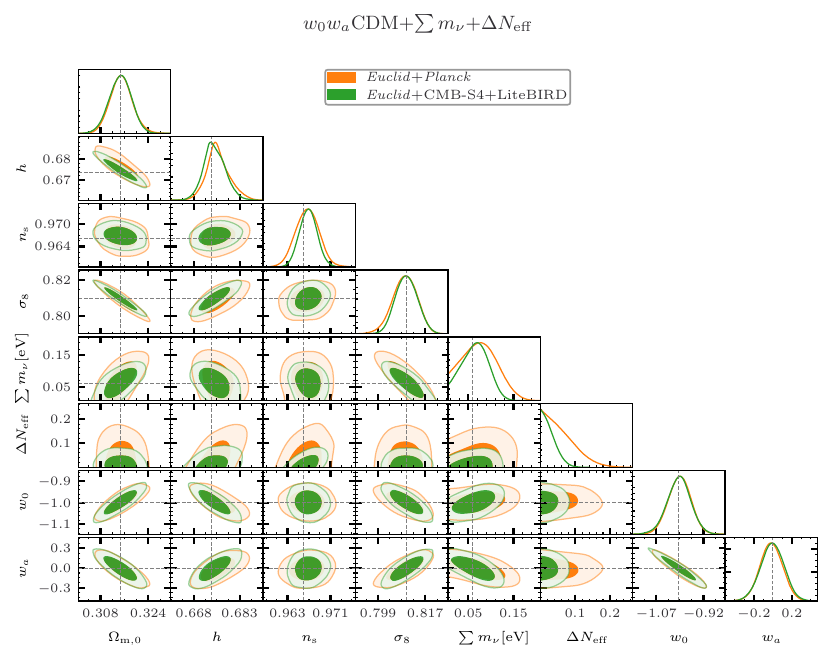}
    \caption{Marginalised 1D distributions (diagonal) and 2D (off-diagonal) 68\% and 95\% contours for a selection of parameters of the $w_0w_a$CDM+$\smnu$+$\neff$ model obtained by fitting \Euclid{}+\Planck (orange), and \Euclid{}+CMB-S4+LiteBIRD (green). The horizontal and vertical dashed lines mark the fiducial values.}
    \label{fig:res_w0waCDM_cmb}
\end{figure}

\begin{figure}[ht]
    \centering
    \includegraphics[width=0.7 \linewidth]{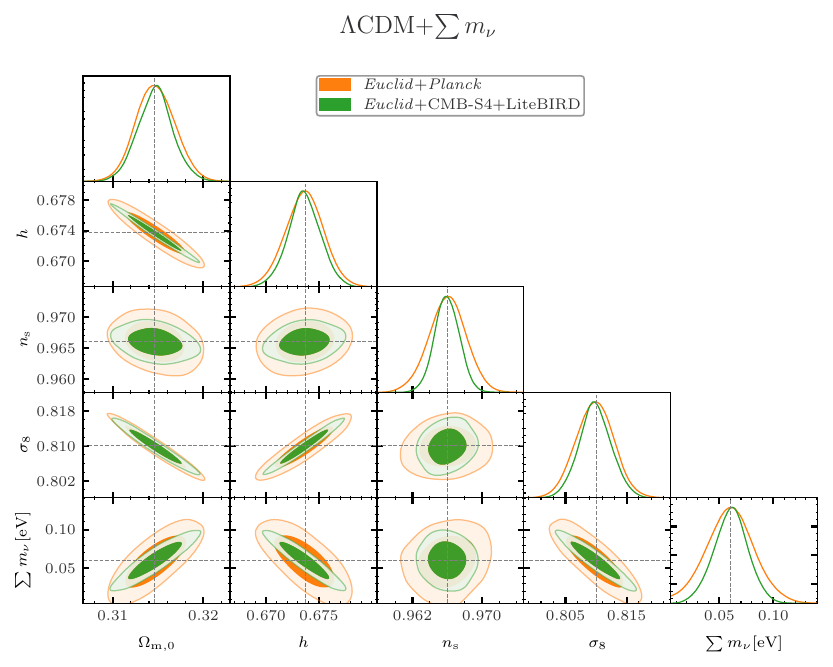}
    \caption{Same as Fig.~\ref{fig:res_w0waCDM_cmb} for the $\Lambda$CDM+$\smnu$ model.}
    \label{fig:res_lcdmM}
\end{figure}

\begin{figure}[ht]
    \centering
    \includegraphics[width=0.50 \linewidth]{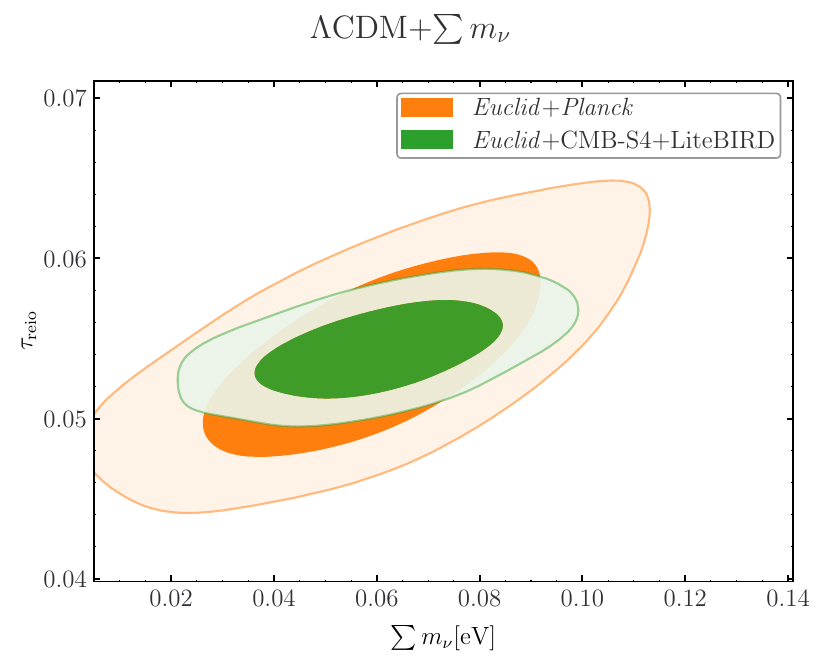}
    \caption{Marginalised 2D contours of $\smnu$ and $\tau_{\rm reio}$ for the $\Lambda$CDM+$\smnu$ model.}
    \label{fig:tau}
\end{figure}

\begin{figure}[ht]
    \centering
    \includegraphics[width=0.55\textwidth]{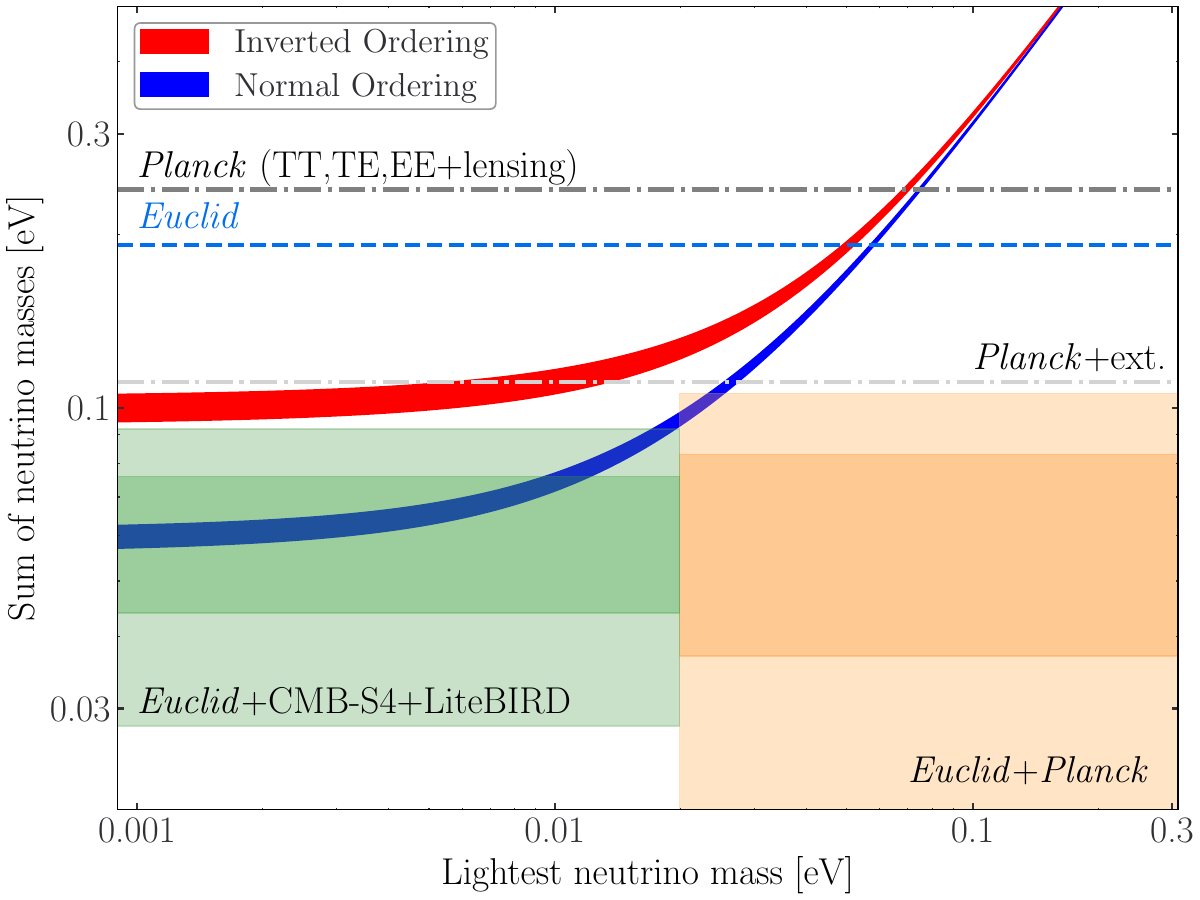}
    \caption{Sum of neutrino masses \smnu{} in the inverted ordering (red solid line) scenarios and in the normal ordering (blue solid line), including the uncertainties on the mass squared differences from oscillation experiments, as a function of the lightest neutrino mass. The grey dot-dashed lines show the 95\% CL upper bounds from \Planck only (dark grey), and in combination with external data (\Planck{}+ext., light grey). For the latter we use the value quoted in \cite{eBOSS:2020yzd} for \Planck{}+Pantheon-SNe+SDSS(BAO+RSD)+DES~$3\times2$pt, which includes \Planck, Pantheon type Ia supernovae, baryon acoustic oscillations and redshift-space distortions from the Sloan Digital Sky Survey, and weak lensing measurements from the Dark Energy Survey. The region above the blue dashed line is excluded at 95\% CL by \Euclid only primary probes. The orange shading shows the 68\% and 95\% CL limits on \smnu{} from \Euclid{}+\Planck in the baseline $\Lambda$CDM+$\smnu$ model, assuming a fiducial value $\smnu=60\,{\rm meV}$. The 68\% and 95\% CL constraints from \Euclid{}+CMB-S4+LiteBIRD are shown in green. We note that we split the two constraints into two different ranges of the lightest neutrino mass only to avoid an overlap of the two shadings. Assuming that the true value of the neutrino mass sum is indeed the minimum allowed by neutrino oscillation experiments in normal ordering, the combination \Euclid{}+CMB-S4+LiteBIRD will rule out the inverted ordering at $2.5\,\sigma$.}
    \label{fig:hierarchy}
\end{figure}

\begin{figure}[ht]
    \centering
    \includegraphics[width=0.55\textwidth]{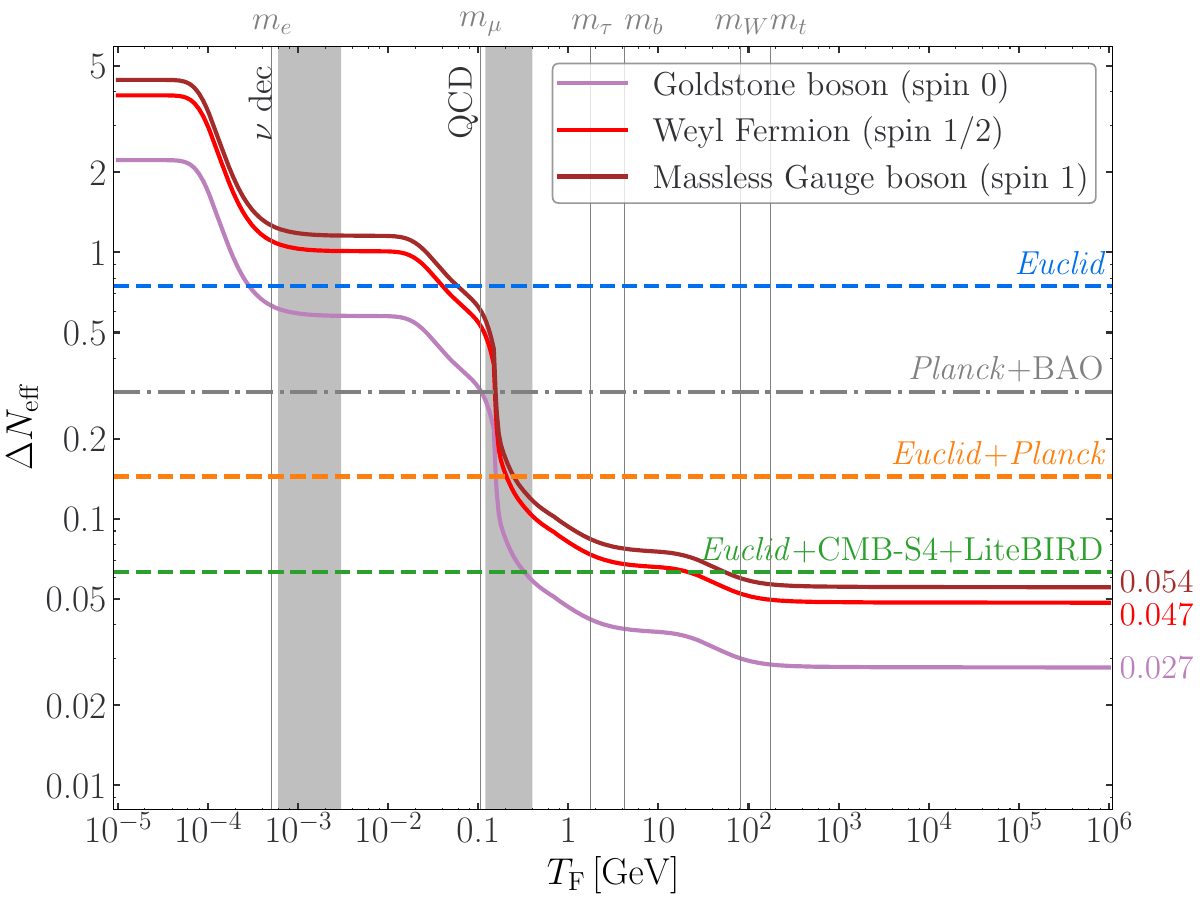}
    \caption{Contribution of new light particles beyond the Standard Model to $\dneff$ as a function of their decoupling temperature. Assuming natural units, we report the temperatures in GeV. As a reference we show the contribution of a Goldstone boson (solid brown line), a Weyl fermion (solid red line), and a massless Gauge boson (solid purple line). The grey vertical bands denote the range of temperatures of the QCD phase transition (around $200\, {\rm MeV}$) and neutrino decoupling (around $1 \,{\rm MeV}$), while the thin grey vertical lines mark the annihilation temperature of Standard Model particles. The grey horizontal dot-dashed line represents the current 95\% CL limit from \Planck TT,TE,EE+lensing+BAO \citep{Planck:2018vyg}. The horizontal dashed lines mark the 95\% upper bounds on $\dneff$ from \Euclid only (in blue) and in combination with current (\Planck) and future (CMB-S4+LiteBIRD) CMB surveys (in orange and green, respectively). 
    As explained in \cref{sec:cmb}, we choose to adopt more conservative assumptions regarding the CMB-S4 sensitivity and foreground removal than in a similar figure in the CMB-S4 white paper, Fig.~23 of \cite{CMB-S4:2016ple}.
}
    \label{fig:neff}
\end{figure}

{\bf Combination with CMB.}
In \cref{fig:res_w0waCDM_cmb,fig:res_lcdmM} we show constraints on the parameters of the $w_0w_a$CDM+$\smnu$+$\dneff$ and $\Lambda$CDM+$\smnu$ models for the combination of \Euclid probes with CMB experiments, \Euclid{}+\Planck (orange contours), and \Euclid{}+CMB-S4+LiteBIRD (green contours).
The same figure for the other cosmological models can be found in \cref{app:res}. 
These figures show that the \Euclid probes will play a crucial role in improving constraints on the cosmological parameters considered in this work. As a matter of fact, we can identify several cases in which future CMB data sets allow for parameter degeneracies that are broken by \Euclid probes.

This is particularly true for the degeneracy between $\smnu$ and $h$. We already discussed above why these parameters are correlated in a CMB-only analysis, and why the \Euclid probes can break this degeneracy, as can be checked again by comparing the $\{ \smnu, h\}$ panel of \cref{fig:res_w0waCDM} with the same panel in \cref{fig:res_w0waCDM_cmb} (see also \cref{fig:res_lcdmM}). Additionally, when CMB data are taken into account, the observable impact of the total neutrino mass is best described by the right panel of \cref{fig:Pk_Mnu_linear_suppression}, and amounts mainly to an overall (redshift-dependent) suppression of the matter and cold-plus-baryonic power spectra. Then, the role of CMB data is to fix the amplitude of the primordial spectrum of fluctuations at high redshift (modulo some uncertainty on the optical depth to reionization that depends on CMB polarisation measurements on large angular scales), while the \Euclid WL probe fixes the amplitude of the matter and cold-plus-baryonic power spectra at low redshift.
Therefore, the combination of \Euclid and CMB complementary data yields much stronger bounds on $\smnu$ than either of the two data sets taken separately, allowing eventually for a neutrino mass detection. Even in the minimal hierarchy scenario, which we assume as our fiducial model, the joint \Euclid plus \Planck forecast returns a $2.6\,\sigma$ evidence for a non-zero mass with $\sigma(\smnu=60\,{\rm meV})=23 \,{\rm meV}$ (\Euclid{}+\Planck). Finally, replacing \Planck with a combination of future CMB surveys leads to the detection of a non-zero neutrino mass at almost $4\,\sigma$ with $\sigma(\smnu=60\,{\rm meV})=16 \,{\rm meV}$ (\Euclid{}+CMB-S4+LiteBIRD).
One of the reasons behind this improvement is the tightening of the constraints on the reionization optical depth from LiteBIRD. Indeed, \cref{fig:tau} shows how the joint fit of LSS and CMB data leads to a strong degeneracy between $\smnu$ and $\tau$ \citep{Allison:2015qca, Archidiacono:2016lnv}. Therefore, the sensitivity to the neutrino mass can be further improved by means of an independent measurement of the reionization optical depth. Such measurement can arise from future 21-cm experiments \citep{Sailer:2022vqx}\footnote{However, foregrounds and uncertainties on the assumptions about the astrophysical processes leading to reionization might limit the accuracy of the determination of $\tau$ from 21-cm surveys.}, from probing the kinetic Sunyaev-Zeldovich effect in the CMB \citep{Park:2013mv}, and from CMB polarization \citep{LiteBIRD:2022cnt}.

Given the fantastic ability of \Euclid to constrain the neutrino mass sum in combination with CMB data, it is worth investigating whether \Euclid data will help to reconstruct the neutrino mass hierarchy. \cref{fig:hierarchy} shows the neutrino mass sum required by the two neutrino mass ordering schemes as a function of the mass of the lightest neutrino. We see that, if the neutrino mass sum has indeed the minimum value allowed by neutrino oscillations in normal ordering (NO), which coincides with the fiducial value of our forecast ($\smnu=60\,{\rm meV}$), then the \Euclid{}+CMB-S4+LiteBIRD sensitivity to the neutrino mass sum will also imply a $2.5 \, \sigma$ indication in favour of NO. Of course, the statistical significance of the evidence in favour of NO will decrease if the best-fit neutrino mass turns out to be larger than the minimum value of NO. It is important to stress that this measurement is not a direct probe of the neutrino mass ordering, but rather a consequence of the strong sensitivity to $\smnu$.

Contrary to the case of $\smnu$, the constraints on $\dneff$ are dominated by CMB probes. Nevertheless the \Euclid probes play an important role in breaking some of the degeneracies with other cosmological parameters, and in particular between \dneff{} and $h$. \Euclid improves the \Planck sensitivity to the Hubble constant by a factor 2.5 \cite[$\sigma(h)=0.0021$, \Euclid{}+\Planck, in $\Lambda$CDM+$\smnu$, compared with table 2 of][]{Planck:2018vyg}; as already discussed and shown in \cref{fig:res_w0waCDM}, this is mainly due to $\GCsp$ data. This significantly increases the sensitivity of combined CMB and \Euclid data to \dneff. Additionally, as shown in the right panel of  \cref{fig:Pk_Neff_linear_suppression}, once we include CMB information (thus fixing the angular size of the sound horizon at recombination), an increase of $\neff$ leads to an enhancement of the matter and cold-plus-baryonic power spectra at small scales. This effect provides additional sensitivity to \neff. \Cref{fig:neff} shows that adding \Euclid improves the current limit \citep{Planck:2018vyg} on the presence of new light particles by more than a factor 2 ($\dneff<0.144$, 95\% CL, \Euclid{}+\Planck, in $\Lambda$CDM+$\smnu$+$\dneff$).
On the longer term, the {\it Euclid} probes even have the potential to improve the sensitivity of future CMB data sets like LiteBIRD+CMB-S4 to \neff{} -- once more, not because of a better sensitivity to the effect of \neff{} itself, but thanks to a significantly more accurate determination of the correlated parameter $h$. Under the conservative assumptions described in \cref{sec:cmb}, LiteBIRD+CMB-S4 alone would achieve
$\Delta \neff< 0.076$
(95\% CL, $\Lambda$CDM+\smnu+\neff). The combination {\it Euclid}+LiteBIRD+CMB-S4 would increase the sensitivity to
$\Delta \neff< 0.063$
(95\% CL, $\Lambda$CDM+\smnu+\neff).\footnote{If we started from more aggressive assumptions regarding the sensitivity and foreground removal in CMB-S4 data, we would still find that the combination of \Euclid with LiteBIRD+CMB-S4 improves the sensitivity to \neff.}
Such bounds approach the asymptotic limits for particles of spin 1 and 1$/$2 at temperatures above the electroweak phase transition, and thus can almost exclude their existence, independently of their decoupling temperature (see~\cref{fig:neff}).

The sensitivity of \Euclid to the Hubble parameter will bring strong indications concerning the puzzle raised by the Hubble tension in the near future. We note that the sensitivity of the combination of \Euclid spectroscopic and photometric probes is such that, in this case, future CMB observations will not lead to a significant improvement. For instance, in the $\Lambda$CDM+$\smnu$ model, the sensitivity will marginally increase from $\sigma(h)=0.0021$ with \Euclid{}+\Planck to $\sigma(h)=0.0016$ with \Euclid{}+CMB-S4+LiteBIRD. In the $w_0w_a$CDM+$\smnu$ model, \Euclid data are essential to break the correlation between the dark energy parameters and the other cosmological parameters, including $h$, therefore, there is no improvement at all in the sensitivity to $h$ due to future CMB experiments (from $\sigma(h)=0.0036$ with \Euclid{}+\Planck to $\sigma(h)=0.0035$ with \Euclid{}+CMB-S4+LiteBIRD).

Moreover, \Euclid will provide some key information regarding the $\sigma_8$ tension, having a $1\,\sigma$ error on $S_8=\sigma_8\left(\Omega_{\rm m}/0.3\right)^{0.5}$ of 0.0049 from \Euclid only in the $\Lambda$CDM+$\smnu$ case. This is nearly a factor of 3 improvement (2.8) with respect to the DES Y3 $3\times2$pt result \citep{DES:2021bvc, DES:2021vln}.

Finally, the sensitivity of \Euclid to dynamical dark energy parameters, which was already illustrated in previous forecasts assuming a fixed neutrino mass and no extra relativistic species \citepalias{Euclid:2019clj,EUCLID:2023uep}, is confirmed in the framework of extended cosmologies with varying \smnu{} and \dneff. As a matter of fact, the contours derived from \Euclid probes do not show significant correlations between the dark energy and neutrino parameters (see \cref{fig:res_w0waCDM}).
On the other hand, the combination of \Euclid with CMB brings back the correlation between $\smnu$ and $\{ w_0, w_a \}$, as we shall comment in the next paragraph on the neutrino mass sensitivity in extended cosmologies.

{\bf Neutrino mass sensitivity in extended cosmologies.} 
The constraints on the total neutrino mass from the \Euclid primary probes alone are nearly independent of the assumed underlying model, as can be checked in \cref{tab:results} and in \cref{fig:res_sens}.
However, once combined with external data sets, the constraints become more model dependent. 

Floating the number of relativistic relics $\dneff$ has a limited impact on \smnu{} bounds, but allowing for time-dependent dark energy degrades the mass sensitivity by a significant amount. For instance, the \Euclid{}+\Planck sensitivity decreases from $\sigma(\smnu=60\,{\rm meV})=23\,{\rm meV}$ in the $\Lambda$CDM case to $38\,{\rm meV}$ in the $w_0w_a$CDM case. Similar trends are observed with the \Euclid{}+CMB-S4+LiteBIRD data set. Previous studies have shown that a correlation between $\smnu$ and $\{ w_0, w_a\}$ is already present in the analysis of CMB data alone \citep{Brinckmann:2018owf}. As a matter of fact, the effect of the total neutrino mass and of the dark energy parameters on CMB lensing are difficult to disentangle. The effects of \smnu{} (with fixed sound horizon scale, like in the right panel of \cref{fig:Pk_Mnu_linear_suppression}) and of $\{ w_0, w_a \}$ on the overall amplitude of the matter power spectrum and on the scale of CMB and BAO peaks are also relatively similar, and do not allow us to remove this correlation. It is nevertheless interesting to see that the constraint on \smnu{} never degrades by more than a factor of 2 when the cosmological constant is replaced by dynamical dark energy.

When the dark energy and radiation density parameters $\{w_0, w_a, \dneff \}$ are varied simultaneously, the sensitivity remains close to that in the $w_0w_a$CDM+\smnu{} model. In the $w_0w_a$CDM+\smnu+\dneff{} model, it is still possible to get about $2\,\sigma$ evidence for a non-zero mass from the \Euclid{}+CMB-S4+LiteBIRD combination.

\section{Conclusions\label{sec:conclusions}}

Building on the work of \cite{EUCLID:2023uep}, we have implemented and validated a pipeline to consistently forecast the constraints on the neutrino parameters (such as mass and abundance) that will be measurable from the upcoming \Euclid mission of the European Space Agency. We have studied how to model the observables of the photometric survey (allowing for cosmic shear and galaxy clustering measurements) and of the spectroscopic survey (allowing for a reconstruction of the galaxy clustering power spectrum) for cosmologies including massive neutrinos, additional relativistic degrees of freedom, and variations in the dark energy equation of state. 

We find that the Fisher matrices computed with the \texttt{CosmicFish} and \texttt{MontePython} pipelines agree to a 10\% level even after marginalizing over all corresponding nuisance parameters, independently of which EBS (\camb{} or \class) is used. For the purpose of this implementation, we have also compared various nonlinear treatments. We conclude that for our purpose the predictions of {\tt HMcode} \citep{Mead:2020vgs} are the closest to numerical simulations for the cosmological models considered here, even outperforming some of the emulators that we have considered.

The validated forecast pipeline predicts a great deal of complementarity between the various probes measured by \Euclid, in particular between the photometric and spectroscopic surveys. The combination of these probes will help to break the degeneracies between the parameters $\{H_0, \sigma_8, \dneff ,\smnu, w_0, w_a \}$. There is a strong complementarity with additional probes as well, which can be used to lift the residual degeneracy among the small-scale fluctuation amplitude, the Hubble constant and the neutrino parameters. Once combined with data from CMB anisotropy experiments, \Euclid data will lead to extremely tight constraints on $\{\sigma_8, H_0, \smnu, \neff\}$.

Indeed, we find that while \Euclid data alone will already allow us to constrain $\sigma(\smnu = 60 \,{\rm meV}) = 52\,{\rm meV}$ in the standard $\Lambda$CDM+\smnu{} model, its combination with CMB anisotropy data from {\it Planck} will lead to a sensitivity of $\sigma(\smnu = 60 \,{\rm meV}) = 23\,{\rm meV}$. 
Thus, we can expect \Euclid{}+\Planck data to raise evidence for a non-zero neutrino mass at least at the $2.6\,\sigma$ level. If the combined data turns out to be best fit with $\smnu \simeq 60 \,{\rm meV}$, they would indicate a preference for normal ordering over inverted ordering roughly at the $2\,\sigma$ level. In combination with future CMB data from LiteBIRD and CMB Stage-IV, \Euclid will reach a sensitivity of around $\sigma(\smnu = 60 \,{\rm meV}) = 16\,{\rm meV}$, which will allow for a neutrino mass detection at almost $4\,\sigma$. Under the same assumption, one would obtain a clear disambiguation of the neutrino mass hierarchy at $2.5\,\sigma$.
These constraints are only mildly sensitive to the addition of ultra-relativistic degrees of freedom ($\Delta \neff > 0$) and degrade only by a factor of roughly 2 when considering a flexible model of time-varying dark energy.

\Euclid is also expected to measure the Hubble parameter with exquisite precision in these cosmologies, with an uncertainty expected to reach 0.1--0.3\,\kmsMpc in combination with CMB data. This will bring very strong clues regarding the puzzle raised by the current Hubble tension. 

The determination of the effective neutrino number \neff{} by CMB experiments is limited by a parameter degeneracy between \neff{} and $H_0$. The \Euclid probes alone will be sensitive to \neff{} and, even more importantly, when combined with CMB data, they will partially lift this degeneracy. Assuming a fiducial model with no additional relativistic degrees of freedom beyond standard neutrinos ($\neff=3.044$), the combination of \Planck data and \Euclid primary probes will provide a 95\% CL upper bound $\dneff < 0.144$ in the $\Lambda$CDM+\smnu+\neff{} case. The bound will tighten to $\dneff < 0.063$ with future CMB data from LiteBIRD and CMB Stage-IV. These predictions remain stable when marginalising over dynamical dark energy parameters. Such bounds will constrain many well-motivated models of additional relativistic particles such as Goldstone bosons, Weyl fermions, or massless gauge bosons. Not only will decoupling times after the QCD transition be potentially ruled out by \Euclid alone, but in combination with CMB experiments, decoupling times even slightly before the QCD transition can be constrained. Indeed, in combination with future CMB surveys, \Euclid will come close to excluding the presence of these particles at any decoupling temperature (even far before the QCD transition, see \cref{fig:neff}). 

In summary, \Euclid will put tight constraints on the parameters of the $\Lambda$CDM model and of some of its most commonly considered extensions, while bringing strong indications concerning the Hubble constant tension. In combination with either current or future CMB data, \Euclid will constrain neutrino physics, potentially providing the first detection of a non-zero neutrino mass, and some strong hints about the mass ordering. Moreover, the combined data will have an enhanced sensitivity to the effective number of neutrinos, and a significant potential to rule out many candidates for light relics beyond the Standard Model. As such, \Euclid will consolidate the ability of precision cosmology to probe neutrino physics.

\begin{acknowledgements}
\AckEC
We warmly thank Julien Bel for providing matter power spectra measurements from the DEMNUni simulations.
N.F. is supported by the Italian Ministry of University and Research (MUR) through Rita Levi Montalcini project  ``Tests of gravity at cosmological scales" with reference PGR19ILFGP. N.F. and F.P. also acknowledge the FCT project with ref. number PTDC/FIS-AST/0054/2021.
F.P. acknowledges partial support from the INFN grant InDark and the Departments of Excellence grant L.232/2016 of the Italian Ministry of University and Research (MUR).
S.P.'s simulations were performed with computing resources granted by RWTH Aachen University under project thes1329. Z.S. acknowledges funding from DFG project 456622116 and support from the IRAP and IN2P3 Lyon computing centers.
N.S. acknowledges support from the Maria de Maetzu fellowship grant: CEX2019-000918-M, financiado por MCIN/AEI/10.13039/501100011033.
The DEMNUni simulations were carried out in the framework of ``The Dark Energy and Massive-Neutrino Universe" project, using the Tier-0 IBM BG/Q Fermi machine and the Tier-0 Intel OmniPath Cluster Marconi-A1 of the Centro Interuniversitario del Nord-Est per il Calcolo Elettronico (CINECA). We acknowledge a generous CPU and storage allocation by the Italian Super-Computing Resource Allocation (ISCRA) as well as from the coordination of the ``Accordo Quadro MoU per lo svolgimento di attivit\`a congiunta di ricerca Nuove frontiere in Astrofisica: HPC e Data Exploration di nuova generazione'', together with storage from INFN-CNAF and INAF-IA2.
\end{acknowledgements}

\bibliographystyle{aa}
\bibliography{main}

\begin{appendix}

\section{Effect of the numerical derivative method \label{app:der}}

\begin{figure}[ht]
    \centering
    \includegraphics[width = 0.6\linewidth]
    {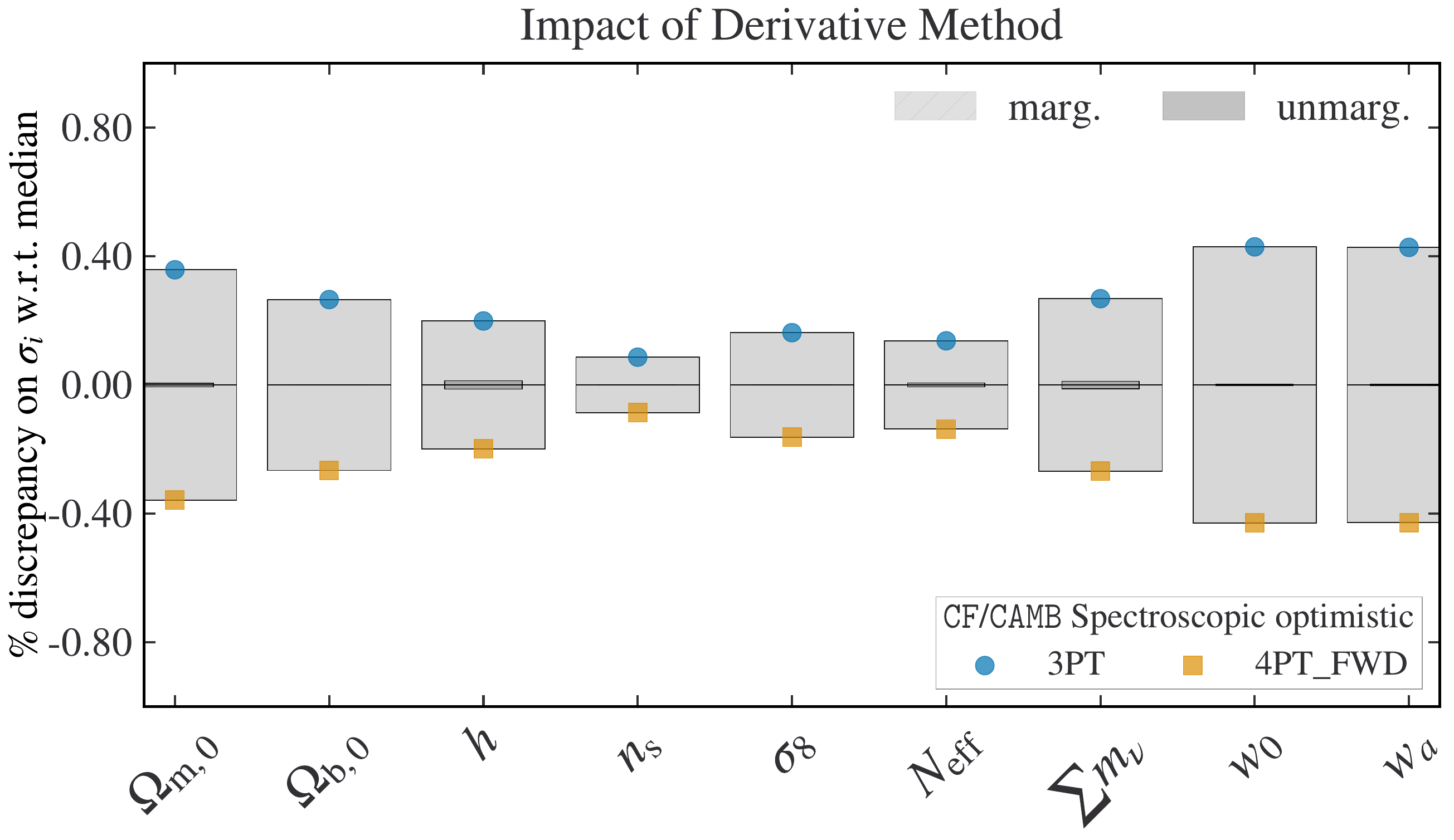}
    \caption{Effect on the marginalised and unmarginalised errors of changing the numerical derivative method from 3PT to 4PT\_FWD in the {\tt CF/CAMB} pipeline. Here we choose the case of the spectroscopic probe with optimistic settings, and we stick to the same assumptions and precision parameters as in the rest of the validation (\cref{sec:validation}). In this test, the nuisance parameters are kept fixed for simplicity. The marginalisation takes place over other cosmological parameters.}
    \label{fig:effect_of_4PT}
\end{figure}

In this work, we set \cosmicfish{} to compute the derivative of observables with respect to cosmological parameters using double-sided derivatives, also called 3-point derivatives (3PT). To check whether our results are stable against the computation of numerical derivatives, we performed a test using a 4-point forward stencil method (4PT\_FWD). We compare the unmarginalised and marginalised errors inferred from the two methods in \cref{fig:effect_of_4PT}. We see that the impact on unmarginalised errors are below 0.1\%, while the marginalised errors are within 0.4\% of each other, that is, well below our validation threshold. We have also checked in 2D contours that the directions of correlation are preserved. Noticeably, the errors from the 4PT\_FWD method are always below the errors from 3PT, showing that our default choice is actually conservative. We conclude that our results are robust against the numerical derivative method.

\section{Settings and accuracy of Einstein--Boltzmann Solvers\label{app:EBS}}

The question of the accuracy settings required by \Euclid forecasts in the \camb{} and \class{} EBSs has been discussed in depth in Section~6 and Appendix~A of \cite{EUCLID:2023uep}. Here we focus on new aspects related to the promotion of \smnu{} and \neff{} as varied parameters. We refer the reader to \cite{EUCLID:2023uep} for more details on other aspects.

Default accuracy settings in the public \camb{} and \class{} versions have been established in order to ensure unbiased MCMC parameter inference from current CMB and large-scale structure surveys. As a general trend, the estimation of Bayesian credible intervals requires less accurate theory predictions when running MCMC than when computing Fisher matrices. Indeed, it is well-known that the excursion in parameter space performed by MCMC tends to smooth out the effect of random numerical noise (in observables and likelihoods) on the estimation of parameter credible intervals. Instead, in Fisher forecasts, the calculation of numerical derivatives with very small step sizes and the inversion of Fisher matrices into covariance matrices tend to enhance the sensitivity of unmarginalised and marginalised errors to such numerical noise. 

For that reason, like in \cite{EUCLID:2023uep}, we use enhanced accuracy settings in Fisher forecasts, using the agreement between marginalised error bars inferred from the \camb{} and \class{} pipelines as an estimator of convergence. When running the {\tt MP/MCMC} pipeline, we instead stick to \class{} default precision. The fact that our {\tt CF/CAMB} and {\tt CF/CLASS} forecasts agree so well (see \cref{fig:1,fig:2,fig:3}) and remain stable against small changes in the choice of derivative step sizes (see \cref{app:step}) proves that our high accuracy settings are sufficient to maintain numerical noise below the \Euclid sensitivity level. Finally, the fact that our Fisher ellipses agree so well with our MCMC 2D contours whenever the posterior is Gaussian with respect to pairs of parameters (see \cref{fig:ValidationTriangle_w0waCDM,fig:ValidationTriangle_mnu+Neff,fig:ValidationTriangle_w0+Mnu}) proves that default precision is sufficient for the purpose of fitting \Euclid data with MCMC.

To be precise, this test proves that our accuracy settings are sufficient to keep random numerical noise (which is by construction independent in \camb{} and \class{}) at a sufficiently low level. It does not prove that the modelling of physical effects is accurate enough, since both codes could in principle rely on the same incorrect modelling of some physical mechanism, and compute observables with the same systematic error. A potential example of this, related to the modelling of massive neutrinos, is discussed in \cref{app:EBS_cosmo}. The only way to mitigate this risk is to compare the results of fits performed with modified versions of EBSs relying on several different schemes for the treatment of hydrogen recombination, of the neutrino phase-space distribution, of the truncation of the Boltzmann hierarchy, of gauge freedom, etc. This goes beyond the level of this work. Note however that such tests have been done repeatedly in many theoretical papers and within several observational collaborations, while the modelling of the $w_0w_a$CDM+\smnu{}+\neff{} model has been intensely scrutinised over three decades. Thus, the impact of these different schemes is well-known and the implementation of $w_0w_a$CDM+\smnu{}+\neff{} in \camb{} and \class{} can be considered as robust. If \camb{} and \class{} did compute observables with the same systematic error, it would be due to a new and yet unidentified physical ingredient, rather than a wrong numerical modelling of the physical assumptions currently going into this cosmological scenario.
 
\subsection{Cosmological parameters\label{app:EBS_cosmo}}

In this work, we use \camb{} version {\tt 1.3.7}. When using the photometric likelihood, we set {\tt halofit\_version = 9} to compute nonlinear corrections with the latest version of {\tt HMcode} \citep{Mead:2020vgs}. For \class{}, we use a branch forked from {\tt v3.2} that implements the same version of {\tt HMcode}. This branch will soon be released publicly as {\tt v3.3}. In the photometric case, we activate accurate {\tt HMcode} calculations with {\tt non linear = hmcode2020}.

The parameters passed to \camb{} and \class{} to account for our fiducial cosmology are the same as described in Appendix A.1 of \cite{EUCLID:2023uep} up to small shifts in the fiducial values. These shifts aim at matching the {\it Planck} best-fit values and the updated prediction $\neff=3.044$ of the standard neutrino-decoupling model \citep{Froustey:2020mcq,Bennett:2020zkv}. They have no impact whatsoever on our results for predicted errors.

Note that $\sigma_8$ is not implemented as a possible input parameter in \camb. We circumvent this issue within \cosmicfish by calling \camb twice, the first time with an arbitrary primordial amplitude parameter $A_{\rm s}$, and the second time with the exact value of $A_{\rm s}$ needed to obtain the desired $\sigma_8$. In the \class case, $\sigma_8$ is passed as a input parameter and the same steps are performed internally.

There are non-trivial aspects related to the setting of neutrino parameters. The approach of \camb{} and \class{} are quite different as far as massive neutrinos are concerned. As a matter of fact, a very precise treatment of the neutrino sector matching the predictions of the standard neutrino decoupling model, described for instance in \cite{Froustey:2020mcq} and in \cite{Bennett:2020zkv}, would have to take into account flavor-dependent non-thermal distortions and flavor mixing in each mass eigenstate. For the sake of computational speed and simplicity, \camb{} and \class{} use by default some effective approaches that attempt to mimic these effects while sticking to a single Fermi--Dirac phase-space distribution shared by all massive neutrinos. There are several ways to set up such an effective description. \camb{} and \class{} use different ones, each with its own motivation. 

These two different treatments generate sub-percent-level differences in the final matter power spectrum, which are too small to be relevant in the analysis of real \Euclid data. However, for the purpose of Fisher matrix calculations, these tiny differences can be amplified by the calculation of numerical derivatives. Thus, for the purpose of validating our Fisher pipelines against each other, we stick to a precise mapping between \camb{} and \class{} parameters, which ensures that the latter code sticks to the same assumptions as the former. This mapping is very similar to the one used in \cite{EUCLID:2023uep} but allows for arbitrary values of \smnu{} and \neff.

In our Fisher forecasts, we call \camb{} with {\tt num\_nu\_massive} set to {\tt 1}, {\tt num\_nu\_massless} set to the varied parameter \neff{} minus 1, and a neutrino mass set to the varied parameter \smnu{}.
Note that the forecast pipeline varies \neff{} in a range given by the fiducial value 3.044 plus or minus one per cent, such that $3.01 < \neff{} < 3.08$. \camb{} uses by default a neutrino model  corresponding to its internal option {\tt share\_delta\_neff = T}.\footnote{See \url{https://cosmologist.info/notes/CAMB.pdf}}
This option triggers several non-trivial operations. Given our input, \camb{} always assumes two massless and one massive neutrino species, and rescales their respective temperature to a common value such that the total neutrino density in the early universe matches \neff{}. 
For that purpose, the true effective number of ultra-relativistic degrees of freedom (for instance massless neutrinos) actually used in the \camb{} equations is redefined internally from $(\neff-1)$ to a new number:
\begin{equation}
N_\mathrm{ur} = 2 + \frac{2}{3} \Big(\neff - \mathrm{floor}[\neff]\Big)\, ,
\label{eq:Nur}
\end{equation}
where $\mathrm{floor}[\neff]$ denotes the largest integer smaller than $\neff$\,. Massive neutrinos are modelled as a perfect Fermi--Dirac species with a temperature $T_\nu$ and fractional density $\Omega_\nu$ computed as
\begin{align}
\frac{T_\nu}{T_\gamma} &= \left(\frac{4}{11}\right)^{1/3} \left(\frac{\neff}{3}\right)^{1/4} \, , \label{eq:Tratio} \\
\Omega_{\nu,0} &= \frac{\smnu}{94.07\,\mathrm{eV}} \left(\frac{\neff}{3}\right)^{3/4} h^{-2}\,.
\label{eq:Omeganu}
\end{align}
Note that, for the sake of consistency, the last equation uses on purpose the numerical value of the ratio $\smnu/[\Omega_{\nu,0} h^2]$ computed in the instantaneous decoupling limit, $94.07\,{\rm eV}$, and not the more realistic value $93.12\,{\rm eV}$ \citep{Froustey:2022sla}. To mimic exactly the same settings in \class, we fix {\tt N\_ncdm} to {\tt 1}, and we set the three input parameters ({\tt N\_ur}, {\tt T\_ncdm}, {\tt Omega\_ncdm}) to the three number calculated respectively with \cref{eq:Nur,eq:Tratio,eq:Omeganu}.

As explained in \cref{sec:neff}, the final MCMC forecasts of \cref{sec:results_euclid} assume instead three massive neutrinos (degenerate in mass and such that each of the three species contributes to \neff{} as 3.044/3) plus extra relativistic species contributing as $(\neff-3.044)$. We achieve this by setting {\tt N\_ncdm} to {\tt 1}, {\tt deg\_ncdm} to {\tt 3}, and the parameters ({\tt N\_ur}, {\tt T\_ncdm}, {\tt Omega\_ncdm}) to: 
\begin{align}
N_\mathrm{ur} &= \neff - 3.044 \,, \label{eq:Nu2} \\
\frac{T_\nu}{T_\gamma} &= \left(\frac{4}{11}\right)^{1/3} \left(\frac{3.044}{3}\right)^{1/4} \, , \label{eq:Tratio2} \\
\Omega_{\nu,0} &= \frac{\smnu}{94.07\,\mathrm{eV}} \left(\frac{3.044}{3}\right)^{3/4} h^{-2}\,.
\label{eq:Omeganu2}
\end{align}

\subsection{Accuracy parameters\label{app:EBS_accuracy}}

\cite{EUCLID:2023uep} have defined some enhanced accuracy settings for the \camb{} and \class{} codes such that, in the framework of the $w_0 w_a$CDM model, \Euclid Fisher forecasts performed with either code are stable and in good agreement with each other. \cite{EUCLID:2023uep} also explained that in the case of \camb{}, on top of increasing some accuracy parameters, it is necessary to adapt the code in order to increase the precision of a bisection algorithm used by both {\tt HALOFIT} and {\tt HMcode}.\footnote{The \camb{} file {\tt fortran/halofit.f90} contains a function {\tt THalofit\_GetNonLinRatios()} that computes a characteristic radius {\tt rmid} with a bisection method. The bisection accuracy is set by the line:
{\tt if (abs(diff).le.0.001) then \ldots}
It is essential to substitute 0.001 with a smaller tolerance equal to (or smaller than) $10^{-6}$, otherwise, the error on {\tt rmid} is way too large given the \Euclid sensitivity, and leads to inaccurate derivatives in the Fisher matrix calculation. In this work we use a tolerance of $10^{-6}$.} This adaptation is still relevant in our context and we use it throughout our validation process.

From the point of view of accuracy setting, the difference between \cite{EUCLID:2023uep} and this work is that we now need to compute some derivatives with respect to the parameters \smnu{} and \neff. 
We found that derivatives with respect to \neff{} are stable and accurate enough with the settings used in \cite{EUCLID:2023uep}. Instead, getting accurate derivatives with respect to \smnu{} requires particular care. Even after solving the issue of the potentially different modelling of massive neutrinos in the way discussed in the previous subsection, one is confronted with the fact that the effect of massive neutrinos is very small and that \camb{} and \class{} compute it with limited precision in order to avoid prohibitive computational time. We circumvent this problem in two ways. Firstly, in the calculation of two-sided derivatives with respect to \smnu{}, we use a step size given by 10\% of the fiducial value -- instead of 1\% for other parameters. With such a step, it is easier to obtain a variation of the power spectrum dominated by the physical effect of massive neutrinos rather than numerical noise. Secondly, in both codes, we use enhanced accuracy settings specific to the massive neutrino sector.

In general, the accuracy with which EBSs compute massive neutrino perturbations depends essentially on three things: the number of discrete momenta used to sample the neutrino phase-space distribution; the cut-off multipole $\ell_{\rm max}$ at which the Boltzmann hierarchy is truncated; and some optional approximation schemes reducing the number of equations in some regime. The number of momenta is controlled by {\tt accuracy\_boost} in \camb{} and {\tt tol\_ncdm\_synchronous} in \class{}. The cut-off $\ell_{\rm max}$ depends on {\tt l\_accuracy\_boost} in \camb{} and {\tt l\_max\_ncdm} in \class{}. Additional approximations can be de-activated in \camb{} with {\tt accurate\_massive\_neutrino\_transfers} set to true. Finally, \class{} features a neutrino fluid approximation \citep{Lesgourgues:2011rh} that gets switched on when each comoving wavenumber is sufficiently deep inside the sub-Hubble regime. This approximation can either be postponed to later times by increasing {\tt ncdm\_fluid\_trigger\_tau\_over\_tau\_k}, or completely de-activated by setting {\tt ncdm\_fluid\_approximation} to {\tt none} or 3.

In the \camb{} precision settings, the two parameters {\tt accuracy\_boost} and {\tt l\_accuracy\_boost} are very important, since they increase the accuracy of the code in multiple places. \cite{EUCLID:2023uep} already used {\tt accuracy\_boost=3}, which is sufficient for an accurate sampling of massive neutrino momentum, and {\tt l\_accuracy\_boost=3}, which leads to a truncation of the Boltzmann hierarchy for massive neutrinos at a high enough $\ell_{\rm max}=75$.
Thus we can define some \camb{} high precision (\camb:{\small HP}) settings that are identical to those of \cite{EUCLID:2023uep} up to the addition of one flag in the last line:

\begin{lstlisting}[language=Python,caption=\camb:{\footnotesize HP} precision settings]
    do_late_rad_truncation = T
    high_accuracy_default = T
    transfer_interp_matterpower = T
    accurate_reionization = F
    accuracy_boost = 3
    l_accuracy_boost = 3
    # - plus tolerance parameter within THalofit_GetNonLinRatios() decreased manually to 1.e-6
    # - plus one new line specific to massive neutrinos: 
    accurate_massive_neutrino_transfers = T
\end{lstlisting} 

The drawback of these settings is that \camb{} becomes very slow, but this is a necessity for performing accurate Fisher matrix forecasts instead of MCMC forecasts. 

In the \class{} case, we also define our  \class:{\small HP} settings in the same way as in \cite{EUCLID:2023uep}, up to three additional lines to deal with massive neutrino accuracy:

\begin{lstlisting}[language=Python,caption=\class:{\footnotesize HP} precision settings, label=lst:classHP]
    k_per_decade_for_bao = 50
    k_per_decade_for_pk = 50
    l_max_g = 20
    l_max_pol_g = 15
    radiation_streaming_approximation = 2
    radiation_streaming_trigger_tau_over_tau_k = 240.
    radiation_streaming_trigger_tau_c_over_tau = 100.
    background_Nloga = 6000
    thermo_Nz_log = 20000
    thermo_Nz_lin = 40000
    tol_perturbations_integration = 1.e-6
    hmcode_tol_sigma = 1.e-8
    # - plus three new lines specific to massive neutrinos: 
    tol_ncdm_synchronous = 1.e-5
    l_max_ncdm = 25
    ncdm_fluid_trigger_tau_over_tau_k = 100.
\end{lstlisting} 

With the last line, the neutrino fluid approximation is still used, but with a smaller impact than when sticking to default precision, because it gets switched on at a later time for each wavenumber.

\begin{figure}
    \centering
    \includegraphics[width=0.8\linewidth]{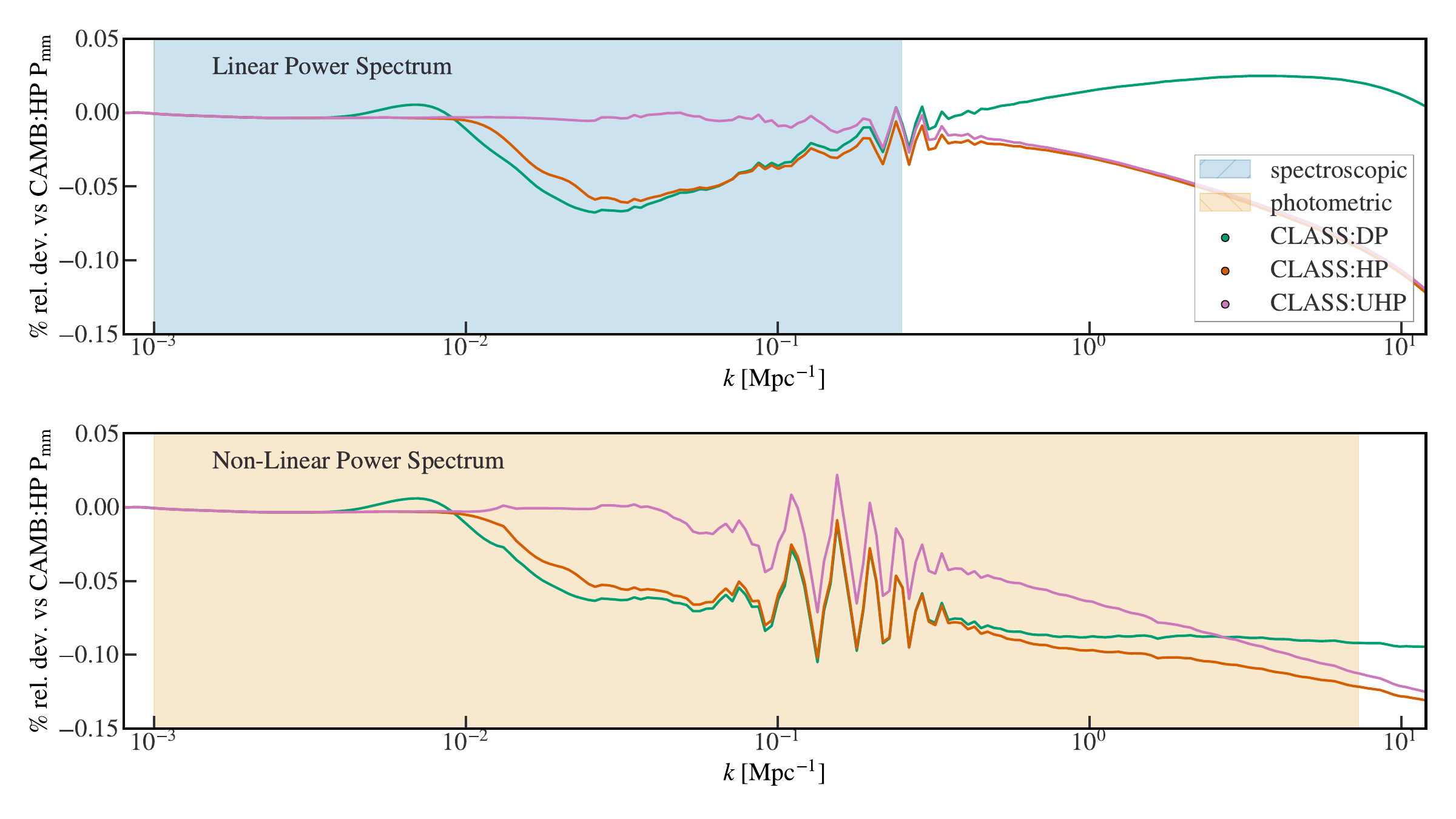}
    \caption{Ratio of the fiducial linear and nonlinear power spectrum $P_\mathrm{mm}(k,z=0)$ computed using \camb{} with high precision settings (\camb{}:{\footnotesize HP}) to that computed using \class{} with default precision (\class:{\footnotesize DP}), high precision (\class:{\footnotesize HP}) or ultra-high precision (\class:{\footnotesize UHP}). In this fiducial case, the summed neutrino mass is $\smnu=60\,{\rm meV}$. The cyan shading denotes the range of wavenumbers probed by the spectroscopic likelihood, while the orange shading shows the range of wavenumbers for which the photometric likelihood is the most sensitive. \label{fig:comparasion_fiducial_spectra}
    }
    
\end{figure}

In \cref{fig:comparasion_fiducial_spectra}, we plot the percentage difference between the linear and nonlinear power spectrum at $z=0$ computed for the fiducial model with a mass of $\smnu=60\,{\rm meV}$ by \camb{} (using \camb:{\small HP}) and \class using either default precision (\class:{\small DP}), high precision (\class:{\small HP}), or an ultra-high precision setting that will be defined below (\class:{\small UHP}). The shading highlights the range of scales used in the spectroscopic likelihood (with the linear spectrum as input) and in the photometric likelihood (with the nonlinear spectrum as input). 

Even with default precision, the difference between \camb:{\small HP} and \class:{\small DP} is at most 0.1\% in the linear and nonlinear spectra, that is, below the \Euclid sensitivity level. Thus, it is not a surprise that MCMC forecasts performed with \class:{\small DP} agree with Fisher forecasts performed with \camb:{\small HP}. 
However, we know that Fisher forecasts require enhanced accuracy to avoid an amplification of numerical errors when computing derivatives and inverting Fisher matrices. In the case of the photometric probes, we find that \camb:{\small HP} and \class:{\small DP} settings are sufficient for computing stable derivatives and obtaining a very good match between the marginalised and unmarginalised errors derived from {\tt CF/CAMB} and {\tt CF/CLASS}. Switching to \class{}:{\small HP} has a marginal impact, but it tends to offer even more stability in the calculation of derivatives, with an even weaker dependence on the step sizes. To stay on the conservative side, we adopt this setting whenever using the photometric likelihood. The excellent agreement between the marginalised and unmarginalised errors obtained with \camb{}:{\small HP} and \class{}:{\small HP} for this probe is illustrated by the upper panels of \cref{fig:1,fig:2,fig:3}.

However, we find a slightly poorer agreement between the marginalised and unmarginalised errors obtained with \camb:{\small HP} and \class:{\small HP} in the spectroscopic case (not shown here for conciseness), with differences marginally exceeding our validation threshold of 10\% in some cases. Looking at various elements in the Fisher matrix, we find that this is caused by small differences in the derivatives with respect to \smnu. After some investigation, we conclude that the below $10^{-3}$ error introduced by the \class{} neutrino fluid approximation is responsible for such small differences. Even with the enhanced setting in the last line of \class:{\small HP}, this approximation leads to a suppression of the matter power spectrum by 0.07\% on scales between $k=10^{-2}h\,{\rm Mpc}^{-1}$ and $k=2\times 10^{-1}h\,{\rm Mpc}^{-1}$, which is clearly visible in the upper plot of \cref{fig:comparasion_fiducial_spectra}. This suppression propagates to the derivative of the linear power spectrum $\partial P(k,z)/\partial {\sum m_\nu}$ and finally to the Fisher matrix.

We fix this issue by defining some ultra-high precision settings, \class:{\small UHP}, in which the neutrino fluid approximation is never used. Switching off the fluid approximation has a potential drawback. In this case, the Bolztmann hierarchy is integrated even deep inside the Hubble scale regime. As a consequence, the evolution of perturbations can be more affected by artefacts coming from the truncation of the hierarchy at $\ell_{\rm max}$. These artefacts can be seen as an artificial reflection of power on the boundary \citep[see for instance][]{Hu:1995fqa,Lesgourgues:2011rh}. Thus, the cut-off multipole must be increased accordingly. In \class:{\small UHP}, we set it to $\ell_{\rm max}=40$. Finally, in the absence of a neutrino fluid approximation, \class{} becomes much slower. We speed it up by 20\% by integrating the differential equations with the evolver {\tt rkck} (a Runge-Kutta evolver more optimised for highly oscillatory massive neutrino equations) instead of the default {\tt ndf15} \citep[an implicit evolver more optimised for stiff systems, see][]{Blas:2011rf}. All in all, this amounts to replacing the last two lines of \class:{\small HP} by three new lines:

\begin{lstlisting}[language=Python,caption=\class:UHP precision settings, label=lst:classUHP]
    k_per_decade_for_bao = 50
    k_per_decade_for_pk = 50
    l_max_g = 20
    l_max_pol_g = 15
    radiation_streaming_approximation = 2
    radiation_streaming_trigger_tau_over_tau_k = 240.
    radiation_streaming_trigger_tau_c_over_tau = 100.
    background_Nloga = 6000
    thermo_Nz_log = 20000
    thermo_Nz_lin = 40000
    tol_perturbations_integration = 1.e-6
    halofit_tol_sigma = 1.e-8
    # - plus four new line specific to massive neutrinos: 
    tol_ncdm_synchronous = 1.e-5
    l_max_ncdm = 40
    ncdm_fluid_appoxmation = 3.
    evolver = 0
 \end{lstlisting}

With such \class:{\small UHP} settings, we are able to validate the Fisher matrix pipelines for the spectroscopic probe with either pessimistic or optimistic settings (see the lower panels in \cref{fig:1,fig:2,fig:3}). 

In summary, when using the {\tt CF/CAMB} pipeline, we always use \camb:{\small HP} settings. For the {\tt CF/CLASS} and {\tt MP/Fisher} pipelines, we use \class:{\small HP} for the photometric probe and \class:{\small UHP} for the spectroscopic probe. Finally, for the {\tt MP/MCMC} pipeline, we always use \class:{\small DP}.

\section{Effect of the step size for \texorpdfstring{$\sum m_\nu$}{the neutrino mass sum} \label{app:step}}

\begin{figure}[ht]
    \centering
    \includegraphics[width= 0.6 \linewidth]{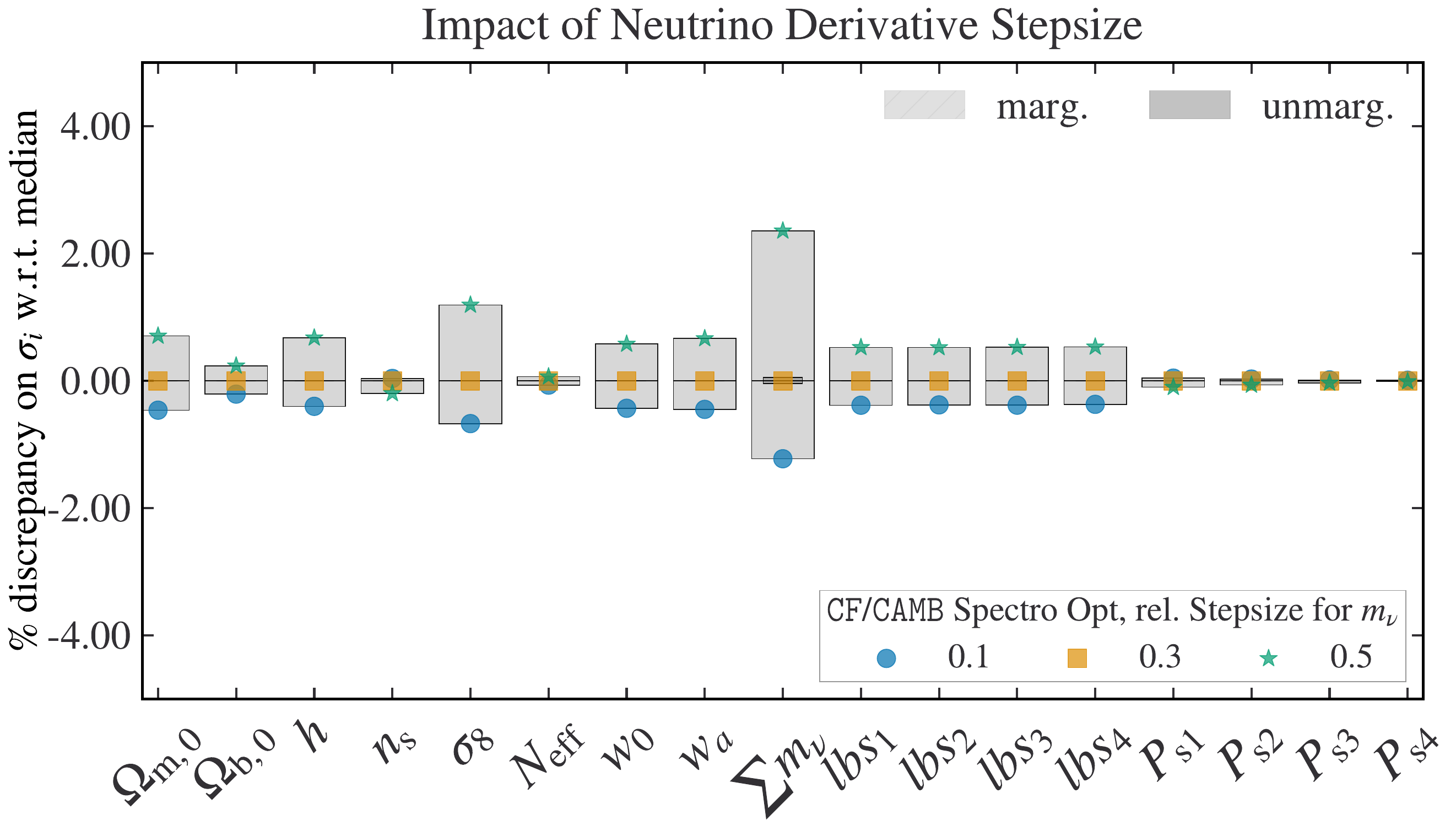}
    \caption{Effect on the marginalised and unmarginalised errors of changing the step size in the calculation of numerical derivatives with respect to the summed neutrino mass \smnu{} in the {\tt CF/CAMB} pipeline. We choose here the case of the spectroscopic probe with optimistic settings, and we stick to the same assumptions and precision parameters as in the rest of the validation \cref{sec:validation}. The step sizes (0.1, 0.3, 0.5) are expressed relative to the fiducial value $\smnu=60\,{\rm meV}$, and thus correspond to ($6\,\mathrm{meV}$, $18\,\mathrm{meV}$, $30\,\mathrm{meV}$).}
    \label{fig:effect_of_stepsize}
\end{figure}

Due to the smallness of the effect of the summed neutrino mass on the power spectrum, we have to use larger step sizes for \smnu{} than for the other parameters, to ensure that the numerical derivatives are dominated by physical effects rather than numerical noise from the EBS. For other parameters our step size is set to 1\% of the fiducial value. Instead, for \smnu, our default relative stepsize is of 10\%. To check whether this choice is numerically stable, we test here the effect of using larger step sizes in the {\tt CF/CAMB} pipeline, for the spectroscopic probe and with optimistic settings. The results can be seen in \cref{fig:effect_of_stepsize}. We find that the errors do not change by more than 3\% when going from a 10\% step size to a 50\% step size (corresponding to step sizes of $6\,\mathrm{meV}$ to $30\,\mathrm{meV}$). Additionally, in the 2D contour plots we checked that the directions of correlation with other cosmological parameters are preserved. This speaks again to the robustness of the derivatives. Using the {\tt CF/CLASS} pipeline we find similar results, but in that case the errors remain stable even when the relative step size is lowered to a few percent only. 

\section{Effect of the nonlinear prescription\label{app:halofit}}

\begin{figure}[ht]
    \centering
    \includegraphics[width=0.6\textwidth]{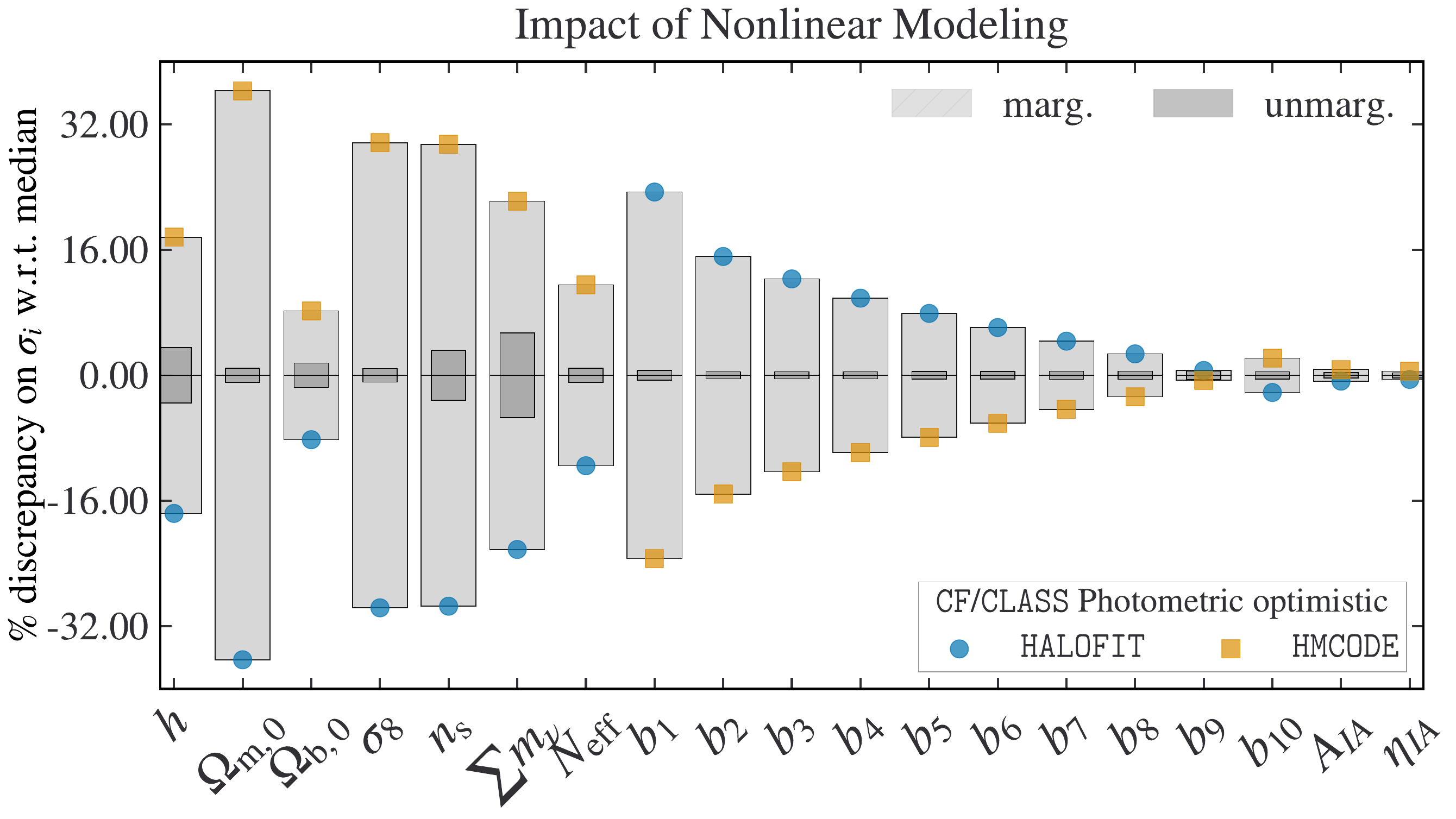}\\
        \caption{Difference between the marginalised and unmarginalised errors inferred from the {\tt CF/CAMB} pipeline using either {\tt HALOFIT} or {\tt HMcode}.}
    \label{fig:effect_of_nonlinear}
\end{figure}

\begin{figure}[ht]
\centering
\includegraphics[width=0.6\textwidth]{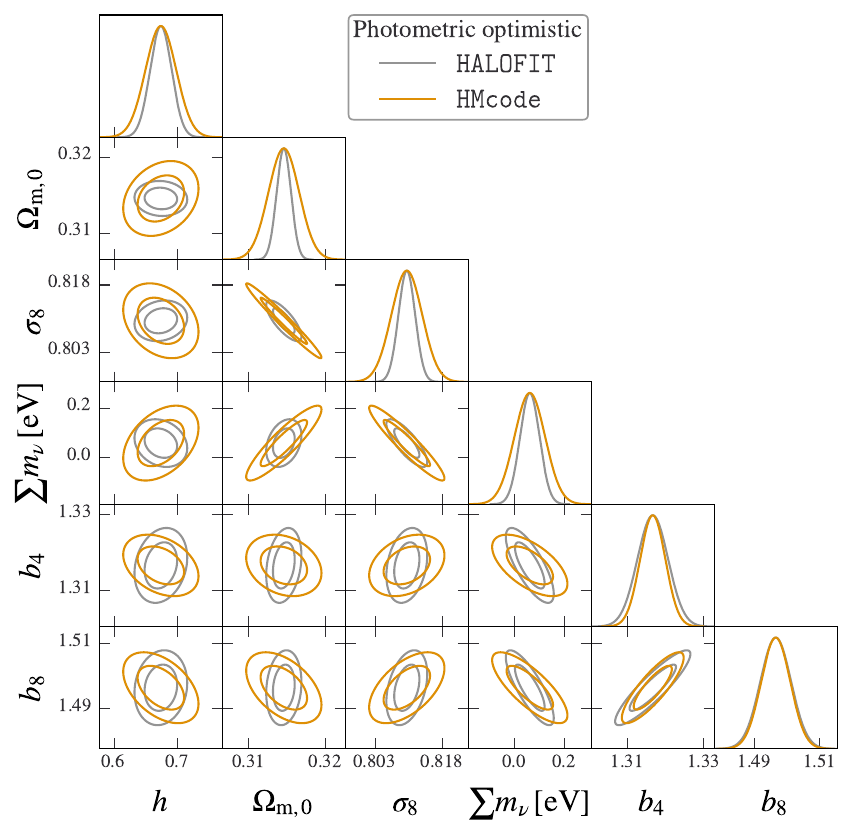}
    \caption{Marginalised posteriors and confidence contours for the same Fisher forecasts as in \cref{fig:effect_of_nonlinear}, that is, when switching from {\tt HMcode} to {\tt HALOFIT}.
    }
    \label{fig:triangle_halofit}
\end{figure}

\begin{figure}[ht]
    \centering
    \includegraphics[width=0.6\textwidth]{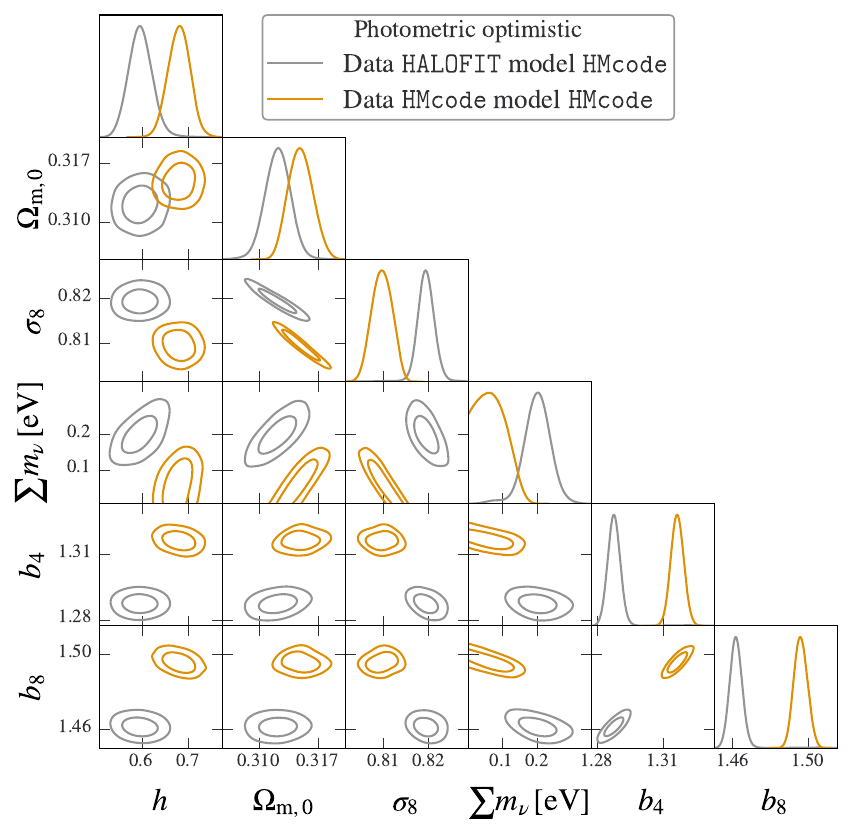}
    \caption{Biasing test, where the fiducial model is generated using {\tt HMcode} and fitted using {\tt HALOFIT}.
   }
    \label{fig:triangle_halofit_bias}
\end{figure}

A difference between \cite{EUCLID:2023uep} and this work resides in the choice of algorithm for the computation of nonlinear corrections to the power spectrum. As explained in \cref{sec:nl}, in the current analysis, we switched from {\tt HALOFIT} to {\tt HMcode} \citep{Mead:2020vgs}. As a matter of fact, the latter is closer to the results of $N$-body simulations, especially in the presence of massive neutrinos. 
In \cref{fig:effect_of_nonlinear} we can see the effect induced by this change on the predicted marginalised and unmarginalised errors of the $\Lambda$CDM+\smnu+\neff{} model. While the unmarginalised errors remain stable up to 10\% variations, we see a strong difference -- of as much as 70\% -- in the marginalised errors on $\Omega_{{\rm m},0}$, $\sigma_8$, $n_{\rm s}$, $h$, and $\smnu$. 
The error on cosmological parameters predicted with {\tt HALOFIT} are systematically lower than those obtained with {\tt HMcode}. 
In the 2D contours of \cref{fig:triangle_halofit} we also see that the directions of correlation between pairs of parameters are different with {\tt HALOFIT}. 
Finally, \cref{fig:triangle_halofit_bias} shows that in the case of optimistic specifications if {\tt HMcode} is used to generate the fiducial model -- or, in other words, is assumed to account for the truth -- but the fit is performed using {\tt HALOFIT}, the cosmological parameters are recovered incorrectly, with up to $3\,\sigma$ bias in the cosmological parameters and even more in the nuisance parameters, in agreement with \cite{Euclid:2020tff}. 
These tests stress the high sensitivity of the \Euclid pipeline to the method used to predict nonlinear corrections. 
In order to measure cosmological parameters, it is essential to fit the data with an extremely accurate model of nonlinear corrections.

\section{Importance of using the CDM+baryons power spectrum \label{app:cb}}

In \cref{sec:euclid-observables}, we stressed the importance of modelling the galaxy power spectrum as a biased tracer of the cold dark matter plus baryonic power spectrum \citep{Castorina:2013wga,DEMNUni1}, such that on linear scales $P_{\delta\delta}^{\rm GG}(k;z) \simeq b^2(z) P_\mathrm{cc}(k;z)$. Here we illustrate the importance of this assumption by deliberately adopting instead the inaccurate prescription $P_{\delta\delta}^{\rm GG}(k;z)=b^2(z) P_\mathrm{mm}(k;z)$, where the total matter power spectrum $P_\mathrm{mm}(k;z)$ includes neutrino density fluctuations. At the level of a forecast, the difference between the two treatments arises not from different assumptions concerning the fiducial model, but from the different dependence of the spectrum on cosmological parameters, and particularly on the neutrino mass. Indeed, we expect that $P_\mathrm{cc}$ is less suppressed by neutrino free-streaming effects, such that
\begin{equation}
\left| \partial_{\Sigma m_\nu} \ln P_{\delta\delta}^{\rm GG}(k;z) \right|
= \left| \partial_{\Sigma m_\nu} \ln P_\mathrm{cc}(k;z) \right|
< \left| \partial_{\Sigma m_\nu} \ln P_\mathrm{mm}(k;z) \right|\,.
\end{equation}
Thus, adopting the wrong prescription should artificially increase the sensitivity to \smnu{} (and to all correlated parameters), as argued in more details in \cite{Vagnozzi:2018pwo}.

\begin{figure}
    \centering
    \includegraphics[width=0.6\textwidth]{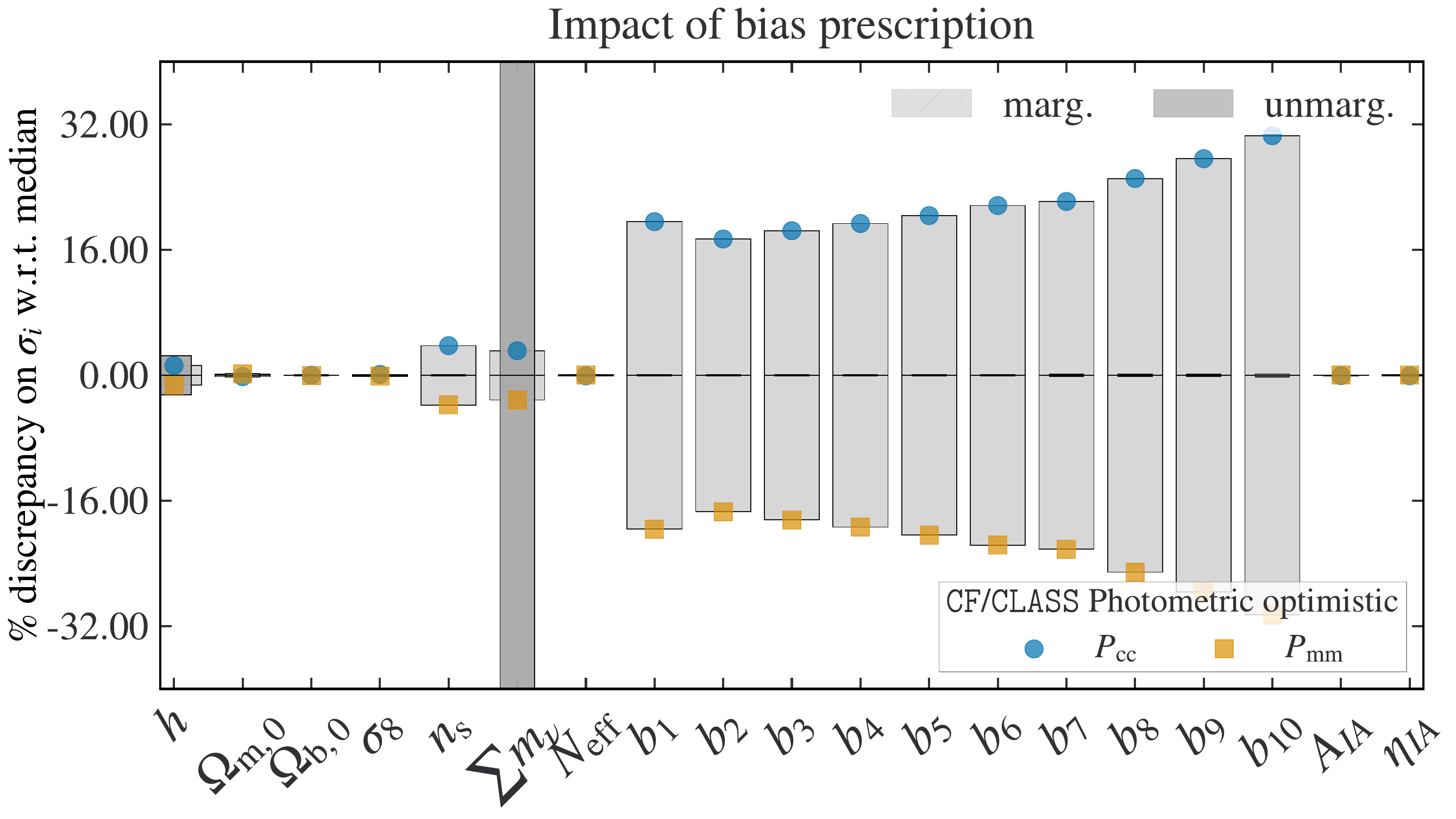}
    \caption{For the photometric probe with optimistic settings, comparison of the one-dimensional marginalised and unmarginalised errors obtained under the assumption that the galaxy power spectrum is a biased tracer of either $P_\mathrm{mm}$ (incorrect modelling) or $P_{\rm cc}$ (correct modelling). For the parameter \smnu{} the relative difference in the unmarginalised error goes outside the frame and is of the order of 43\%.}
    \label{fig:A_dots_PcbvPmm}
\end{figure}

\begin{figure}
\centering
    \includegraphics[width=0.6\textwidth]{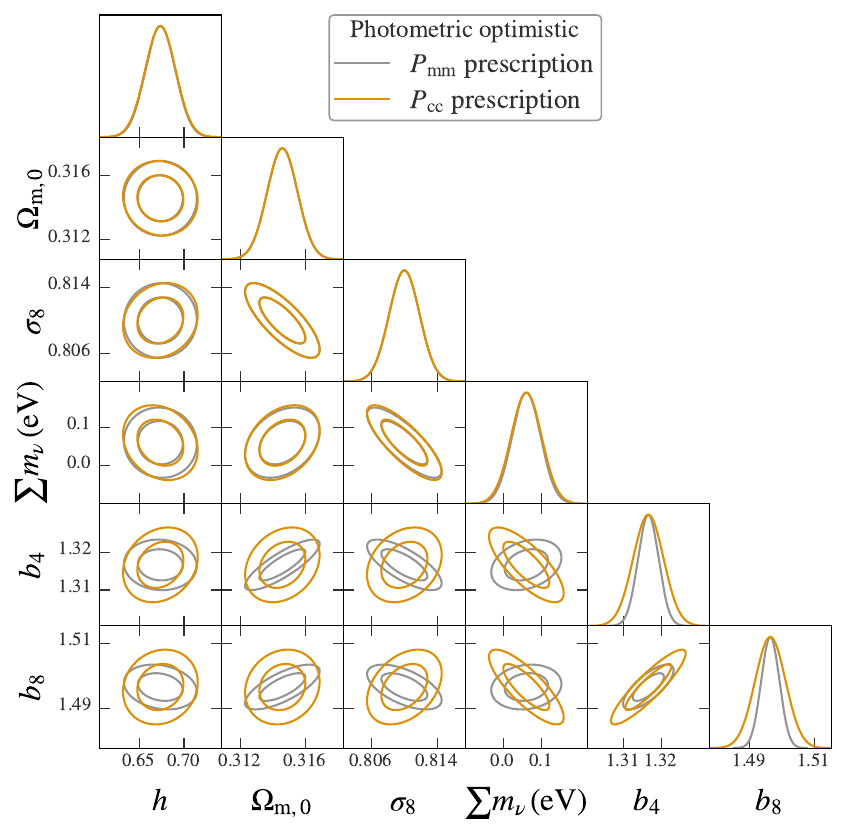}
    \caption{Marginalised posteriors and confidence contours for the same Fisher forecasts as in \cref{fig:A_dots_PcbvPmm}.
    }
    \label{fig:A_triangle_PcbvPmm}
\end{figure}

\begin{figure}
    \centering
    \includegraphics[width=0.6\textwidth]{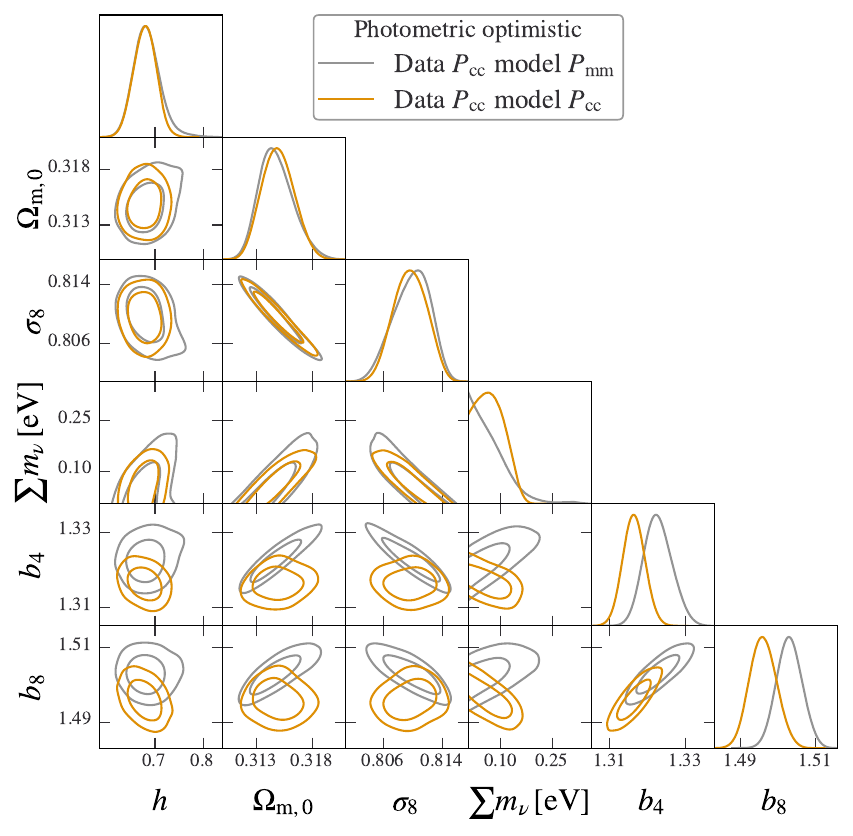}
    \caption{For the photometric probe with optimistic settings, marginalised posteriors and confidence contours from two MCMC forecasts based on the same fiducial model generated with the $P_\mathrm{cc}$ prescription. The mock data are then fitted either with the $P_\mathrm{mm}$ (incorrect modelling) or $P_\mathrm{cc}$ (correct modelling) prescription.
    }
    \label{fig:bias_pcb}
\end{figure}

This issue is relevant both for the galaxy clustering component of the photometric probe and for the spectroscopic probe. Here we choose to illustrate its importance in the case of the photometric probe with optimistic settings. In \cref{fig:A_dots_PcbvPmm,fig:A_triangle_PcbvPmm}, we present a comparison between two Fisher forecasts for this probe, the first one adopting the correct approach, and the second one switching to the wrong prescription both for the fiducial and fitted models. In \cref{fig:bias_pcb} we present a second test based on MCMC forecasts in which we stick to the correct approach for the fiducial model, but we fit it using either the correct or the incorrect prescription.

In the first test, the unmarginalised error on \smnu{} differs by 43\% between the two cases (see \cref{fig:A_dots_PcbvPmm}). This is a direct consequence of the different derivatives of $P_\mathrm{cc}$ and $P_\mathrm{mm}$ with respect to \smnu. This discrepancy gets mitigated by the marginalisation over all other parameters, which reduces the difference on $\sigma(\smnu)$ to about 3\%. However, in the marginalised case, the error bars on the parameters that are correlated with \smnu{} also increase -- most noticeably for the bias parameters. As a matter of fact, the response of $P_\mathrm{cc}(k;z)$ to a variation of the neutrino mass is more degenerate with a change of amplitude (and thus of linear bias) than the response of $P_\mathrm{mm}(k;z)$. Thus, when using the wrong prescription, one underestimates the correlation between \smnu{} and the bias parameters (see \cref{fig:A_triangle_PcbvPmm}), and artificially reduces the error not only on \smnu{} but also on each $b_i$\,.

In the second test, we can check that fitting the correct fiducial model with the wrong prescription leads to artificially strong constrains on \smnu{} (see \cref{fig:bias_pcb}). In this case, the theoretical model based on $P_\mathrm{mm}$ cannot fit perfectly the fiducial model based on $P_\mathrm{cc}$: there is always a residual difference. This test is particularly interesting because it shows that the incorrect model fits this residual by a combination of a wrong \smnu{} and wrong biases. This could be expected, since the difference between $P_\mathrm{cc}$ and $P_\mathrm{mm}$ is close to a constant factor over a wide range of scales, as can be checked in \cref{fig:Pk_Mnu_linear_suppression}. This explains the shifted means in the \smnu{} and $b_i$ posteriors visible in \cref{fig:bias_pcb}. Looking at the \smnu{} posteriors, we see that if such a mistake occurred in the analysis of real data, one would incorrectly conclude that the data prefer a vanishing neutrino mass, while in reality $\smnu=60\,{\rm meV}$ is preferred.

\section{Parameter correlations in different cosmologies \label{app:res}}

\begin{figure}[ht]
    \centering
    \includegraphics[width=0.7 \linewidth]{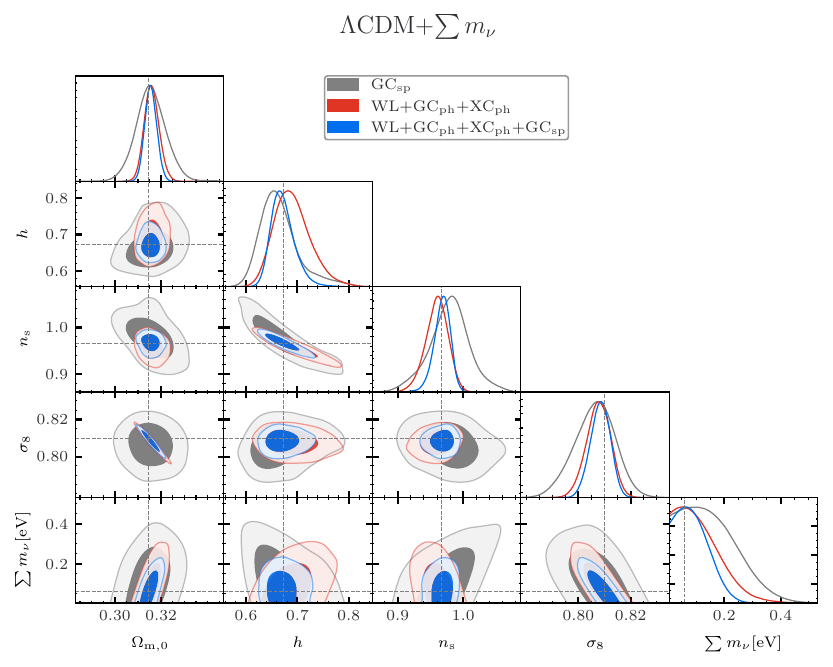}
    \caption{Same as \cref{fig:res_w0waCDM}, but for the baseline $\Lambda$CDM+$\smnu$ model.}
    \label{fig:app_lcdmM}
\end{figure}

\begin{figure}[ht]
    \centering
    \includegraphics[width=0.83 \linewidth]{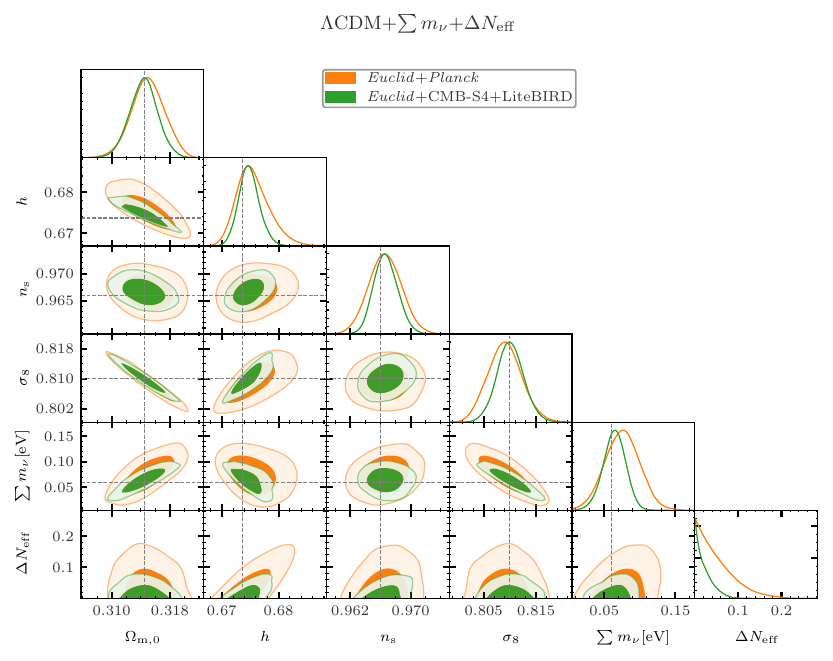}
    \caption{Same as \cref{fig:res_lcdmM}, but for the $\Lambda$CDM+$\smnu$+$\dneff$ model.}
    \label{fig:app_lcdmMN_cmb}
\end{figure}

\begin{figure}[ht]
    \centering
    \includegraphics[width=0.9 \linewidth]{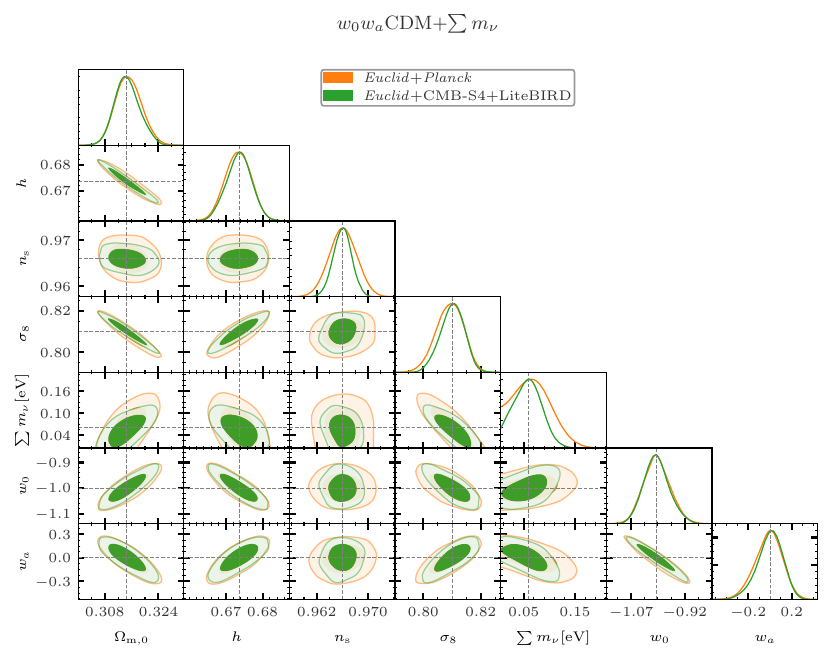}
    \caption{Same as \cref{fig:res_lcdmM}, but for the $w_0w_a$CDM+$\smnu$ model.}
    \label{fig:app_wcdmM_cmb}
\end{figure}

In this Appendix we show additional triangle plots with the 1D marginalised posteriors and 2D confidence contours that are not shown in \cref{sec:results_euclid}.
\Cref{fig:app_lcdmM} shows the results obtained in the baseline $\Lambda$CDM+$\smnu$ model from each of the \Euclid primary probes and from their combination.
\Cref{fig:app_lcdmMN_cmb,fig:app_wcdmM_cmb} show the results obtained in the intermediate models ($\Lambda$CDM+$\smnu$+$\dneff$, and $w_0w_a$CDM+$\smnu$, respectively) with the data combinations \Euclid{}+\Planck (orange), and \Euclid{}+CMB-S4+LiteBIRD (green).
Note that, when comparing \cref{fig:app_lcdmM} with \cref{fig:res_w0waCDM}, the photometric probes (red countours) show a different direction of degeneracy between $\smnu$ and $h$, which are positively correlated in the baseline $\Lambda$CDM+$\smnu$ model.
On the other hand, when combining \Euclid with CMB (\cref{fig:app_lcdmMN_cmb,fig:app_wcdmM_cmb}), it is difficult to identify any significant change in the direction of the degeneracies in parameter space between one model and the other.

\section{Correlation with bias parameters \label{app:nuisance}}

\begin{figure}[ht]
    \centering
    \includegraphics[width=0.9 \linewidth]{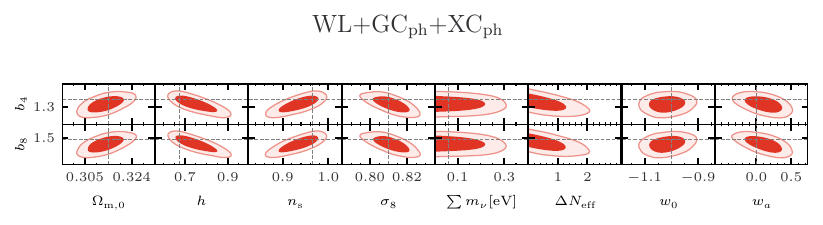}
    \caption{2D confidence contours showing the correlation between the bias parameters of the fifth, and of the ninth redshift bins ($b_4$ and $b_8$) and the cosmological parameters, obtained by fitting photometric observables (WL$+\GCph+\XCph$) to the $w_0w_a$CDM+$\smnu$ model.}
    \label{fig:app_bias_photo}
\end{figure}

\begin{figure}[ht]
    \centering
    \includegraphics[width=0.9 \linewidth]{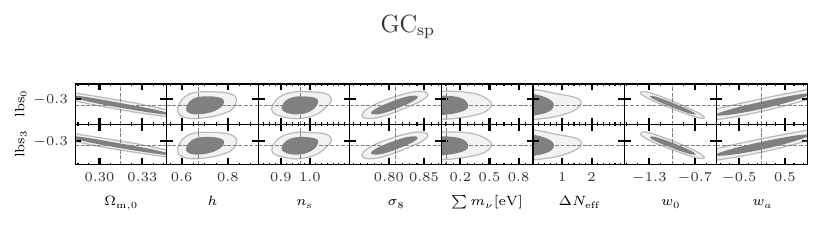}
    \caption{2D confidence contours showing the correlation between the bias parameters of the first and of the fourth redshift bin [$\ln{b_0 \sigma_8 (z_0)}$, and $\ln{b_3 \sigma_8 (z_3)}$] and the cosmological parameters, obtained by fitting spectroscopic observables (\GCsp) to the $w_0w_a$CDM+$\smnu$ model.}
    \label{fig:app_bias_spectro}

\end{figure}

\Cref{fig:app_bias_photo,fig:app_bias_spectro} show the correlation between the bias parameters of photometric and spectroscopic observables, respectively, and the cosmological parameters in the extended $w_0w_a$CDM+$\smnu$+$\dneff$ model. We show only two redshift bins for each observable noticing that the correlation is the same in all the redshift bins. It appears that there is no correlation between the neutrino mass and the bias parameters, either in WL+$\GCph+\XCph$ or in $\GCsp$. On the other hand, a mild anti-correlation is found between the bias parameters and $\dneff$ in WL+$\GCph+\XCph$, thus arising from \GCph. The reason why this correlation appears in \GCph while it is not present in $\GCsp$ is that the projection of the matter power spectrum in angular harmonics and the convolution with the window function redistribute the power over a wider range of scales; the feature induced by a variation of $\dneff$ on the matter power spectrum (see \cref{fig:Pk_Neff_linear_suppression}) is smoothed out and thus can be mimicked by a variation of the bias.

Besides $\dneff$, the bias parameters show correlations with most cosmological parameters, highlighting the importance of a careful study of bias from simulations.

\section{Impact of baryonic feedback \label{app:bf}}

\begin{figure}[ht]
    \centering
    \includegraphics[width=0.9 \linewidth]{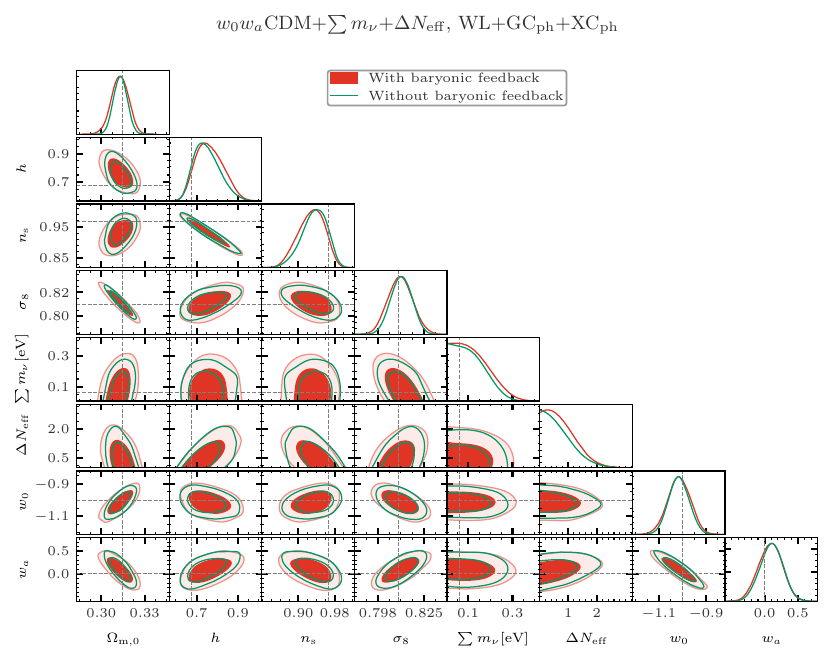}
    \caption{1D marginalised posteriors and 2D confidence contours for cosmological parameters and for baryonic feedback parameters, obtained by fitting photometric observables (WL+\GCph+\XCph) to the $w_0w_a$CDM+$\smnu$ model, with and without varying baryonic feedback parameters (red-filled and green-empty contours, respectively). The contours for baryonic feedback parameters are also shown when they are varied.}
    \label{fig:app_bf}
\end{figure}

An accurate modelling of large scale structure observables in the nonlinear regime requires accounting for baryonic feedback effects \citep{Semboloni:2011fe, Parimbelli:2018yzv, Schneider:2019snl, Schneider:2019xpf, Euclid:2020tff, SpurioMancini:2023mpt}. Here, we test the robustness of our results on the neutrino mass sensitivity against such effects. To this end, we multiply the matter power spectrum $P_{\rm mm}(k_\ell,z)$ of \cref{eq:photo_Cl} in \cref{sec:photo} by the boost factor obtained with the emulator \bcemu{} \citep{Giri:2021qin}. Note that we apply the boost factor only to $P_{\rm mm}(k_\ell,z)$, thus, accounting for baryonic feedback effects only in weak lensing. We vary two \bcemu{} parameters. The first one, $\log_{10}M_{\rm c}$, is related to the critical mass $M_{\rm c}'$ such that halos with a virial mass smaller than $M_{\rm c}'$ have a gas profile shallower than the Navarro-Frenk-White profile. $M_{\rm c}$ is the dimensionless parameter such that $M_{\rm c}=M_{\rm c}'/(1\,h^{-1}\,M_\odot)$. The second parameter is the index $\nu_{\sfont{{M_{\rm c}}}}$ describing the power-law redshift dependence of $M_{\rm c}$. This choice is motivated by the fact that $\log_{10}M_{\rm c}$ is the parameter governing the suppression of $P_{\rm mm}(k_\ell,z)$, thus, it is expected to be the most correlated with the neutrino mass. Moreover, given the cut-off at $\ell=1500$ in our pessimistic specifications, including more than one free parameter for baryonic feedback might lead to an overfitting, leaving the additional parameters unconstrained.
In \cref{fig:app_bf}, we compare the constraints from the \Euclid photometric probe on cosmological parameters obtained with and without baryonic feedback (red-filled and green-empty contours, respectively) in the $w_0w_a$CDM+$\smnu$+$\dneff$ model. We assume pessimistic specifications as in our final results of \cref{sec:results_euclid}. We see that baryonic feedback does not have a dramatic impact on the constraints, and the critical mass parameter does not show a well defined correlation with either of the neutrino parameters. This outcome was expected since baryonic feedback effects are mostly relevant at $\ell>1500$, thus, beyond the range fitted in the pessimistic scenario. 
Nevertheless, the variation of the $\log_{10}M_{\rm c}$ parameter slightly degrades the sensitivity to the neutrino mass: more specifically, the 95\% upper bound on $\smnu$ increases by about 10\%. As already pointed out in \cite{SpurioMancini:2023mpt}, if the analysis was extended to $\ell=3000,\,5000$, the degradation in sensitivity to the neutrino mass due to the inclusion of baryonic feedback effects would be more dramatic.

\end{appendix}

\end{document}